%% file: arXiv_v2.tex
\documentclass[%
twocolumn,
 superscriptaddress,
preprintnumbers,
nofootinbib,
 amsmath,amssymb,
 aps,
longbibliography,
notitlepage
]{revtex4-1}

\usepackage{graphicx}
\usepackage{hyperref}
\usepackage{color}
\usepackage{braket}
\usepackage{bm}
\usepackage[normalem]{ulem} 
\usepackage[dvipsnames]{xcolor}

\usepackage{array}

\renewcommand{\arraystretch}{2.0}
\usepackage{makecell}

\usepackage{tikz}
\pgfdeclarelayer{bg}
\pgfsetlayers{bg,main}
\usetikzlibrary{decorations.pathmorphing}
\usetikzlibrary{arrows.meta}
\tikzset{snake it/.style={decorate, decoration=snake}}
\usetikzlibrary{calc}
\usepackage{tikz-3dplot} 
\tdplotsetmaincoords{70}{120}  
\tdplotsetrotatedcoords{0}{0}{0}

\newcommand{\green}{\color [rgb]{0,0,0}}

\begin{document}

\preprint{UT-Komaba/24-2}

\title{
 Anomaly inflow, dualities, and quantum simulation \\ of abelian lattice gauge
theories induced by measurements
}

\author{Takuya Okuda}

\email{takuya@hep1.c.u-tokyo.ac.jp}

\affiliation{Graduate School of Arts and Sciences, University of Tokyo\\
Komaba, Meguro-ku, Tokyo 153-8902, Japan}

\author{Aswin Parayil Mana}

\email{aswin.parayilmana@stonybrook.edu}

\affiliation{C.N. Yang Institute for Theoretical Physics \& Department of Physics and Astronomy, Stony Brook University, Stony Brook, NY 11794, USA}

\author{Hiroki Sukeno}

\email{hiroki.sukeno@gmail.com}

\affiliation{C.N. Yang Institute for Theoretical Physics \& Department of Physics and Astronomy, Stony Brook University, Stony Brook, NY 11794, USA}

\begin{abstract}
Previous work~[SciPost Phys. {\bf14}, 129 (2023)] has demonstrated that quantum simulation of abelian lattice gauge theories (Wegner models including the toric code in a limit) in general dimensions can be achieved by local adaptive measurements on symmetry-protected topological (SPT) states with higher-form generalized global symmetries. The entanglement structure of the resource SPT state reflects the geometric structure of the gauge theory.
In this work, we explicitly demonstrate the anomaly inflow mechanism between the deconfining phase of the simulated gauge theory on the boundary and the SPT state in the bulk, by showing that the anomalous gauge variation of the boundary state obtained by bulk measurement matches that of the bulk theory.
Moreover, we construct the resource state and the measurement pattern for the measurement-based quantum simulation of a lattice gauge theory with a matter field (Fradkin-Shenker model), where a simple scheme to protect gauge invariance of the simulated state against errors is proposed. We further consider taking an overlap between the wave function of the resource state for lattice gauge theories and that of a parameterized product state, and we derive precise dualities between partition functions with insertion of defects corresponding to gauging higher-form global symmetries, as well as measurement-induced phases where states induced by a partial overlap possess different (symmetry-protected) topological orders. 
Measurement-assisted operators to dualize quantum Hamiltonians of
lattice gauge theories and their non-invertibility are also presented.
\end{abstract}

\def\thefootnote{}\footnotetext{The author names are listed alphabetically.}

\maketitle

\def\thefootnote{\arabic{footnote}}
\setcounter{footnote}{0}

\tableofcontents

\section{Introduction}

There has been growing recognition that adaptive mid-circuit measurement of a quantum state is a valuable technique.
The applications include quantum computation~\cite{raussendorf2001one, raussendorf2003mbqc-on-CS,NIELSEN2006147, briegel2009measurement}, the preparation of topologically ordered ground states~\cite{piroli2021quantum, https://doi.org/10.48550/arxiv.2112.01519, verresen2021efficiently, 2022arXiv220501933B, 2022PRXQ....3d0337L, 2023PRXQ....4b0339T}, and triggering phase transitions in quantum many-body systems~\cite{2019PhRvX...9c1009S, 2018PhRvB..98t5136L,2019PhRvB..99v4307C}.

On the other hand, the simulation of dynamics in  quantum many-body systems such as lattice gauge theories is deemed an important application of quantum computers~\cite{feynman2018simulating, lloyd1996universal, 2012JordanLeePreskill, 2013AnP...525..777W, 2016ConPh..57..388D, 2016RPPh...79a4401Z, 2016Natur.534..516M, 2020EPJD...74..165B,  2022RSPTA.38010069Z,2023arXiv231012201H,2023PRXQ....4b7001B, 2023CmPhy...6..127H}. 
Gauge theories are important in both high-energy and condensed matter physics, describing the dynamics of elementary particles in the former and providing effective descriptions of phases of matter in the latter.
The work~\cite{2023ScPP...14..129S} presented a quantum simulation scheme of abelian lattice gauge theories where the discrete time evolution is driven by the adaptive measurement of an entangled resource state.
The resource state for the measurement-based quantum simulation (MBQS) is a cluster state~\cite{2001PhRvL..86..910B,raussendorf2003mbqc-on-CS, 2004PhRvA..69f2311H} whose entanglement pattern reflects the geometric structure of the underlying classical statistical model, or the Euclidean path integral of the simulated theory.
The number of necessary qubits in MBQS is proportional to the number of time steps in the simulation.
The measurement-based simulation scheme may have an advantage over the circuit-based scheme if one-qubit measurements can be implemented faster than entangling gates and if sufficiently many qubits are available.

The resource state has a symmetry-protected topological (SPT) order~\cite{gu2009tensor,pollmann2012symmetry,pollmann2010entanglement,chen2011classification,schuch2011classifying, chen2011complete, chen2011two, chen2013symmetry} with higher-form generalized global symmetries~\cite{Gaiotto:2014kfa, Yoshida:2015cia}.
In general, an SPT state has a close connection to the physics of anomalies in quantum field theories.
Namely, when an SPT state is placed on a spacetime with a boundary, the quantum field theory on the boundary may have a quantum anomaly that is canceled by the SPT state via a cancellation mechanism known as anomaly inflow~\cite{Callan:1984sa}.

The aim of this paper is two-fold.  
First, we wish to deepen our understanding of how measurement relates simulated gauge theories and their MBQS resource states. 
We study the relations in terms of the anomaly inflow and dualities. 
Second, we generalize the idea of the MBQS to broader models beyond pure lattice gauge theories.

We demonstrate an anomaly inflow between the deconfining phase of the simulated gauge theory (Wegner model~$M_{d,n}$~\cite{Wegner,kogut1979introduction}) with anomalous symmetries and the resource state used for the MBQS of the gauge theory by relating the bulk and boundary systems by measurement.
We also reproduce, within Hamiltonian lattice formulations, the result that the bulk and boundary partition functions defined as functionals of defect world-volumes are given in terms of the intersection and linking numbers, respectively.\footnote{%
See~\cite{Putrov:2016qdo} for a continuum approach to the study of linking invariants in topological phases of matter.}

Moreover, we extend the simulation scheme of~\cite{2023ScPP...14..129S} to the MBQS of the Fradkin-Shenker model~\cite{PhysRevD.19.3682} which is a gauge theory that includes matter degrees of freedom.
We propose a method to protect the gauge invariance of the Fradkin-Shenker model in the presence of noise, where the correction method is based on properties of the resource state.
 Our method uses the idea of quantum error correction, where measurement outcomes in the MBQS are used to construct correction operators. The Wegner model~$M_{d,n}$ is a lattice model in $d$ spacetime dimensions with an $(n-2)$-form gauge symmetry for $n\geq 2$.  It includes such models as Kitaev's toric code, gauge theories in $3+1$ dimensions, and Ising models~$(n=1)$ in general dimensions, which are important in condensed matter and high-energy physics.
The Fradkin-Shenker model exhibits a rich phase structure some of whose features are shared by gauge-matter systems with more general gauge symmetries~and representations~\cite{fradkin_2013}.
The model is also relevant for the description of quantum liquid crystals~\cite{BEEKMAN20171}.
Thus, our results on anomaly inflow and MBQS for these models provide foundations upon which further progress can be based.

Another topic of this paper is the relation between the MBQS resource state and dualities.
The partition function of the classical spin model underlying $M_{d,n}$ can be expressed as an overlap involving the MBQS resource state~\cite{2023ScPP...14..129S}.
We show that dualities in classical abelian lattice gauge theories can be derived by applying the Hadamard transform on the resource state, generalizing~\cite{PhysRevLett.98.117207} that treated the self-dual Ising model in $1+1$ dimensions.
We also study the implementation of the Kramers-Wannier transform on the quantum models in terms of the unitary entangling operator that constructs the resource state and measurements. Our construction provides many lattice examples of non-invertible duality defects as well as condensation defects~\cite{Roumpedakis:2022aik}, generalizing those in the Ising chain and the ($3+1$)-dimensional gauge theory~\cite{2016JPhA...49I4001A,2022PTEP.2022a3B03K}.

We present a formula expressing the statistical partition function (the Euclidean path integral) of the Fradkin-Shenker model as an overlap between the MBQS resource state and a product state with parameters that are related to coupling constants. We also argue that the wave function obtained by partially taking an overlap between the MBQS resource state and the product state in the bulk induces different states at the boundary that belong to different (symmetry-protected) topological phases depending on the parameter in the product state. For certain limits of parameters in the above overlap product state, we can precisely determine the resulting boundary state: the product state, the 2d generalized cluster state, and the 2d toric code state. 
Viewing the bulk product state as measurement bases, our finding can be interpreted as a type of measurement-induced phases.
 
This paper is organized as follows. In Section~\ref{sec:inflow}, we discuss in detail the anomaly inflow mechanism for the Wegner models and their MBQS resource states.
In Section~\ref{sec:dualities}, we study the Kramers-Wannier dualities in the Wegner models.
We re-derive the KW dualities for the classical statistical partition functions using the Hadamard transformation on the MBQS resource state. In Section~\ref{sec:Fradkin-Shenker}, we present the generalization of MBQS to the Fradkin-Shenker model. 
In Section~\ref{sec:duality-phases-FS}, we study dualities in the Fradkin-Shenker model based on a cluster state as well as measurement-induced phases.
We conclude the paper with discussion in Section~\ref{sec:discussion}.
In Appendix~\ref{sec:inflow-continuum}, we review the continuum description of the gauge theories and the SPT theories corresponding to the Wegner models and discuss the anomaly inflow between them. In Appendix~\ref{sec:technical-FS-gauge-protection}, we provide technical details in protecting the gauge invariance of the Fradkin-Shenker model.

\section{Anomaly inflow in Wegner models}
\label{sec:inflow}

In this section, we  demonstrate the anomaly inflow between the deconfining phase of the Wegner model~$M_{d,n}$ and the corresponding MBQS resource state.
 We do this by constructing the partition functions of the two models as functionals of defect world-volumes, which we show are given by the linking and intersection numbers, respectively.

\subsection{Review of the MBQS resource state}
 We begin by reviewing the resource state for the MBQS of the Wegner model~$M_{d,n}$.
We use the machinery of algebraic topology.
 
Let us consider a $(d-1)$-dimensional hypercubic lattice.
Let $\Delta_0$ be the set of 0-cells (vertices), $\Delta_1$ the set of 1-cells (edges), $\Delta_2$ the set of 2-cells (faces), and so on.
The degrees of freedom of $M_{d,n}$ live on ($n-1$)-cells. We write $C_k$ ($k=0,...,d-1$) for the group of $k$-chains $c_k$ with $\mathbb{Z}_2$ coefficients, which are the formal linear combinations
\begin{align}
c_k = \sum_{\sigma_k \in \Delta_k } a(c_k ; \sigma_k) \sigma_k \, , 
\end{align}
with $a(c_k ; \sigma_k) = \{0,1 \text{ mod }2\}$ the incidence number. 
We have the boundary operator, which is a linear map $\partial: C_{i+1} \rightarrow C_{i}$ such that $\partial \sigma_{i+1}$ is the sum of the $i$-cells at the boundary of $\sigma_{i+1}$. 
We have a chain complex
\begin{align}
C_{d-1} \overset{\partial}{\longrightarrow} C_{d-2} \overset{\partial}{\longrightarrow} \cdots \overset{\partial}{\longrightarrow} C_0 \,  ,
\end{align}
with $\partial^2 =0$.
Similarly, we have the dual hypercubic lattice, the dual cells ($\Delta_{d-1}\simeq \Delta^*_0$, $\Delta_{d-2}\simeq \Delta^*_1$, etc.), and the dual chain complex
\begin{align}
C^*_{d-1} \overset{\partial^*}{\longrightarrow} C^*_{d-2} \overset{\partial^*}{\longrightarrow} \cdots \overset{\partial^*}{\longrightarrow} C^*_0 \,  ,
\end{align}
with $(\partial^*)^2 =0$.
 Between $c_i \in C_i$ and $c^*_{d-1-i} \in C^*_{d-1-i}$, we define the intersection number
\begin{equation}
 \# (c_i \cap c^*_{d-1-i}) := \sum_{\sigma_i \in \Delta_i} a(c_i;\sigma_i) a({\green c^*_{d-1-i}; \sigma_{d-1-i}^*}) \,,   
\end{equation}
where {\green the cells} $\sigma_i$ and  {\green $\sigma_{d-1-i}^*$ on the original and the dual lattices} are identified. Generalization to the $G=\mathbb{Z}_N$ modules can be done by an appropriate definition of boundary operators which takes into account the orientation of cells, see Ref.~\cite{2023ScPP...14..129S}.

We will often place qubits on $i$-cells $\sigma_i \in \Delta_i$ ($i=0,...,d-1$).
We denote an operator $A$ on the qubit at $\sigma_i$ as $A(\sigma_i)$. 
As a useful shorthand notation, we define for a chain $c_i \in C_i$
\begin{align}
A(c_i) = \prod_{\sigma_i \in \Delta_i} \big(A(\sigma_i) \big)^{a(c_i;\sigma_i)} \, . 
\end{align}
This notation is also used for dual chains ($c^*_i \in C^*_i$ etc.) as well as for bulk chains ($\bm c_i \in \bm C_i$ or $\bm c^*_i \in \bm C^*_i$ etc.) below.

The Wegner model $M_{d,n}$ in the quantum Hamiltonian formulation is defined with the Hamiltonian
\begin{equation}\label{eq:Wegner-Hamiltonian}
H_\text{Weg}= 
- \sum_{\sigma_{n-1}} X(\sigma_{n-1})
- \lambda\sum_{\sigma_n} Z(\partial \sigma_n) 
\end{equation}
and the Gauss law constraint%
\footnote{%
\green Here $\sigma_{n-2}$ is identified with $\sigma^*_{d+1-n}$.
}
\begin{equation}
    X( \partial^* \sigma_{n-2}) = 1 \,.
\end{equation}
 The real parameter~$\lambda$ is a coupling constant.

To describe the resource state for the MBQS of the Wegner model~$M_{d,n}$,
we introduce the $d$-dimensional ({\it i.e.}, one-higher dimensional) hypercubic lattice and a chain complex on it. 
In MBQS, the value of the $d$-th coordinate $x_d$ specifies the layer where the simulated state is defined. 
We use the notation where the bold font denotes an object in the bulk. 
(See Table~\ref{tab:cell-notations} for a summary of font conventions.)
A cell $\bm\sigma_i$ inside a layer $x_d = j$ is of the form
\begin{align}
\bm\sigma_i = \sigma_i \times \{j\} \, ,
\end{align}
while a cell $\bm\sigma_i$ extending in the $d$-th direction takes the form
\begin{align}
\bm\sigma_i = \sigma_{i-1} \times [j,j+1] \, , 
\end{align}
where $\{j\}$ is a point in the $d$-th direction and $[j,j+1]$ is an interval in the $d$-th direction.
Using similar definitions, we have the bulk chain complexes:
\begin{align}
&\bm C_{d} \overset{\bm \partial}{\longrightarrow} \bm C_{d-1} \overset{\bm \partial}{\longrightarrow} \cdots \overset{\bm \partial}{\longrightarrow} \bm C_0 \,  ,
\label{eq:bulk-chain-complex-primal}
\\
&\bm C^*_{d} \overset{\bm \partial^*}{\longrightarrow} \bm C^*_{d-1} \overset{\bm \partial^*}{\longrightarrow} \cdots \overset{\bm \partial^*}{\longrightarrow} \bm C^*_0 \,  . \label{eq:bulk-chain-complex-dual}
\end{align}
The bulk chain complex is related to the boundary chain complex as $\bm{C}_n=C_n\otimes C_0(M_d)\oplus C_{n-1}\otimes C_1(M_d)$, where $M_d$ is the one-dimensional space parametrized by $x_d$.
The differentials are related as $\bm\partial = \partial\otimes \mathrm{id} + \mathrm{id}\otimes \partial$.

\begin{table*}[]
    \centering
\renewcommand{\arraystretch}{1.3}
    
    \begin{tabular}{c||c|c|c|c|c}
cell complex & dimension & coordinates & chain complex & font for chains and cycles & differential \\ 
\hline
boundary (space)    & $d-1$ & $x_1,\ldots,x_{d-1}$ & $C$ & italic ($c_i,z_i$) & $\partial$\\
bulk (space) & $d$ & $x_1,\ldots,x_{d-1},x_{d}$ &  $\bm C = C\otimes C(M_d)$ & bold italic ($\bm c_i,\bm z_i$) & $\bm\partial$\\
boundary (spacetime)  & $d$ & $x_1,\ldots,x_{d-1},\tau$ & $C\otimes C(M_\tau)$ & roman ($\mathrm{c}_i,\mathrm{z}_i$) & $\partial_\text{tot}$ \\
bulk (spacetime)  & $d+1$ & $x_1,\ldots,x_{d-1},x_{d},\tau$ &  
$ \bm C \otimes C(M_\tau)$ & bold roman ($\mathbf{c}_i,\mathbf{z}_i$) & $\bm\partial_\text{tot}$
    \end{tabular}
   
\caption{ Summary of notations for the cell complexes.
The four chain complexes are related by tensor multiplication.
}
    \label{tab:cell-notations}
\end{table*}

The resource state (called the generalized cluster state in \cite{2023ScPP...14..129S}) $|\text{gCS}_{(d,n)}\rangle$ is defined explicitly as
\begin{align} \label{eq:gCS-def}
&|\text{gCS}_{(d,n)}\rangle = \mathcal{U}_{CZ}
(|+\rangle^{\otimes \bm \Delta_{n}} |+\rangle^{\otimes \bm\Delta_{n-1} }
)
\, , \\
& \mathcal{U}_{CZ} = \prod_{ \substack{ \bm\sigma_n  \in \bm\Delta_{n}   \\ \bm\sigma_{n-1}\in \bm\Delta_{n-1}} } CZ^{a(\bm \delta \bm\sigma_n; \bm\sigma_{n-1})}_{\bm \sigma_n, \bm \sigma_{n-1}} \, .
\end{align}
 Here $CZ_{ab} = |0\rangle\langle 0|_a \mathrm{id}_b +|1\rangle\langle 1|_a Z_b$ denotes the controlled Z gate, and $|+\rangle$ the eigenstate of $X$ with eigenvalue $+1$.
The stabilizers for $|\text{gCS}_{(d,n)}\rangle$ are then given by
\begin{align} \label{eq:stabilizers-d-n-1}
K(\bm \sigma_n) &= X(\bm \sigma_n) Z(\bm \partial \bm\sigma_n) \qquad  (\bm\sigma_{n} \in \bm \Delta_{n}) \, , \\
K(\bm \sigma_{n-1}) &= X(\bm \sigma_{n-1}) Z(\bm \partial^* \bm\sigma_{n-1}) \qquad  (\bm \sigma_{n-1} \in \bm \Delta_{n-1}) \, . \label{eq:stabilizers-d-n-2}
\end{align}
For instance, when $(d,n)=(3,2)$, this is the  Raussendorf-Bravyi-Harrington state~\cite{RBH} with qubits on 2-cells (faces) and 1-cells (edges) of the 3d cubic lattice.

\subsection{Defects in the boundary theory}
\label{sec:defects-lattice}

The low-energy limit of the Wegner model $M_{d,n}$ for large $\lambda>0$ (deconfining phase) is  described by what we call the generalized toric code $TC_{d,n}$ and is defined by the Hamiltonian
\begin{equation} \label{eq:Hamiltonian-generalized-toric}
    H= 
- \sum_{\sigma_n} Z(\partial \sigma_n)  -  \sum_{\sigma_{n-2}} X( \partial^* \sigma_{n-2})
\,.
\end{equation}

In the current notation, the original toric code~\cite{Kitaev_2003} is $TC_{3,2}$, and $TC_{d,2}$ is the $(d-1)$-dimensional toric code.
Let us denote by $G^{[q]}$ a $q$-form $G$ symmetry~\cite{Gaiotto:2014kfa}.
The symmetries of~$TC_{d,n}$ are generated by 
\begin{equation} \label{eq:sym-generators-Wegner}
\begin{aligned}
Z(z_{n-1}) \quad&:\quad \text{$\mathbb{Z}_2$ $(d-n)$-form symmetry $\mathbb{Z}_2^{[d-n]}$,
}
\\
  X( z^*_{d-n}) \quad &:\quad \text{$\mathbb{Z}_2$ $(n-1)$-form symmetry $\mathbb{Z}_2^{[n-1]}$.
  }
\end{aligned}
\end{equation}

They are supported on an $(n-1)$-cycle and a dual $(d-n)$-cycle
with $\partial z_{n-1}=0$ and $\partial^* z^*_{d-n} =0$, respectively, and are nothing but the logical operators when the model is regarded as a quantum code.
In the Wegner model with finite couplings, the first symmetry is violated by the first term in the Hamiltonian~(\ref{eq:Wegner-Hamiltonian}), but it appears as an emergent symmetry at low energies in the deconfining phase $\lambda\gg 1$.
The two symmetry generators do not commute when the cycles $z_{n-1}$ and $ z^*_{d-n}$ intersect.
This signifies a mixed anomaly between the symmetries.

One can consider excited states, {\it i.e.,} the simultaneous eigenstates of $Z(\partial \sigma_n)$ and $X(\partial^* \sigma_{n-2})$ (when $n\geq2$), some of whose eigenvalues are $-1$:
\begin{align}
Z(\partial \sigma_n) |\mathcal{E}\rangle &= (-1)^{\#(\sigma_n\cap z^*_{d-n-1})}|\mathcal{E}\rangle 
\,, \label{eq:Z-partial-sigma-n-E}\\
X(\partial^* \sigma_{n-2}) |\mathcal{E}\rangle &= (-1)^{a(z_{n-2},\sigma_{n-2})}|\mathcal{E}\rangle \, \label{eq:Z-partial-star-sigma-n-2-E}
\end{align}
for all $\sigma_n\in\Delta_n$, $\sigma_{n-2}\in\Delta_{n-2}$.
Here an $(n-2)$-chain~$z_{n-2}$ and a dual $(d-n-1)$-chain~$z^*_{d-n-1}$ specify the spatial profiles of the excitations in the state, which we write $|\mathcal{E}(z_{n-2},z^*_{d-n-1})\rangle \equiv |\mathcal{E}\rangle$ keeping in mind that the excitations alone do not specify the state uniquely.
(On a topologically non-trivial spatial manifold, the ground state is not unique even without excitations.)
It follows from these equations (using, for example, $1=Z(\partial^2 c_{n+1})=\prod_{\sigma_n} Z(\partial \sigma_n)^{a(\partial c_{n+1};\sigma_n)}$) that $\partial z_{n-2}=\partial^* z^*_{d-n-1} =0$,  
{\it i.e.}, $z_{n-2}$ and $z^*_{d-n-1}$ are a cycle and a dual cycle.
(Moreover, $z_{n-2}$ and $z^*_{d-n-1}$ are homologically trivial if we impose the periodic boundary conditions due to the non-degeneracy of the intersection pairing.)
Equation~(\ref{eq:Z-partial-sigma-n-E}) also implies that when we modify the support of the symmetry generator in~(\ref{eq:sym-generators-Wegner}) to $z_{n-1}+\partial c_n$, we get an extra phase $(-1)^{\#(c_n\cap z^*_{d-n-1})}$.
A similar statement holds for~(\ref{eq:Z-partial-star-sigma-n-2-E}).

The spatial profile of an excitation can be modified by $X$ and $Z$ operators supported on chains:
\begin{widetext}
\begin{align}
Z(c_{n-1}) |\mathcal{E}(z_{n-2},z^*_{d-n-1})\rangle
&\propto
|\mathcal{E}(z_{n-2}+\partial c_{n-1} ,z^*_{d-n-1})\rangle \,,
\\
X(c^*_{d-n}) |\mathcal{E}(z_{n-2},z^*_{d-n-1})\rangle
&\propto
|\mathcal{E}(z_{n-2} ,z^*_{d-n-1}+\partial^* c^*_{d-n})\rangle \,.
\end{align}
\end{widetext}
These properties imply that the excitations and the symmetry generators are different versions of the same objects, where the former contain the time direction in their world-volumes while the latter are purely spatial.
The two versions of defects can be combined into cycles and dual cycles in a larger chain complex, as we will see in Section~\ref{subsec:anomaly-inflow}.

We summarize the boundary and bulk models in Table~\ref{tab:models}.

\begin{table*}[t]
    \centering 
\begin{tabular}{c||c|c|c|c|c}
& lattice model & cells with qubits & symmetries & partition function & order
\\
\hline
boundary & $TC_{d,n}$ (\ref{eq:Hamiltonian-generalized-toric}) & $\sigma_{n-1}\in \Delta_{n-1}$ & $\mathbb{Z}_2^{[d-n]} \times \mathbb{Z}_2^{[n-1]}$ (\ref{eq:sym-generators-Wegner}) & $\text{GSD}\cdot(-1)^\text{linking number}$ (\ref{eq:Zbdry-linking}) & {\green topological}  \\
bulk & $\mathrm{gCS}_{d,n}$ (\ref{eq:gCS-Hamiltonian}) & $\bm\sigma_{n-1}\in\bm\Delta_{n-1}$, $\bm\sigma_n\in\bm\Delta_n$ & $\mathbb{Z}_2^{[d-n]} \times \mathbb{Z}_2^{[n-1]}$ (\ref{eq:gCS-symm-generators}) & $(-1)^\text{intersection number}$ (\ref{eq:partition-function-gCS})(\ref{eq:bulk-rel-intersection}) & {\green SPT}
    \end{tabular}    
\caption{ Summary of the boundary and bulk models.  
The partition functions are given in terms of the ground state degeneracy (GSD) as well as the linking and intersection numbers of defect world-volumes.}
    \label{tab:models}
\end{table*}

\subsection{Symmetry protected topological state
}\label{sec:SPT}

We now turn to the lattice description of the bulk SPT phase.
In~\cite{2023ScPP...14..129S}, two of the current authors claimed that the generalized cluster state $|{\rm gCS}_{(d,n)}\rangle$ defined in~(\ref{eq:gCS-def}) is in an SPT phase\footnote{%
See~\cite{Yoshida:2015cia,2022arXiv220811699L} for earlier studies on special values of $(d,n)$.}.
We now show that $|{\rm gCS}_{(d,n)}\rangle$
provides a lattice realization of a continuum SPT theory, defined by the classical action~(\ref{eq:Z-SPT-Wegner-partition-function}) in Appendix~\ref{sec:inflow-continuum}.
The state $|{\rm gCS}_{(d,n)}\rangle$ 
is the ground state of the cluster Hamiltonian\footnote{%
This type of Hamiltonian was studied in~\cite{2022arXiv220811699L} as an example of the decorated domain wall construction in a related context.}
\begin{equation}\label{eq:gCS-Hamiltonian}
\begin{aligned}
H_{\rm gCS} &= -\sum_{\bm{\sigma}_n\in\bm{\Delta}_n} X(\bm{\sigma}_n) Z(\bm\partial \bm{\sigma}_n) 
\\
&\qquad
-\sum_{\bm{\sigma}_{n-1}\in\bm{\Delta}_{n-1}} X(\bm{\sigma}_{n-1}) Z(\bm\partial^* \bm{\sigma}_{n-1}) \,.
\end{aligned}
\end{equation}
The symmetries $ \mathbb{Z}_2^{[d-n]}$ and $\mathbb{Z}_2^{[n-1]}$ of the system are respectively generated by
\begin{equation} \label{eq:gCS-symm-generators}
 X(\bm z_n)    \qquad \text{and} \qquad X(\bm z^*_{d-n+1}) \,,
\end{equation}
where $\bm z_n$ and $\bm z^*_{d-n+1}$ are $n$- and dual $(d-n+1)$-cycles satisfying $\bm\partial \bm z_n=\bm  \partial^*\bm z^*_{d-n+1}=0$.
The operators in~(\ref{eq:gCS-symm-generators}) can be viewed as topological defects oriented purely in the space-like directions.

One piece of evidence for the SPT order
is the appearance of a projective representation
when defects are restricted to the boundary~\cite{2023ScPP...14..129S}.
Another piece of evidence is provided by the torus partition function. For this, we define the symmetric excited state
\begin{equation}\label{eq:excitation-def}
    |\mathcal{E}(\bm z_{n-1},\bm z^*_{d-n}) \rangle := Z(\bm z_{n-1}) Z(\bm z^*_{d-n}) |\text{gCS}_{(d,n)}\rangle \,,
\end{equation}
where $\bm z_{n-1}$ and $\bm z^*_{d-n}$ are $(n-1)$- and dual $(d-n)$-cycles satisfying $\bm \partial \bm z_{n-1}=0$ and $\bm \partial^* \bm z^*_{d-n}=0$.
This state is invariant under the action of the symmetry generators of the forms $X(\bm\partial^*\bm c^*_{d-n+2})$ and $X(\bm\partial\bm c_{n+1})$, {\it i.e., } the generators supported on (dual) boundary chains.
Any excited state with this symmetry property is of the form~(\ref{eq:excitation-def}).
(See~\cite{Roberts:2018ruo} for a discussion of such excitations in the $d=3$, $n=2$ case.)
The excitations along these cycles are a manifestation of the same defects as represented by~(\ref{eq:gCS-symm-generators}).
Indeed the locations $\bm z_{n-1}$ and $\bm z^*_{d-n}$ of the excitations can be changed to $\bm z_{n-1}+\bm \partial \bm c_n$
and $\bm z^*_{d-n}+\bm \partial^* \bm c^*_{d-n+1}$ by applying $X(\bm c_n) $  and $X(\bm c^*_{d-n+1}) $ to the state, where $\bm c_n$ and $\bm c^*_{d-n+1}$ are $n$- and dual $(d-n+1)$-cochains. 
This means that the symmetric excitations~(\ref{eq:excitation-def}) naturally combine with the symmetry generators to form defects with general world-volumes in spacetime.

Let us consider the partition function of the system in spacetime $T^{d+1}$ in the presence of spatial and temporal defects.
Since we are interested in the low-energy SPT state, the trace defining the partition function is taken in the 1-dimensional space spanned by $  |\bm{\mathcal{E}}(\bm z_{n-1},\bm z^*_{d-n}) \rangle$.
The partition function, or more precisely its sign,
is then
\begin{align}
& \quad Z_\text{gCS}[\mathbf{z}_n,\mathbf{z}^*_{d-n+1}]
\nonumber \\
&
  = 
\langle \bm{\mathcal{E}}(\bm z_{n-1},\bm z^*_{d-n}) |  X(\bm z_n)   X(\bm z^*_{d-n+1})  
\nonumber \\
&\qquad\qquad\qquad\qquad\qquad \times
|\bm{\mathcal{E}}(\bm z_{n-1},\bm z^*_{d-n}) \rangle \label{eq:partition-function-gCS-insertion}
\\
&= (-1)^{\# (\mathbf{z}_n \cap \mathbf{z}^*_{d-n+1})} \,,\label{eq:partition-function-gCS}
\end{align}
where 
\begin{align}\label{eq:bfz-bmz}
    \mathbf{z}_n &= \bm z_n \times \{\text{point}\} + \bm z_{n-1} \times S^1  \,,\\   
    \quad
    \mathbf{z}^*_{d-n+1} &= \bm z^*_{d-n}  \times S^1 
    +\bm z^*_{d-n+1}
    \times \{\text{point}\}   
    \,.
\end{align}
The intersection number~$\# (\mathbf{z}_n \cap \mathbf{z}^*_{d-n+1})$ in~(\ref{eq:partition-function-gCS}) arises from the $X$ operators commuting through the $Z$ operators in~(\ref{eq:excitation-def}).
See~\cite{Seifnashri:2023dpa} for a formulation of defects similar to~(\ref{eq:partition-function-gCS-insertion}).
In Section~\ref{eq:bulk-partition-function-with-boundary}, we will generalize the expressions for $\mathbf{z}_n$ and $\mathbf{z}^*_{d-n+1}$ in~(\ref{eq:bfz-bmz}) by dividing $S^1$ into several segments, corresponding to more general $X$ operator insertions in~(\ref{eq:partition-function-gCS-insertion}) without changing the result (\ref{eq:partition-function-gCS}).

 See Appendix~\ref{sec:inflow-continuum} where we show the match between the lattice and continuum descriptions of the model.

\subsection{Anomaly inflow in the lattice model}\label{subsec:anomaly-inflow}

Here we exhibit the anomaly inflow using  defects.\footnote{%
The work~\cite{Roberts:2018ruo} contains a related discussion in the $(d,n)=(3,2)$ case with a focus on error correction.
The paper~\cite{Fidkowski2015RealizingAA} discusses a phenomenon similar to anomaly inflow involving a bulk long-range entangled state.
}
We use coordinates $(x_1,\ldots,x_d)$ for space and $\tau$ for the Euclidean time.
The bulk is the region~$0\leq x_d\leq L_d$
and the boundaries are at $x_d=0$ and $x_d=L_d$.
In the bulk, we have qubits on $n$-cells $\bm\sigma_n$ and ($n-1$)-cells $\bm\sigma_{n-1}$.
On the boundary, we have qubits on $n$-cells $\sigma_n$ and $(n-1)$-cells $\sigma_{n-1}$.
We will introduce cycles as the world-volumes of defects extended in the spacetime directions in the bulk and the boundary.
Then we will define the bulk and boundary partition functions.
We will compute their variations under deformations of the defects and show that they are identical, exhibiting anomaly inflow.
We will also explain that the bulk and boundary systems are related by measurement and post-selection.

\subsubsection{Defects in terms of a chain complex}

For the generalized cluster state $\mathrm{gCS}_{d,n}$, qubits are placed on the cells $\bm\sigma_n\in\bm\Delta_n$ and $\bm\sigma_{n-1}\in\bm\Delta_{n-1}$ of the $d$-dimensional hypercubic lattice.
We model the time by a circle~$M_\tau$ divided into $L_\tau$ intervals $[k,k+1]=\{\tau | k\leq\tau\leq k+1\}$ ($k=0,1,\ldots,L_\tau$).
The total spacetime $M_\text{tot}$ is modeled by a ($d+1$)-dimensional hypercubic lattice with periodic boundary conditions in directions $(\tau,x_1,\ldots,x_{d-1})$ and an open boundary condition in the $x_d$ direction ($0\leq x_d\leq L_d$).
We propose that the world-volume of the first type of defect is now given by a relative cycle, {\it i.e.}, a chain $\mathbf{z}_n^\text{rel} \in 
\bm{C}_{n}\otimes C_0(M_\tau)\oplus \bm{C}_{n-1}\otimes C_1(M_\tau)$ such that $\bm\partial_\text{tot} \mathbf{z}_n^\text{rel}$ is a chain on the boundary: 
\begin{equation}\label{eq:der-z-rel}
\bm\partial_\text{tot} \mathbf{z}_n^\text{rel} = \mathrm{z}_{n-1} \otimes \{x_d=0\} + \mathrm{z}'_{n-1}\otimes \{x_d=L_d\} 
\end{equation}
with
\begin{equation}
\mathrm{z}_{n-1} , \mathrm{z}'_{n-1}
 \in C_{n-1}\otimes C_0(M_\tau)\oplus C_{n-2}\otimes C_1(M_\tau) 
 \,,
\end{equation}
where $\bm\partial_\text{tot}=\bm\partial\otimes\text{id}+\text{id}\otimes\partial$ is the boundary operator on $M_\text{tot}$.
Since $\bm\partial_\text{tot}^2=0$, we have $( \partial\otimes\text{id}+\text{id}\otimes\partial)\mathrm{z}_{n-1}=( \partial\otimes\text{id}+\text{id}\otimes\partial) \mathrm{z}'_{n-1}=0$.
The dual cycle can be decomposed as~$\mathbf{z}_{d-n+1}^* = \mathrm{z}_{d-n}^*\otimes \{x_d=0\} + \mathrm{z}_{d-n}^{\prime*}\otimes \{x_d=L_d\}+\ldots$ .
We also propose that the world-volume of the second type of defect is given by a chain (dual cycle) $\mathbf{z}_{d-n+1}^*\in
 \bm{C}_{n}\otimes C_0(M_\tau)\oplus \bm{C}_{n-1}\otimes C_1(M_\tau)
$ such that $\bm\partial_\text{tot}^*\mathbf{z}_{d-n+1}^*=0$.
We will define the partition function of the bulk theory as a functional of $\mathbf{z}_n^\text{rel} $ and $\mathbf{z}_{d-n+1}^*$.
We will also define the partition function of the boundary theory at $x_d=0$ as a functional of $\mathrm{z}_{n-1} $ and $ \mathrm{z}_{d-n}^*$ (and do the same for the boundary theory at $x_d=L_d$).

The relative cycle can be decomposed into the space- and time-like components as
\begin{equation}
\begin{aligned}
\mathbf{z}_n^\text{rel} &= \sum_{k=0}^{L_\tau-1} \bm{c}_n^{(k)}\otimes\{\tau=k\}
\\
&\qquad\qquad
+ \sum_{k=0}^{L_\tau-1} \bm{z}_{n-1}^{\text{rel}(k)} \otimes \{k\leq \tau \leq k+1\} \,.
\end{aligned}
\end{equation}
Writing $\mathrm{z}_{n-1}=\sum_k c_{n-1}^{(k)}\otimes \{\tau=k\}+ \sum_k z_{n-2}^{(k)}\otimes\{k\leq \tau\leq k+1\}$ with $\partial c_{n-1}^{(k)} = z_{n-2}^{(k)}-z_{n-2}^{(k-1)}$  and similarly for $\mathrm{z}'_{n-1}$, we obtain from (\ref{eq:der-z-rel})
\begin{equation}
\bm\partial \bm z_{n-1}^{\text{rel}(k)} = z_{n-2}^{(k)} \otimes\{x_d=0\} + z_{n-2}^{\prime(k)} \otimes\{x_d=L_d\} \,,
\end{equation}
\begin{equation}\label{eq:z-rel-change}
\begin{aligned}
\bm z_{n-1}^{\text{rel}(k)} 
&= \bm z_{n-1}^{\text{rel}(k-1)} + \bm\partial \bm c_n^{(k)} 
\\
&\quad + c_{n-1}^{(k)}\otimes \{x_d=0\} + c_{n-1}^{\prime(k)}\otimes \{x_d=L_d\}  \,.
\end{aligned}
\end{equation}

The dual cycle can also be decomposed:
\begin{align}
\mathbf{z}^*_{d-n+1} &= \sum_{k=0}^{L_\tau-1} \bm{z}_{d-n}^{*(k)}\otimes \{\tau=k\} \\\nonumber
&\qquad + \sum_{k=0}^{L_\tau-1} \bm{c}_{d-n+1}^{*(k+1/2)} \otimes \{k\leq\tau\leq k+1\} \,.
\end{align}
Here we identify the point $\{\tau=k\}$ and the interval $\{k\leq\tau\leq k+1\}$ on the original lattice with the interval $\{k-1/2 \leq \tau \leq k+1/2\}$ and the the point $\{\tau=k+1/2\}$ on the dual lattice, respectively.
We have
\begin{equation}\label{eq:z-star-change}
\bm\partial^*\bm z_{d-n}^{*(k)}=0 \,,
\qquad
\bm z_{d-n}^{*(k+1)} = \bm z_{d-n}^{*(k)} + \bm\partial^* \bm c_{d-n+1}^{*(k+1/2)} \,.
\end{equation}
We interpret the cycles $\bm{z}_{n-1}^{\text{rel}(0)}$ and $\bm{z}_{d-n}^{*(0)}$ as specifying the initial state, while the chains lead us to insert
\begin{align}
\begin{split}
&X(\bm c_n^{(k)})\,, \  Z(c_{n-1}^{(k)}\otimes\{x_d=0\})\,,  \text{ and }\\
&\qquad Z(c_{n-1}^{\prime(k)}\otimes\{x_d=L_d\} )\text{ at } \tau = k \,,
\end{split}
\end{align}
as well as
\begin{equation}
X(\bm c_{d-n+1}^{*(k+1/2)}) \text{ at } \tau = k+1/2 \,.
\end{equation}
See Figure~\ref{fig:defect-fig} for an illustration.
As we will see below, such operator insertions induce time evolution and modify the excitations of the state.

\begin{figure*}
   \includegraphics[scale=1]{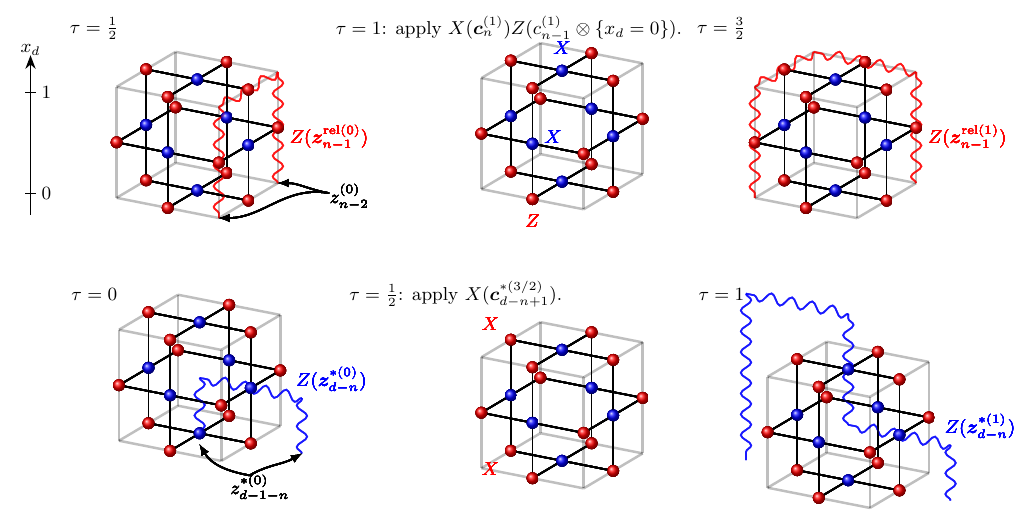}
    \caption{Operator insertions and motion of defects in the generalized cluster state. The figure is drawn for $(d,n)=(3,2)$. 
    }
    \label{fig:defect-fig}
\end{figure*}

\subsubsection{Bulk partition function in the presence of boundary}
\label{eq:bulk-partition-function-with-boundary}

 Let us define the state with excitations as 
\begin{equation}
 |\bm{\mathcal{E}}(\bm z_{n-1}^{\text{rel}},\bm z^{*}_{d-n}) \rangle := Z(\bm z_{n-1}^{\text{rel}}) Z(\bm z^{*}_{d-n}) |\text{gCS}_{(d,n)}\rangle    
\end{equation} 
for a general relative cycle~$ z_{n-1}^{\text{rel}}$ and a dual cycle $\bm z^{*}_{d-n}$, generalizing~(\ref{eq:excitation-def}).
We define the bulk partition function as
\begin{align}
&\quad\ Z_\text{bulk}[\mathbf{z}_n^\text{rel} ,\mathbf{z}_{d-n+1}^*] 
\nonumber \\
:&=
\left \langle \bm{\mathcal{E}}(\bm z_{n-1}^{\text{rel}(0)},\bm z_{d-n}^{*(0)}) \right|
\mathcal{O}_\text{ins} 
\left |\bm{\mathcal{E}}(\bm z_{n-1}^{\text{rel}(0)},\bm z_{d-n}^{*(0)}) \right\rangle
\\
&=
{\rm Tr} \left[
\mathcal{O}_\text{ins} \mathbb{P}(\bm{z}_{n-1}^{\text{rel}(0)},\bm{z}_{d-n}^{*(0)})
\right] \,,
\label{eq:bulk-partition-function-def-relative}
\end{align}
where
\begin{align}
\begin{split}
\label{eq:O-ins}
\mathcal{O}_\text{ins}  :=& \prod_{k=1}^{L_\tau} \left(X(\bm{c}_n^{(k)})   Z(c_{n-1}^{(k)}\otimes\{x_d=0\})\right. \\
&\left. \qquad \times Z(c_{n-1}^{\prime(k)}\otimes\{x_d=L_d\} )X(\bm c_{d-n+1}^{*(k-1/2)}) \right)
\end{split}
\end{align}
with $k$ increasing from right to left and
\begin{equation}\mathbb{P}(\bm{z}_{n-1}^{\text{rel}(0)},\bm{z}_{d-n}^{*(0)})
:=
\left |\bm{\mathcal{E}}(\bm z_{n-1}^{\text{rel}(0)},\bm z_{d-n}^{*(0)}) \right\rangle\left \langle \bm{\mathcal{E}}(\bm z_{n-1}^{\text{rel}(0)},\bm z_{d-n}^{*(0)}) \right| \,.
\end{equation}
 The definition~(\ref{eq:bulk-partition-function-def-relative}) of the partition function is natural because the operator~$\mathcal{O}_\text{ins}$ describes the time evolution of the system and the trace resembles the path integral for a quantum system.
Let us evaluate the partition function.
The inserted operators pick up signs due to anti-commutation of $X$ and $Z$, and modify the excitations of the state according to~(\ref{eq:z-rel-change}) and~(\ref{eq:z-star-change}).
The state comes back to the initial state, and the overall sign is given by the intersection number:
\begin{equation} \label{eq:bulk-rel-intersection}
Z_\text{bulk}[\mathbf{z}_n^\text{rel} ,\mathbf{z}_{d-n+1}^*]  = (-1)^{\#(\mathbf{z}_n^\text{rel}\cap \mathbf{z}_{d-n+1}^*)} \,.
\end{equation}
This generalizes the result~(\ref{eq:partition-function-gCS}), and is not a topological invariant due to the boundary.
Under the deformations
\begin{equation}
\mathbf{z}_n^\text{rel} \rightarrow \mathbf{z}_n^\text{rel} + \partial_\text{tot} \mathbf{c}_{n+1} +c_n\otimes\{x_d=0\} +c'_n\otimes\{x_d=L_d\}  
\end{equation}
and
\begin{equation}
\mathbf{z}_{d-n+1}^* \rightarrow \mathbf{z}_{d-n+1}^* + \bm\partial_\text{tot}^* \mathbf{c}^*_{d-n+2} \,,
\end{equation}
the partition function changes as
\begin{equation}\label{eq:variation-bulk}
\begin{aligned}
&\quad\ Z_\text{bulk}[\mathbf{z}_n^\text{rel} ,\mathbf{z}_{d-n+1}^*]  
\\
&\rightarrow
(-1)^{\#((c_n\otimes\{x_d=0\} +c'_n\otimes\{x_d=L_d\}  )\cap \mathbf{z}_{d-n+1}^*)}
\\
&\qquad\times
(-1)^{\#((\mathrm{z}_{n-1}\otimes\{x_d=0\} + \mathrm{z}'_{n-1}\otimes\{x_d=L_d\})\cap \mathbf{c}^*_{d-n+2})}
\\
&\qquad \times
Z_\text{bulk}[\mathbf{z}_n^\text{rel} ,\mathbf{z}_{d-n+1}^*]   \,.
\end{aligned}
\end{equation}

\subsubsection{Boundary partition function}
\label{sec:variation-boundary}

We focus on  the boundary at $x_d=0$ and consider the boundary theory (the generalized toric code $TC^{(2)}_{d,n}$) there.
Recall that we have the decomposition 
\begin{equation}\label{eq:zn-1-decomposition}
\begin{aligned}
\mathrm{z}_{n-1} &=\sum_k c_{n-1}^{(k)}\otimes \{\tau=k\} \\
&\qquad + \sum_k z_{n-2}^{(k)}\otimes\{k\leq \tau\leq k+1\}
\end{aligned}
\end{equation}
with 
\begin{equation}\label{eq:pczz}
\partial c_{n-1}^{(k)} = z_{n-2}^{(k)}-z_{n-2}^{(k-1)} \,.
\end{equation}\label{pscszszs}
We also decompose the dual cycle as $\mathrm{z}_{d-n}^*=\sum_k z_{d-1-n}^{*(k)}\otimes\{\tau= k\}+\sum_k c_{d-n}^{*(k+1/2)}\otimes\{k\leq\tau\leq k+1\}$ with
\begin{equation}
\partial^* c_{d-n}^{*(k+1/2)} = z_{d-1-n}^{*(k+1)} - z_{d-1-n}^{*(k)} \,.
\end{equation}
We define the partition function as a functional of defect world-volumes $\mathrm{z}_{n-1}$ and $\mathrm{z}_{d-n}^*$ 
\begin{align}
\label{eq:Zbdry-trace}
\begin{split}
&Z_\text{bdry}[\mathrm{z}_{n-1}, \mathrm{z}_{d-n}^*] :=  {\rm Tr}\left[
Z(c_{n-1}^{(L_\tau)}) X(c_{d-n}^{*(L_\tau-1/2)})\ldots \right.\\
&\left.\qquad\times  Z(c_{n-1}^{(1)})X(c_{d-n}^{*(1/2)}) \mathbb{P}(z_{n-2}^{(0)}, z_{d-1-n}^{*(0)})
\right] \,,
\end{split}
\end{align}
where the projector
\begin{equation}
\begin{aligned}    
& \quad\ \mathbb{P}(z_{n-2}^{(0)}, z_{d-1-n}^{*(0)})
\\
&:=
  \prod_{\sigma_{n-2} \in \Delta_{n-2}} \frac{1+ (-1)^{a(z_{n-2}^{(0)};\sigma_{n-2})} X(\partial^* \sigma_{n-2})}{2}
  \\
  &\quad\times
  \prod_{\sigma_n\in\Delta_n} \frac{1+(-1)^{\#(\sigma_n\cap z_{d-1-n}^{*(0)})}Z(\partial\sigma_n)}{2} 
\end{aligned}
\end{equation}
 specifies the initial configuration of excitations and the time-ordered insertion of operators inside the trace induce the motion of excitations.
 
 We claim that the boundary partition function transforms as
 \begin{align}
&\frac{
Z_\text{bdry}[\mathrm{z}_{n-1}+\partial_\text{tot} \mathrm{c}_n,\mathrm{z}_{d-n}^* ] 
}{
Z_\text{bdry}[\mathrm{z}_{n-1},\mathrm{z}_{d-n}^* ] }
= (-1)^{\#(\mathrm{c}_n\cap\mathrm{z}_{d-n}^*)} \, , \label{eq:bdry-anomaly-defect-1} \\
&\frac{
Z_\text{bdry}[\mathrm{z}_{n-1},\mathrm{z}_{d-n}^* +\partial^*_\text{tot} \mathrm{c}^*_{d-n+1}] 
}{
Z_\text{bdry}[\mathrm{z}_{n-1},\mathrm{z}_{d-n}^* ] 
}
= (-1)^{\#(\text{z}_{n-1} \cap \mathrm{c}^*_{d-n+1})} \, .
\label{eq:bdry-anomaly-defect-2} 
\end{align}
Here the differentials $\partial_\text{ tot}$ and $\partial^*_\text{ tot}$ are for the product of the spatial and temporal complexes, {\it i.e.}, they act as $\partial\otimes {\rm id}+{\rm id}\otimes \partial$ and $\partial^*\otimes {\rm id}+{\rm id}\otimes \partial^*$, respectively.
 For the first equality~(\ref{eq:bdry-anomaly-defect-1}), it is enough to consider separately the cases $\mathrm{c}_n = \sigma_n\otimes \{k\}$ and $\mathrm{c}_n = \sigma_{n-1}\otimes [k,k+1]$ by linearity.
 In the first case, the insertion changes by $Z(c_{n-1}^{(k)}) \rightarrow Z(\partial\sigma_n) Z(c_{n-1}^{(k)})$,  and the extra operator $Z(\partial\sigma_n)$ picks up the sign $(-1)^{\#(\sigma_n\cap z_{d-1-n}^{*(k)})}= (-1)^{\#(\mathrm{c}_n \cap \mathrm{z}_{d-n}^*)}$.
 In the second case, the insertion changes by $X(c_{d-n}^{*(k+1/2)}) \rightarrow Z(\sigma_{n-1} )X(c_{d-n}^{*(k+1/2)} )Z(\sigma_{n-1} ) $, giving an extra sign $(-1)^{\#(\sigma_{n-1}\cap c_{d-n}^{*(k+1/2)})}= (-1)^{\#(\mathrm{c}_n \cap \mathrm{z}_{d-n}^*)}$.
 The second equality~(\ref{eq:bdry-anomaly-defect-2}) can be shown similarly.
The anomalous variations (\ref{eq:bdry-anomaly-defect-1}) and (\ref{eq:bdry-anomaly-defect-2}) as well as those at the boundary $x_d=L_d$ are identical to the variations~(\ref{eq:variation-bulk}) of the bulk partition function.
This completes the demonstration of anomaly inflow.

While most of the above computations are valid with general boundary conditions in the $(x_1,\ldots,x_{d-1})$-directions, an even stronger statement holds for the periodic boundary conditions; for the boundary partition function to be non-vanishing, the cycles need to be homologically trivial ($\mathrm{z}_{n-1}=\partial_\text{ tot} \mathrm{h_n}$, $\mathrm{z}^*_{d-n} = \partial^*_\text{ tot} \mathrm{h}^*_{d-n+1}$ for some chains $\mathrm{h_n}$ and $\mathrm{h}^*_{d-n+1}$) and the partition function is given in terms of the linking number as
\begin{equation} \label{eq:Zbdry-linking}
Z_\text{bdry}[\mathrm{z}_{n-1},\mathrm{z}_{d-n}^* ] 
=
\text{GSD} \cdot
(-1)^{\#(\mathrm{h}_n\cap\partial^*_\text{ tot}\mathrm{h}^*_{d-n+1})}
\,,
\end{equation}
where the
ground state degeneracy is given by
\begin{equation}
\log_2 \text{GSD} =  
\frac{(d-1)!}{(n-1)!(d-n)!} \,.
\end{equation}

To prove the formula~(\ref{eq:Zbdry-linking}) for the boundary partition function, we recall from the discussion below~(\ref{eq:Z-partial-star-sigma-n-2-E}) that we can write 
\begin{equation}
z^{(k)}_{n-2} = \partial \tilde c^{(k)}_{n-1}   
\end{equation}
for some chains $\tilde c^{(k)}_{n-1}$.
From the relation~(\ref{eq:pczz}) we get 
\begin{equation}
c^{(k)}_{n-1} = \tilde{c}^{(k)}_{n-1} - \tilde{c}^{(k-1)}_{n-1} + z^{(k)}_{n-1}   
\end{equation}
for some $z^{(k)}_{n-1}$ with $\partial z^{(k)}_{n-1}=0$.
For the trace~(\ref{eq:Zbdry-trace}) to be non-zero, we need the product of the Pauli operators to 
be proportional to the identity.
This requires that
\begin{equation}
\sum_k z^{(k)}_{n-1} = 0 \,.    
\end{equation}
Now, let us define
\begin{equation}
\mathrm{h}_n := \sum_k (\tilde{c}^{(k)}_{n-1} - \sum_{j=1}^k z^{(j)}_{n-1})\otimes \{k\leq\tau\leq k+1\}\,.
\end{equation}
By an explicit computation, we find that
$ \partial_\text{ tot}\mathrm{h}_n=\mathrm{z}_{n-1}$ using~(\ref{eq:zn-1-decomposition}).
Similarly, we have $\mathrm{z}^*_{d-n} = \partial^*_\text{ tot} \mathrm{h}^*_{d-n+1}$ for some $\mathrm{h}^*_{d-n+1}$.
The value of the logarithm of GSD is the ($n-1$)-th Betti number for $T^{d-1}$.
The claim~(\ref{eq:Zbdry-linking}) follows from~(\ref{eq:bdry-anomaly-defect-1}) and~(\ref{eq:bdry-anomaly-defect-2}).

The bulk and boundary are related by $X$-measurements and post-selection as we now show.
From the bulk state $ |\bm{\mathcal{E}}_0\rangle := \left |\bm{\mathcal{E}}(\bm z_{n-1}^{\text{rel}(0)},\bm z_{d-n}^{*(0)}) \right\rangle$ and the bulk operator~$\mathcal{O}_\text{ins}$ in (\ref{eq:O-ins}),
\begin{equation}\label{eq:Psi-0-Psi}
|\Psi_0\rangle_\text{bdry} := \langle+|' |\bm{\mathcal{E}}_0\rangle \,,
\quad
|\Psi\rangle_\text{bdry} := \langle+|' \mathcal{O}_\text{ins} |\bm{\mathcal{E}}_0\rangle \,,
\end{equation}
where $\langle+|'$ is the product of $\langle+|$ states on all the qubits of the cluster state gCS$_{d,n}$ (those on $\bm\Delta_n \cup\bm\Delta_{n-1}$) except at those at $\sigma_{n-1}\otimes\{x_d=0\}$ and $\sigma_{n-1}\otimes\{x_d=L_d\}$ for $\sigma_{n-1}\in \Delta_{n-1}$.
Projecting onto $|+\rangle'$ is equivalent to $X$-measurements and post-selecting the $X=+1$ outcomes.
We have 
\begin{equation}\label{eq:O-Psi-boundary}
|\Psi\rangle_\text{bdry}  = \mathcal{O}(x_d=0)\mathcal{O}(x_d=L_d)|\Psi_0\rangle_\text{bdry} \,,
\end{equation}
where
\begin{equation}
\begin{aligned}
\mathcal{O}(x_d=0)
&:=Z(c^{(L_\tau)}_{n-1}) X(c^{*(L_\tau -1/2)}_{d-n}) \ldots 
\\
&\qquad \times
Z(c^{(1)}_{n-1}) X(c^{*(1/2)}_{d-n}) \,.    
\end{aligned}
\end{equation}
Here we omitted ``$\otimes\{x_d=0\}$'' to avoid clutter, and we define $\mathcal{O}(x_d=L_d)$ similarly.
 The operator~$\mathcal{O}(x_d=0)$ is precisely what appears in the definition of the boundary partition function~(\ref{eq:Zbdry-trace}).
Thus, the state and the operator insertion in the bulk induces the corresponding state and operator insertion on the boundaries.

\section{Dualities in Wegner models}
\label{sec:dualities}

\subsection{Wegner's dualities in classical statistical models and strange correlators}
\label{sec:dualities-classical}

In this subsection, we reproduce the Kramers-Wannier-Wegner duality~\cite{Wegner} between $M_{d,n}$ and (a gauged version of) $M_{d,d-n}$ viewed as {\it classical statistical models}, using the cluster state $|{\rm gCS}_{(d,n)}\rangle$ and the Hadamard transform~\cite{PhysRevLett.98.117207}.
We focus on the binary case for simplicity.

We consider a $d$-dimensional hypercubic lattice with periodic identifications in all $d$ directions.
Let us define a state by projecting the cluster state $|\text{gCS}_{d,n}\rangle$ as
\begin{equation}
|\Phi_{d,n}\rangle := \langle +|^{\otimes \bm{\Delta}_{n-1}}
\cdot 
|\text{gCS}_{d,n}\rangle \,.
\end{equation}
Up to a Hadamard transformation, the state~$|\Phi_{d,n}\rangle $ is a special state in the generalized toric code $TC_{d+1,n+1}$~\cite{2001PhRvL..86..910B,RBH,https://doi.org/10.48550/arxiv.2112.01519}.
Indeed, it is stabilized by the stabilizers that define the code,
\begin{equation}
Z(\bm\partial^* \bm{\sigma}_{n-1}) |\Phi_{d,n}\rangle  
= 
X(\bm\partial \bm{\sigma}_{n+1}\rangle|\Phi_{d,n}\rangle 
=|\Phi_{d,n}\rangle  \,,
\end{equation}
and is also stabilized by some of the logical operators,
\begin{equation}
X(\bm z_n) |\Phi_{d,n}\rangle = |\Phi_{d,n}\rangle  \,,
\end{equation}
where $\bm z_n$ is a homologically non-trivial $n$-cycle.
Let us also introduce a product state
\begin{equation}\label{eq:omega-product-state}
\langle\omega(J)| := 
\bigotimes_{\bm\sigma_n}
{}_{\bm\sigma_n}\!\langle 0|
e^{J X(\bm\sigma_n)}   \,,
\end{equation}
where $J$ is a real parameter.
The overlap between the two states computes the classical partition function of $M_{d,n}$~\cite{Wegner} up to a constant~\cite{2023ScPP...14..129S}:
\begin{equation}\label{eq:Omega-Phi-Z}
\langle\omega(J)|\Phi_{d,n}\rangle
= 2^{-|\bm\Delta_{n-1}|}2^{-|\bm\Delta_n|/2} Z_{(d,n)}(J) \,,
\end{equation}
where
\begin{equation}
Z_{(d,n)}(J) = 
\hspace{-3mm}
\sum_{\{s(\sigma_{n-1})= 0,1\}}
\exp\Big(
J \sum_{\sigma_n} \prod_{\sigma_{n-1}\subset\sigma_n} (-1)^{s(\sigma_{n-1})}
\Big) \,.
\end{equation}

Following~\cite{PhysRevLett.98.117207}, we consider the simultaneous Hadamard transform
\begin{equation}
\mathsf{H}:= \prod_{\bm\sigma_n} H_{\bm\sigma_n}
\end{equation}
in the overlap.
Let $|\Phi^*_{d,d-n}\rangle$ be the state constructed in the same way as $|\Phi_{d,d-n}\rangle$ but on the dual lattice rather than the original lattice.
The state $\mathsf{H}|\Phi_{d,n}\rangle$ almost coincides with $|\Phi^*_{d,d-n}\rangle$ because both are stabilized by $
X(\bm\partial^* \bm{\sigma}_{n-1})$ and $Z(\bm\partial \bm{\sigma}_{n+1})$ (defined on the original lattice).
There is a slight difference because $\mathsf{H}|\Phi_{d,n}\rangle$ is stabilized by logical operators $Z(\bm z_n)$ while  $|\Phi^*_{d,d-n}\rangle$ is stabilized by logical operators $X(\bm z_{d-n}^*)$, where $z_n$ and $z_{d-n}^*$ are homologically non-trivial (dual) cycles.
The difference can be accounted for by summing over logical operators:
\begin{align}
& \quad \ \mathsf{H}|\Phi_{d,n}\rangle  \nonumber
\\
&= 
\frac{1}{| H_n(T^d,\mathbb{Z}_2)|}
\bigg(
\sum_{[\bm z_n] \in H_n(T^d,\mathbb{Z}_2)}Z(\bm z_n) 
\bigg) |\Phi^*_{d,d-n}\rangle
\,, 
\label{eq:H-Phi-Phi-star}
\\
&\quad \ |\Phi^*_{d,d-n}\rangle  \nonumber
\\
&= 
\frac{1}{| H_{d-n}(T^d,\mathbb{Z}_2)|}
\bigg(
\sum_{[\bm z^*_{d-n}] \in H_{d-n}(T^d,\mathbb{Z}_2)}X(\bm z^*_{d-n}) 
\bigg) 
\nonumber\\
&\quad\times
\mathsf{H}|\Phi_{d,n}\rangle 
\,,
\label{eq:Phi-star-H-Phi}
\end{align}
where $H_i(T^d,\mathbb{Z}_2)$ denotes the $i$-th homology group of the $d$-dimensional torus $T^d$ with $\mathbb{Z}_2$ coefficients, and $[\bullet]$ is the homology class represented by the cycle $\bullet$.

We note that
\begin{align}
\langle \omega(J)| \mathsf{H} = (\sinh 2J)^{|\bm\Delta_n|/2} 
\langle \omega(J^*)|
\,,
\label{eq:Omega-H-Omega}
\end{align}
where
\begin{equation} \label{eq:J-relation}
 J^* =-\frac12 \log \tanh J \,.
\end{equation}
From the identity $\langle\omega(J)|\Phi_{d,n}\rangle = \langle\omega(J)|\mathsf{H}\cdot \mathsf{H} |\Phi_{d,n}\rangle$, we get the relation 
\begin{equation}
\begin{aligned}    
 Z_{(d,n)}(J)
&= 
\frac{2^{|\bm\Delta_{n-1}|}}{2^{|\bm\Delta_{n+1}|}}
\frac{(\sinh 2 J )^{|\bm\Delta_n|/2}}{| H_{d-n}(T^d,\mathbb{Z}_2)|}
\\
&\qquad \times
\sum_{[ \bm z_n 
] \in H_{n}(T^d,\mathbb{Z}_2)}
Z_{(d,d-n)}^\text{twisted}(J^*,z_n)
\,.
\end{aligned}
\end{equation}
Here, we defined the twisted partition function
\begin{align}
\begin{split}
Z^\text{twisted}_{(d,d-n)}(J^*,z_n) &= 
\hspace{-2mm}
\sum_{\{s(\sigma^*_{d-n-1})\}}
\exp\Big(
J^* \sum_{\sigma_n} (-1)^{\# (z_n\cap \sigma^*_{d-n})}\\
&\quad \times
\prod_{\sigma^*_{d-n-1}\subset\sigma^*_{d-n}} (-1)^{s(\sigma^*_{d-n-1})}
\Big) 
\end{split}
\end{align}
with the twist defined by the defect on the cycle $z_n$.
The summation over defect configurations is equivalent to gauging the symmetries generated by the defects.
Thus we derived the precise statement that the Kramers-Wannier dual of the model $M_{d,n}$ is the model $M_{d,d-n}$ whose ($d-1-n$)-form symmetry is gauged: schematically, we can write
\begin{equation}\label{eq:Mdn-Mdd-n-duality}
M_{d,n}  (J) \simeq M_{d,d-n} (J^*)/\mathbb{Z}_2^{[d-1-n]} \,.
\end{equation}
 We made explicit the dependence on the parameters that are related as in~(\ref{eq:J-relation}).
For a path integral explanation of the necessity of gauging, see the appendix of~\cite{Kapustin:2014gua}.

\begin{figure}
    \centering
\includegraphics[scale=0.9]{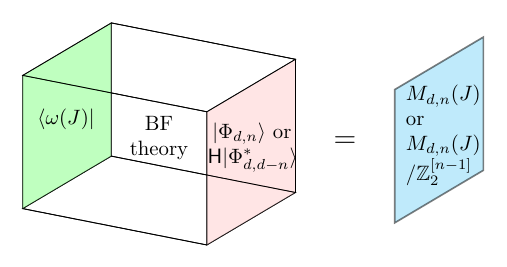}
\caption{Pictorial illustration of the relations~(\ref{eq:PhidntoMdn}) and~(\ref{eq:Phistardd-ntoMdn}).
On the LHS, the states $|\Phi_{d,n}\rangle$ and $ \mathsf{H}|\Phi^*_{d,d-n}\rangle$ define topological boundary conditions for the symmetry topological field theory, which in this case is the BF theory in Appendix~\ref{sec:inflow-continuum} with replacement $(d,n,N)\rightarrow (d+1,n+1,2)$.
The other boundary is defined by the product state~$\langle\omega(J)|$ in~(\ref{eq:omega-product-state}).
The two boundary conditions give rise to two classical statistical models $M_{d,n}(J)$ and $M_{d,n}(J)/\mathbb{Z}_2^{[n-1]}$ on the RHS.
For $(d,n)=(2,1)$, we get Ising and Ising$/\mathbb{Z}_2$ from two states in the toric code.
}
    \label{fig:SymTFT}
\end{figure}

We can interpret the overlaps~$\langle \omega(J)|\Phi_{d,n}\rangle$ and~$\langle \omega(J)| \mathsf{H}|\Phi^*_{d,d-n}\rangle$\footnote{%
An overlap like these is sometimes called a strange correlator~\cite{2014PhRvL.112x7202Y, 2018arXiv180105959B}.
}
in terms of the so-called symmetry topological field theory~\cite{Gaiotto:2020iye,Apruzzi:2021nmk,Freed:2022qnc,Kaidi:2022cpf}.
By~(\ref{eq:Omega-Phi-Z}), the former overlap is the classical partition function of $M_{d,n} (J)$ up to normalization.
By a consideration similar to the previous paragraph, the relation~(\ref{eq:Phi-star-H-Phi}) 
implies that
$\langle \omega(J)|  \mathsf{H}|\Phi^*_{d,d-n}\rangle$ is,   again up to normalization, the classical partition function of $M_{d,n} (J)$ whose ($n-1$)-form symmetry is gauged.
The states $|\Phi_{d,n}\rangle$ and $ \mathsf{H}|\Phi^*_{d,d-n}\rangle$, which are both states in the generalized toric code $TC_{d+1,n+1}$  but are characterized by  different logical operators that stabilize them, define different boundary conditions for the same $(d+1)$-dimensional topological field theory that underlies the code.
 (The continuum description of the topological field theory is the BF theory in Appendix~\ref{sec:inflow-continuum} with replacement $(d,n,N)\rightarrow (d+1,n+1,2)$.)
Thus different boundary conditions lead to theories that are locally the same but are different in their global properties:
\begin{align}
&\text{topological boundary condition defined by }\nonumber\\
&\qquad|\Phi_{d,n}\rangle  \longrightarrow M_{d,n} (J)
\,, \label{eq:PhidntoMdn}
\\
&\text{topological boundary condition defined by }\nonumber\\  &\qquad\mathsf{H}|\Phi^*_{d,d-n}\rangle  \longrightarrow M_{d,n} (J)/\mathbb{Z}_2^{[n-1]} \,.\label{eq:Phistardd-ntoMdn}
\end{align}
 See Figure~\ref{fig:SymTFT} for an illustration.

For $(d,n)=(2,1)$,  the two states   $|\Phi_{1,1}\rangle$ and $\mathsf{H}|\Phi^*_{1,1}\rangle$ in the usual toric code $TC_{3,2}$ are stabilized by $X(\bm z_1)$ and $Z(\bm z_1)$ respectively, where $\bm z_1$ is an arbitrary 1-cycle.
The symmetry topological field theory is the  three-dimensional BF theory that underlies the original toric code~$TC_{3,2}$, namely the $U(1)^2$ Chern-Simons theory whose level matrix is 
\begin{equation}
K=
\left(
\begin{array}{cc}
0&2\\
2&0
\end{array}
\right)\,.
\end{equation}
Each topological boundary condition of such a theory is indeed characterized by the set of Wilson lines that can be absorbed into the boundary~\cite{Kapustin:2010hk}.
 The relations~(\ref{eq:PhidntoMdn}) and (\ref{eq:Phistardd-ntoMdn}) imply that the two boundary conditions defined by  $|\Phi_{1,1}\rangle$ and  $\mathsf{H}|\Phi^*_{1,1}\rangle$ give rise to the 2-dimensional Ising model $M_{2,1}(J)$ and its gauged version $M_{2,1}(J)/\mathbb{Z}_2$, respectively.
The latter is dual to $M_{2,1}(J^*)$ by the KW duality~(\ref{eq:Mdn-Mdd-n-duality}).

\subsection{Measurement-assisted Kramers-Wannier transformations}

\label{sec:non-inv}

The Wegner model $M_{d,n}$ is related to $M_{d,d-n}$ via the Kramers-Wannier transformation  \cite{Wegner} as we saw above in the classical case.
 In the quantum Hamiltonian formulation, let us realize the transformation by measurement following~\cite{https://doi.org/10.48550/arxiv.2112.01519,2023PRXQ....4b0339T}.
The linear transformation can be regarded as an operator representation of the duality defect.
This method allows us to compute the fusion of duality defects in a straightforward manner.

We consider a $(d-1)$-dimensional hypercubic lattice and place qubits
on $n$- and $(n-1)$-cells.\footnote{%
The generalization to $N(\geq 3)$-dimensional qudits is straightforward.
}
 We impose the periodic boundary conditions.
The degrees of freedom of the original model $M_{d,n}$ are the qubits on $(n-1)$-cells.
The qubits on $n$-cells, or equivalently $(d-n-1)$-cells on the dual lattice, are the degrees of freedom of the dual model $M_{d,d-n}$.
We denote the total Hilbert space on the $i$-cells by $\mathcal{H}_i$ for $i=n, n-1$.
Let us consider the entangler for the 
generalized cluster state 
$|{\rm gCS}_{(d-1,n)}\rangle$~\cite{2023ScPP...14..129S}
\begin{equation}
\mathcal{U}_{CZ}:= \prod_{\sigma_{n-1}\subset \sigma_n} CZ_{\sigma_{n-1},\sigma_n} 
\,,
\end{equation}
where the product is over all the pairs $(\sigma_{n-1},\sigma_n) \in \Delta_{n-1}\times \Delta_n$ such that $\sigma_{n-1}\subset\sigma_n$.
Following~\cite{https://doi.org/10.48550/arxiv.2112.01519,2023PRXQ....4b0339T}, let us define the linear map
\begin{equation}\label{eq:KWweg}
{\sf KW} := \langle +|^{\otimes \Delta_{n-1}} 
\,
\mathcal{U}_{CZ}
\,
 | +\rangle^{\otimes \Delta_n} 
 :
\quad
\mathcal{H}_{n-1} 
\rightarrow 
\mathcal{H}_{n} 
\end{equation}
that implements the Kramers-Wannier~\cite{PhysRev.60.252} (or Wegner~\cite{Wegner}) duality transformation.
The $n$- and $(n-1)$-cells on the original lattice can be respectively identified with the ($d-n-1$)- and ($d-n$)- cells on the dual lattice.
To be more explicit, let us write computational basis states on  $(n-1)$- and $n$-cells as $|s\rangle_{\Delta_{n-1}}$ with $s=\{s_{\sigma_{n-1}} \in \mathbb{Z}_2\}_{\sigma_{n-1}\in\Delta_{n-1}}$ and $|t\rangle_{\Delta_{n}}$ with $t=\{t_{\sigma_{n}} \in \mathbb{Z}_2\}_{\sigma_{n}\in\Delta_{n}}$, respectively.
Then
\begin{align}
{\sf KW} |s\rangle_{\Delta_{n-1}}
&=2^{-\frac{|\Delta_n|}{2}-\frac{|\Delta_{n-1}|}{2}}
\\\nonumber
&\times\sum_t\prod_{\sigma_{n-1},\sigma_n} 
(-1)^{s_{\sigma_{n-1}}t_{\sigma_n}a(\partial\sigma_n;\sigma_{n-1})}|t\rangle_{\Delta_n} \,.
\end{align}
 Compare, for example, with \cite{2016JPhA...49I4001A,Li:2023ani} that study the $(d,n)=(2,1)$ case.

The contraction with $\langle +|^{\otimes \Delta_n}$ can be interpreted as the ideal result of measurement in the $X$-basis with all the outcomes $+$.
Non-ideal measurement outcomes other than $+$ can be corrected by applying appropriate Pauli $X$ operators if the state to be transformed satisfies a certain symmetry condition~\cite{https://doi.org/10.48550/arxiv.2112.01519} as we now show.
 Consider the transformation $|\Psi\rangle_{\Delta_{n-1}}\rightarrow \mathsf{KW}|\Psi\rangle_{\Delta_{n-1}} $.
We define the pre-measured state
\begin{align}
    \ket{\Psi_\mathrm{pre}}
    &= \mathcal{U}_{CZ}
    \ket{\Psi}_{\Delta_{n-1}}\ket{+}_{\Delta_n}\\\nonumber
    &=\prod_{\sigma_{n-1},\sigma_n}CZ^{a(\partial\sigma_n,\sigma_{n-1})}_{\sigma_{n-1},\sigma_n}\ket{\Psi}_{\Delta_{n-1}}\ket{+}_{\Delta_n} \,,
\end{align}
which is the state before any measurement is performed. 
We assume that $\ket{\Psi}_{\Delta_{n-1}}$ is symmetric with respect to the $(n-1)$-form symmetry, {\it i.e.,} 
\begin{align}
    X(z^*_{d-n})\ket{\Psi}_{\Delta_{n-1}}=\ket{\Psi}_{\Delta_{n-1}} 
    \label{eq:XPsi}
\end{align}
for any dual $(d-n)$-cycle $z^*_{d-n}$.
Then the state $\ket{\Psi_\mathrm{pre}}$ 
is also symmetric: 
\begin{subequations}
    \begin{align}
    X(z^*_{d-n})\ket{\Psi_\mathrm{pre}}&=
    \mathcal{U}_{CZ} 
    \prod_{\sigma_n,\sigma_{n-1}}Z_{\sigma_n}^{a(\partial\sigma_n,\sigma_{n-1})a(z^*_{d-n},\sigma_{n-1})}\\\nonumber
    &\qquad\times X(z^*_{d-n})\ket{\Psi}_{\Delta_{n-1}}\ket{+}_{\Delta_n}
    \\
    &=\ket{\Psi_\mathrm{pre}} \,,
\end{align}
\end{subequations}
where we noted that
$\sum_{\sigma_{n-1}} a(\partial\sigma_n,\sigma_{n-1})a(z^*_{d-n},\sigma_{n-1}) = \#(\partial\sigma_n\cap z^*_{d-n}) = \#(\sigma_n\cap\partial^* z^*_{d-n})=0 $.
Suppose that the measurement outcome is $\bra{+}^{\otimes \Delta_{n-1}}Z(c_{n-1})$.
The state after the measurement is
\begin{align}
&\quad\    \bra{+}^{\otimes \Delta_{n-1}}Z(c_{n-1})\ket{\Psi_\mathrm{pre}}
\\
     &=\bra{+}^{\otimes \Delta_{n-1}}Z(c_{n-1}) 
    X(z^*_{d-n})X(z^*_{d-n})\ket{\Psi_\mathrm{pre}}
    \\
    &=(-1)^{\#(z^*_{d-n}\cap c_{n-1})}
    \bra{+}^{\otimes \Delta_{n-1}}Z(c_{n-1})\ket{\Psi_\mathrm{pre}}
\end{align}
where the last line follows from moving one of the $X(z^*_{d-n})$'s to the left past through $Z(c_{n-1})$  and using~(\ref{eq:XPsi}). 
Thus the measurement outcomes specified by $c_{n-1}$ must obey the constraint $\#(z^*_{d-n}\cap c_{n-1})=0$ mod $2$. 
If $z^*_{d-n}$ is a dual boundary, {\it i.e.,} $z^*_{d-n}=\partial^*c^*_{d-n+1}$ for some dual chain $c^*_{d-n+1}$, we have
$\#(z^*_{d-n}\cap c_{n-1})=\#(\partial^*c^*_{d-n+1}\cap c_{n-1})=\#(c^*_{d-n+1}\cap \partial c_{n-1})$. 
Since this is true for any
$c^*_{d-n+1}$ 
we have $\partial c_{n-1}=0$, {\it i.e.,} $c_{n-1}$ is a cycle. 
Now letting $z^*_{d-n}$ be an arbitrary cycle
and from the fact that intersection pairing for homology is non-degenerate on $T^d$, we find that the cycle $c_{n-1}$ is a representative of the trivial homology class. Hence $c_{n-1}=\partial c_n$ for some $n$-chain $c_n$. 
Now the non-ideal measurement outcomes can be corrected by countering with $X(c_n)$:
 \begin{align}
 &\quad \ X(c_n) \langle +|^{\otimes \Delta_{n-1}} 
     Z(c_{n-1})
 \,
 \mathcal{U}_{CZ}
 \,
  | +\rangle^{\otimes \Delta_n} |\Psi\rangle_{\Delta_{n-1}} 
  \nonumber\\
  &= {\sf KW} \, |\Psi\rangle_{\Delta_{n-1}} \,.
 \end{align}

The Kramers-Wannier transformation from $n$-cells to $(n-1)$-cells can be written as
\begin{equation}
{\sf KW}' :=  
\langle +|^{\otimes \Delta_{n}} 
\,
\mathcal{U}_{CZ}
\,
 | +\rangle^{\otimes \Delta_{n-1}} 
 :
\quad
\mathcal{H}_{n} 
\rightarrow 
\mathcal{H}_{n-1} \,.
\end{equation}
Note that  $CZ_{\sigma_{n-1},\sigma_{n}}$ and $CZ_{\sigma_{n},\sigma_{n-1}}$ are equal.
We wish to compute the composition ${\sf KW'\circ KW}$.
For this purpose, on each $(n-1)$-cell we introduce a copy of qubit and indicate the corresponding states by a prime.
 We write $s_{n-1} = \sum_{\sigma_{n-1}\in\Delta_{n-1}} s_{\sigma_{n-1}}\sigma_{n-1} \in \Delta_{n-1}$ and $|s_{n-1}\rangle = \bigotimes_{\sigma_{n-1}} |s_{\sigma_{n-1}}\rangle_{\sigma_{n-1}}$, etc.
Then
\begin{align}
\begin{split}
&\langle t_{ n-1}| {\sf KW'\circ KW} | s_{ n-1}\rangle =
\langle t_{ n-1}|+\rangle^{ \otimes \Delta_{n-1}} \langle +|^{ \otimes \Delta_{n-1}} |s_{ n-1}\rangle
\\
&\times\langle +|^{ \otimes \Delta_n} 
\prod_{\sigma_n\in\Delta_n} Z_{\sigma_n}^{\sum_{\sigma_{n-1}} a(\partial\sigma_n;\sigma_{n-1})(s_{\sigma_{n-1}}-t_{\sigma_{n-1}})} |+\rangle^{ \otimes \Delta_n}
\end{split}\\
&=\qquad
2^{-|\Delta_{n-1}|}
\prod_{\sigma_n} \delta_{\sum_{\sigma_{n-1}} a(\partial\sigma_n;\sigma_{n-1})(s_{\sigma_{n-1}}-t_{\sigma_{n-1}}),0}
 \,.
\end{align}
After simple manipulations and up to normalization, the result can be written as
\begin{equation}\label{eq:KW-prime-KW}
{\sf KW'\circ KW}  =  \sum_{z^*_{d-n}\in Z^*_{d-n}} X(z^*_{d-n}) \,,
\end{equation}
where $Z^*_{d-n}$ is the group of dual $(d-n)$-cycles, {\it i.e.}, the dual chains annihilated by $\partial^*$.
Note that we identified $(n-1)$-chains $c_{n-1}$ and dual $(d-n)$-chains $c^*_{d-n}$ via $a(c_{n-1},\sigma_{n-1}) = \# (c^*_{d-n} \cap \sigma_{n-1})$.
The RHS of~(\ref{eq:KW-prime-KW}), which is a sum of networks of symmetry generators localized on the constant time slice, implements the gauging of the ($n-1$)-form $\mathbb{Z}_2$ symmetry on the codimension-1 hypersurface in spacetime.
Such a localized object is known as a condensation defect~\cite{Roumpedakis:2022aik}.
Thus our construction provides a large class of examples of fusion rules that involve condensation defects and non-invertible symmetries.

Let us now specialize to $d=2$ and $n=1$, corresponding to the $(1+1)$-dimensional Ising model.
In this case ${\sf KW}'={\sf KW}$, and
\begin{equation}
\langle t|{\sf KW}^2 |s\rangle = 2^{-|\Delta_0|} \prod_{e\in\Delta_1} \delta_{\sum_{v\subset e}(s_e+t_e),0}^\text{mod 2}  \,.
\end{equation}
When the lattice is periodic, the product of  Kronecker deltas equals a sum of two products of Kronecker deltas: $ \prod_{e\in\Delta_1} \delta_{\sum_{v\subset e}(s_v+t_v),0}^\text{mod 2}  = \prod_v \delta_{s_v,t_v}^\text{mod 2} + \prod_v \delta_{s_v,t_v+1}^\text{mod 2} $.
Thus we have, up to normalization,
\begin{equation} \label{eq:fusion1}
{\sf KW}^2  = {\sf 1} + \prod_v X_v \,.
\end{equation}
Using the definition of ${\sf KW}$ and the identity $X_1 CZ_{12}=CZ_{12} X_1 Z_2$, it is easy to see that
\begin{equation} \label{eq:fusion2}
{\sf KW}(\prod_v X_v) =( \prod_e X_e) {\sf KW} = {\sf KW} \,.
\end{equation}
We also have
\begin{equation} \label{eq:fusion3}
(\prod_v X_v )^2 = {\sf 1} \,.
\end{equation}
Equations~(\ref{eq:fusion1})-(\ref{eq:fusion3}) are precisely the fusion rules for the defects in the Ising model.
See (3.27) of~\cite{2016JPhA...49I4001A}.
In particular ${\sf KW}$ is identified with the duality defect (as it should be) and it generates a non-invertible (0-form) symmetry.

Similar self-dual cases ${\sf KW}'={\sf KW}$ arise more generally when $d$ is even and $n=d/2$.
They include the $\mathbb{Z}_2$ gauge theory in $3+1$ dimensions~\cite{2022PTEP.2022a3B03K} and the 6-dimensional self-dual theory of a $\mathbb{Z}_2$ 2-form gauge field.

\section{Measurement-based quantum simulation for a gauge theory with matter}
\label{sec:Fradkin-Shenker}

As we mentioned several times, the work~\cite{2023ScPP...14..129S}, presented the MBQS scheme for Wegner's model $M_{d,n}$. 
We recall that $M_{3,1}$ corresponds to the $(2+1)$d transverse-field Ising model, and $M_{3,2}$ to the $(2+1)$d pure $\mathbb{Z}_2$ lattice gauge theory.
In this section, we construct an MBQS scheme for the Fradkin-Shenker model~\cite{PhysRevD.19.3682}, {\it i.e.,} the $\mathbb{Z}_2$ gauge theory coupled to an Ising matter in $(2+1)$~dimensions.

Let us consider the two-dimensional square lattice with qubits introduced on 0-cells $\sigma_0\in\Delta_0$ and 1-cells $\sigma_1\in \Delta_1$ (the Lieb lattice), where the qubits correspond to matter degrees of freedom on sites and the gauge field on links, respectively.
In this section, the tilde and the double bracket are used for the operators and the states associated with gauge degrees of freedom, respectively.
States in the physical Hilbert space of the gauge theory must satisfy the Gauss law constraints
\begin{align}\label{eq:FS-Gauss}
X(\sigma_0) \widetilde{X}(\partial^* \sigma_0) = 1  
\end{align}
for any $\sigma_0\in\Delta_0$.
The Hamiltonian of the Fradkin-Shenker model is given by
\begin{align} \label{eq:FS-Hamiltonian}
H_{\rm FS} 
&= 
- \frac{1}{\lambda} \sum_{\sigma_0 \in \Delta_0} X(\sigma_0)
- g \sum_{\sigma_1 \in \Delta_1} \widetilde{X}(\sigma_1) 
\nonumber\\
&\quad - \lambda \sum_{\sigma_1 \in \Delta_1} \widetilde{Z}(\sigma_1) Z(\partial \sigma_1) 
- \frac{1}{g} \sum_{\sigma_2 \in \Delta_2} \widetilde{Z}(\partial \sigma_2) \, ,
\end{align}
where $\widetilde{Z}(\sigma_1) Z(\partial \sigma_1)$ is the minimal gauge coupling term.
The quantum Hamiltonian above results via the transfer matrix formalism from the continuum time limit~\cite{Fradkin:1978th} of the Euclidean path integral in 3d given by
\begin{align}\label{eq:FS-partition-function}
&Z_{\rm FS}(J,K) = \nonumber\\
&\sum_{ \substack{ \{ s({\bm \sigma_0})=0,1 ; \\ U({\bm \sigma_1}) = 0,1 \} } }  
\exp \Big[ J \sum_{ \bm \sigma_1 \in \bm \Delta_1 } (-1)^{U ({\bm \sigma_1} ) }\prod_{\bm \sigma_0 \subset \bm \sigma_1 } (-1)^{s({ \bm \sigma_0})}   \nonumber \\
&\qquad+ K \sum_{{\bm \sigma_2 \in \bm \Delta_2}} \prod_{\bm \sigma_1 \subset \bm \sigma_2 }  (-1)^{U ({\bm \sigma_1} ) } \Big] \, ,
\end{align}
where $\bm \sigma_1$ and $\bm \sigma_2$ are the 1- and 2-cells in the 3d cubic lattice.
The spacetime structure of the Lagrangian above provides us with a blueprint for the resource state of the MBQS, which we will show in the following subsection.

\subsection{Resource state for the Fradkin-Shenker model
}

\begin{figure*}
\includegraphics[scale=1]{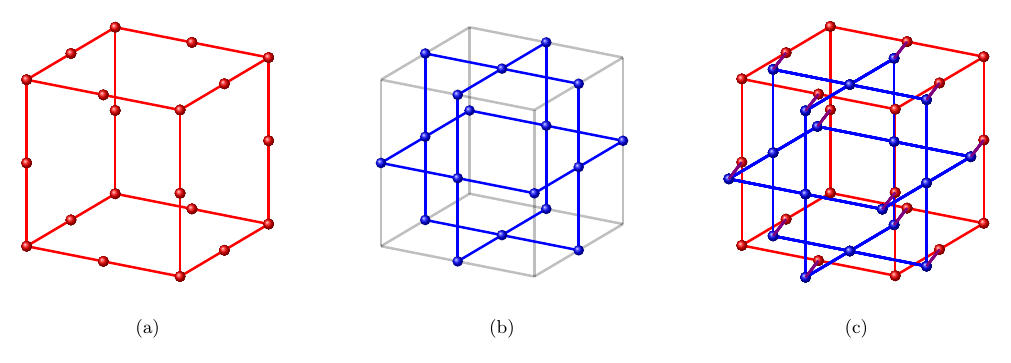}
    \caption{The resource state for simulating (a) the $(2+1)$d transverse-field Ising model $|\text{gCS}_{(3,1)}\rangle$, (b) the $(2+1)$d lattice gauge theory $|\text{gCS}_{(3,2)}\rangle\!\rangle$, and (c) the Fradkin-Shenker model $|\text{CS}_\text{FS}\rangle\!\rangle$. Balls represent the qubits and lines the $CZ$ entanglers. (Gray lines in (b) are to show the underlying 3d cubic lattice.) The state (c) is obtained by overlapping the states (a) and (b) and applying the entangler at the overlaying edge qubits.}
    \label{fig:FS-CS}
\end{figure*}

Let us consider the 3d cubic lattice, where the ``matter" logical+ancillary qubits are introduced on 0-
and 1-cells, and the ``gauge" logical+ancillary qubits are introduced on 1-
and 2-cells.
The resource state for the Fradkin-Shenker model
is the simultaneous $+1$ eigenstate of the stabilizers
\begin{align}
& K({\bm \sigma_1}) = X({\bm \sigma_1}) \widetilde{Z}({\bm \sigma_1}) Z({\bm \partial \bm \sigma_1}) \, , 
\\
&
K({\bm \sigma_0}) = X({\bm \sigma_0}) Z( {\bm \partial^* \bm \sigma_0} ) \, , \quad \\
& \widetilde{K} ({\bm \sigma_2}) = \widetilde{X}({\bm \sigma_2}) \widetilde{Z}({\bm \partial \bm \sigma_2}) \, , 
\\
&
\widetilde{K}({\bm \sigma_1}) = 
\widetilde{X}({\bm \sigma_1}) Z({\bm \sigma_1}) \widetilde{Z} ({\bm \partial^* \bm \sigma_1}) \, .
\end{align} 
Explicitly the cluster state can be written as 
\begin{align} \label{eq:FS-resource-state}
&\quad \ | {\rm CS}_{\rm FS} \rangle \nonumber \\
&= \prod_{ \substack{ \bm \sigma_0 \in \bm \Delta_0 \\ \bm \sigma_1 \in \bm \Delta_1  } } 
CZ^{a(\bm \partial^*\bm \sigma_0; \bm \sigma_1)}_{\bm \sigma_0, \bm \sigma_1}
\prod_{ 
\substack{ {\bm \sigma_1 \in \bm \Delta_1}  }
} 
C\widetilde{Z}_{\bm \sigma_1, \bm \sigma_1}\nonumber \\
&\quad \times \prod_{ \substack{ {\bm \sigma_1 \in \bm \Delta_1} \\ {\bm \sigma_2 \in \bm \Delta_2}  } }
\widetilde{CZ}^{a(\bm \partial\bm \sigma_2; \bm \sigma_1)}_{\bm \sigma_2, \bm \sigma_1}|+\rangle^{ {\bm \Delta_0} } 
|+\rangle^{ {\bm \Delta_1} } 
|+\rangle\!\rangle^{ {\bm \Delta_1} } 
|+\rangle\!\rangle^{ {\bm \Delta_2}} \, \nonumber \\
&= \Big(
\prod_{ 
\substack{ {\bm \sigma_1 \in \bm \Delta_1}  }
} 
C\widetilde{Z}_{\bm \sigma_1, \bm \sigma_1} \Big) \times 
|{\rm gCS}_{(3,1)}\rangle \otimes  
|{\rm gCS}_{(3,2)}\rangle\!\rangle \, .
\end{align}
Writing the entangler in the last line as $\mathcal{U}_{\rm m.c.}$,
indicating the role of the minimal coupling between the Ising model and the gauge theory,  
the stabilizers of $|{\rm CS}_{\rm FS}\rangle$ can be understood as the conjugation of stabilizers of gCS${}_{(3,1)}$ and gCS${}_{(3,2)}$ by $\mathcal{U}_{\rm m.c.}$. 
See Figure~\ref{fig:FS-CS} for illustration.

\subsection{MBQS protocol}
\label{sec:FS-MBQS}

Let us consider the first-order Trotter approximation of the time evolution $e^{-it H_\text{FS}}$. 
We wish to realize 
\begin{align}
&|\phi(t) \rangle = T(t) |\phi (0) \rangle \, ,\\
&T(t) = 
\Bigg(
\prod_{\sigma_1 \in \Delta_1 } e^{-i \delta t  g \widetilde{X}(\sigma_1 ) }
\prod_{\sigma_0 \in \Delta_0} e^{-i \delta t \frac{1}{\lambda} X(\sigma_0 ) }  \nonumber\\
&\times\prod_{\sigma_2 \in \Delta_2 } e^{-i \delta t \lambda   \widetilde{Z}(\partial \sigma_2 ) }
\prod_{\sigma_1 \in \Delta_1 } e^{-i \delta t  \frac{1}{g} \widetilde{Z}(\sigma_1)Z(\partial \sigma_1 ) } 
\Bigg)^{j} \, ,
\label{eq:evolution-FS}
\end{align}
with $t = \delta t \cdot j$ ($j \in \{1,...,L_z\}$) and $\delta t$ the Trotter time step. 
To achieve the time evolution, we use the cluster state $|\text{CS}_\text{FS}\rangle$ with boundaries and perform adaptive, sequential measurements.%
\footnote{%
\green We stress that the time evolution operator~(\ref{eq:evolution-FS}) involves non-Clifford operators because the rotation angles such as $\delta t g$ and $\delta t/\lambda$ are generic real numbers.
Consequently, there is advantage in the     quantum simulation of the Fradkin-Shenker model (and the models in~\cite{2023ScPP...14..129S}) over its classical simulation.
}

For concreteness, we take the open boundary condition for the $z$ direction, and the periodic boundary condition for both $x$ and $y$ directions. 
The lattice has two boundaries at $z=0$ and $z=L_z$.
On the boundary at $z=0$ and in the bulk $0<z<L_z$, the qubits associated with matter are placed on 0-cells $\bm \Delta_0$ and 1-cells $\bm\Delta_1$ as in gCS${}_{(3,1)}$,  and the qubits associated with the gauge field are placed 
on 1-cells $\bm\Delta_1$ and 2-cells $\bm\Delta_2$ as in gCS${}_{(3,2)}$.
Instead of $|+\rangle^{\bm\Delta^{z=0}_0 \cup \bm\Delta^{z=0}_1}$, we set $|\phi\rangle_{\bm\Delta^{z=0}_0 \cup \bm\Delta^{z=0}_1} = |\phi(0)\rangle$ as the initial state loaded at the boundary $z=0$ of the cluster state. 
Ancillary qubits are also placed
at $\bm\Delta^{z=0}_1$ and $\bm\Delta^{z=0}_2$. 
On the other boundary at $z=L_z$, qubits are placed on 0-cells $\Delta_0$ and 1-cells $\Delta_1$, where the final state after the simulation will be induced after measuring the other qubits. (We don't  place
ancillary qubits at $z=L_z$, as a convention.)

We claim that a sequence of adaptive single-qubit measurements on individual qubits (starting from the qubits in $z=0$, continuing up to $z<j$) induces the state $|\phi\rangle_{\bm\Delta_0^{z=j}\cup \bm\Delta_1^{z=j}}= \mathcal{O}^{(j)}_{\rm bp} |\phi(t)\rangle$ at a boundary of the reduced cluster state:
\begin{widetext}
\begin{align} \label{eq:FS-MBQS-resource}
|{\rm CS}^{(j)}_{\rm FS} \rangle 
=  &
\Bigg(
\prod_{{\bm\sigma_1} \in {\bm \Delta}_1} C\widetilde{Z}_{\bm\sigma_1, \bm\sigma_1} 
\times 
\prod_{\substack{ \bm\sigma_0 \in \bm \Delta_0 \\  \bm\sigma_1 \in \bm \Delta_1 }  } CZ^{a(\bm \partial \bm \sigma_1; \bm \sigma_0)}_{\bm\sigma_0, \bm\sigma_1} 
\times
\prod_{\substack{ \bm\sigma_1 \in \bm \Delta_1 \\  \bm\sigma_2 \in \bm \Delta_2 }  } \widetilde{CZ}^{a(\bm \partial \bm \sigma_2; \bm \sigma_1)}_{\bm\sigma_1, \bm\sigma_2} \Bigg) \nonumber \\
& \qquad \times 
|+\rangle^{\bm\Delta^{j < z \leq L_z}_0}
|+\rangle^{\bm\Delta^{j \leq z <L_z}_1}
|+\rangle\!\rangle^{\bm\Delta^{j < z  \leq L_z}_1}
|+\rangle\!\rangle^{\bm\Delta^{j \leq z <L_z}_2}
|\phi\rangle_{\bm\Delta^{z=j}_0 \cup \bm\Delta^{z=j}_1} \ . 
\end{align}
\end{widetext}
Here, the byproduct operator $\mathcal{O}^{(j)}_{\rm bp}$ is a product of Pauli operators, whose specific form depends on outcomes of the measurements in $0 \leq z <j$. 
We also omitted unimportant overall constants.
The state with $j=0$ corresponds to the resource state before any measurement takes place, and that with $j=L_z$ is the state after all the bulk measurements were conducted, $|{\rm CS}^{(L_z)}_{\rm FS} \rangle 
=|\phi\rangle_{\bm \Delta^{z=L_z}_0 \cup \bm\Delta^{z=L_z}_1}$.

\begin{figure*}
    \includegraphics[width=0.9\linewidth]{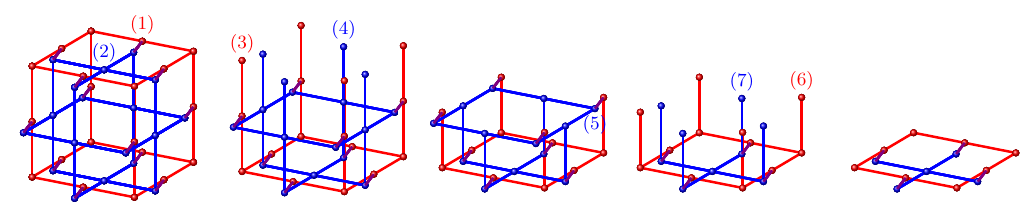}
    \caption{
    The MBQS protocol for the Fradkin-Shenker model. Each number represents the measurement of the corresponding number in the procedure. }
    \label{fig:MBQS-FS}
\end{figure*}

To describe the measurement protocol, we will use the measurement bases
\begin{equation}
\begin{split}\label{eq:MABX-def}
&\mathcal{M}_{(A)} := \big\{ e^{i\xi X}|s\rangle \ \big|\  s=0,1\big\}\,,   \\
&\mathcal{M}_{(B)} :=
\{ e^{i\xi Z} H |s \rangle \ \big| \  s=0,1\} \,,\\
&\mathcal{M}_{(X)} := 
\{ H|s\rangle \,\big| \,
s=0,1\}  \,.
\end{split}
\end{equation} 
Here, $\xi \in \mathbb{R}$ is an adjustable parameter.
For the $A$-type measurement, we will use the formula,
\begin{align}
&\Big(\langle s | e^{i\xi X}\Big)_a 
\prod_{b \in Q(a)} CZ_{a,b} |+\rangle_a |\psi\rangle_{\text{others}} \nonumber\\ 
&= \frac{1}{\sqrt{2}} \Big( \prod_{b \in Q(a)} Z_b \Big)^s 
e^{i \xi \prod_{b \in Q(a)} Z_b } | \psi\rangle_{\text{others}} \, ,
\end{align}
where $Q(a)$ is the set of qubits entangled to the qubit $a$ by the $CZ$ gate and $|\psi\rangle_\text{others}$ is a general wave function that contains the set of qubits $Q(a)$.
For the $B$-type measurement, on the other hand, we will use the formula,
\begin{align}
&\Big( \langle s| H e^{i \xi Z}\Big)_a
CZ_{a,b}
\sum_{z_a=0,1} \psi(z_{a}) |z_a\rangle_a \otimes |+\rangle_b \nonumber\\ 
&= \sum_{z_a = 0,1} 
\psi(z_a)
\langle s | H |z_a \rangle_a \otimes \Big( H e^{i\xi Z} | z_a \rangle \Big)_b \nonumber \\
&= \Big(\frac{1}{\sqrt{2}} H e^{i \xi Z} Z^s \Big)_b \sum_{z_a = 0,1} \psi(z_a) |z_a\rangle_b  
\end{align}
with $\psi(z_a): \{0,1\}\rightarrow  \mathbb{C}$,
which is a statement that a general state on the qubit $a$ written as $\sum_{z_a=0,1} \psi(z_{a}) |z_a\rangle_a$ is teleported to $b$ (as indicated by the subscript in $\sum_{z_a=0,1} \psi(z_{a}) |z_a\rangle_b$) and an additional unitary $H e^{i \xi Z} Z^s$ is applied.
This is also valid when the wave function contains dependency on bases other than $z_a$ ({\it i.e.}, a multi-qubit state).
See, for example, Ref.~\cite{2018AdPhX...361026W} for a pedagogical introduction to Measurement-Based Quantum Computation (MBQC) related to the above formula. 
Finally, for the $X$-type measurement, we obtain a projection:
\begin{align}
&\Big(\langle s | H \Big)_a 
\prod_{b \in Q(a)} CZ_{a,b} |+\rangle_a |\psi\rangle_{\text{others}} \nonumber \\
&= \frac{1}{2} \Big(1 +  (-1)^s\prod_{b \in Q(a)} Z_b \Big)  | \psi\rangle_{\text{others}} \, .
\end{align}
This is commonly used in quantum error corrections, see {\it e.g.}~\cite{2002JMP....43.4452D}.
The series of measurements to obtain the state with $|\phi\rangle_{\bm\Delta_1^{z=j+1}}=\mathcal{O}^{(j+1)}_{\rm bp} |\phi(t+\delta t)\rangle$ in the next step is as follows.
To avoid clutter, we denote all measurement angles by the same symbol $\xi$ and suppress the cell dependence of the measurement outcomes~$\{s\}$.
\begin{itemize} 
    \item[(1)] {Measure matter qubits on ${\bm \sigma_1}=\sigma_1 \times  \{j\} $ with the basis $\mathcal{M}_{(A)}$}. This gives rise to the time evolution 
    \[ 
    \big( \widetilde{Z}(\sigma_1)Z(\partial \sigma_1) \big)^s
    \exp\big[i \xi \widetilde{Z}(\sigma_1)Z(\partial \sigma_1) \big] \]  
    acting on the simulated state. 
    \item[(2)] {Measure gauge qubits on ${\bm \sigma_2}=\sigma_2 \times  \{j\}$ with the basis $\mathcal{M}_{(A)}$}. This gives rise to the time evolution 
    \[
    \big(\widetilde{Z}(\partial \sigma_2) \big)^s
    \exp\big[i \xi \widetilde{Z}(\partial \sigma_2) \big] \] 
    acting on the simulated state.  
    \item[(3)] {Measure matter qubits on ${\bm \sigma_0}=\sigma_0 \times  \{j\} $ with the basis $\mathcal{M}_{(B)}$ ($\xi=0$)}. This leads to the teleportation to the layer $[j,j+1]$ with additional Hadamard gate $H(\sigma_0) Z(\sigma_0)^s$ acting on the simulated state at $\sigma_0 \times  [j,j+1]$. 
    \item[(4)] {Measure gauge qubits on ${\bm \sigma_1}=\sigma_1 \times  \{j\}$ with the basis $\mathcal{M}_{(B)}$ ($\xi=0$)}. This leads to the teleportation to the layer $[j,j+1]$ with additional Hadamard gate $\widetilde{H}(\sigma_1)\widetilde{Z}(\sigma_1)^s$ acting on the simulated state at $\sigma_1 \times  [j,j+1]$. 
    \item[(5)] {Measure gauge qubits on ${\bm \sigma_1}=\sigma_0 \times  [j,j+1]$ with the basis $\mathcal{M}_{(X)}$}. This leads to the projector 
    \[\frac{1+(-1)^{s}Z(\sigma_0)\widetilde{Z}(\partial^* \sigma_0)}{2}\] 
    acting on the simulated state. 
    Combined with the Hadamard transforms that appeared in (3) and (4) and those that will appear below, this becomes a projector 
    \[\frac{1+(-1)^{s}X(\sigma_0)\widetilde{X}(\partial^* \sigma_0)}{2} \ . \]
    The roll of this projector is to check the eigenvalue of the simulated state with respect to the Gauss law operator $X(\sigma_0)\widetilde{X}(\partial^* \sigma_0)$.
     \item[(6)] {Measure matter qubits on ${\bm \sigma_1}=\sigma_0 \times  [j,j+1] $ with the basis $\mathcal{M}_{(B)}$}. This gives rise to the time evolution 
     \[ H(\sigma_0) \big( Z(\sigma_0) \big)^s \exp\big[i \xi Z(\sigma_0) \big] \]  
     acting on the simulated state and teleportation of the simulated state to the layer at $z=j+1$. 
    \item[(7)] {Measure gauge qubits on ${\bm \sigma_2}=\sigma_1 \times  [j,j+1]$ with the basis $\mathcal{M}_{(B)}$}. This gives rise to the time evolution 
    \[H(\sigma_1) \big( \widetilde{Z}(\sigma_1) \big)^s\exp\big[i \xi \widetilde{Z}(\sigma_1) \big] \] 
    acting on the simulated state and teleportation of the simulated state to the layer at $z=j+1$. 
\end{itemize}
See Figure~\ref{fig:MBQS-FS} for illustration.

Combining everything, we find that the new state is
\begin{widetext}
\begin{align}
&|\phi \rangle_{\bm \Delta^{z=j+1}_0 \cup \bm \Delta^{z=j+1}_1}
= 
\Bigg( 
\prod_{\sigma_1 \in \Delta_1 }
\widetilde{X}(\sigma_1 )^s e^{-i \xi \widetilde{X}(\sigma_1 ) }
\prod_{\sigma_0 \in \Delta_0} X(\sigma_0 )^s e^{-i \xi  X(\sigma_0 ) }  
\prod_{\sigma_0 \in \Delta_0} \frac{1+ (-1)^s X(\sigma_0) \widetilde{X}(\partial^*\sigma_0)}{2}
\, 
\nonumber \\ 
& \times 
\prod_{\sigma_0 \in \Delta_0} Z(\sigma_0)^s
\prod_{\sigma_1 \in \Delta_1} \widetilde{Z}(\sigma_1)^s
\prod_{\sigma_2 \in \Delta_2 } 
\widetilde{Z}(\partial \sigma_2 )^s
e^{-i \xi  \widetilde{Z}(\partial \sigma_2 ) }
\prod_{\sigma_1 \in \Delta_1 }
\big( \widetilde{Z}(\sigma_1)Z(\partial \sigma_1 ) \big)^s
e^{-i \xi\widetilde{Z}(\sigma_1)Z(\partial \sigma_1 ) } 
\Bigg)
|\phi \rangle_{\bm \Delta^{z=j}_0 \cup \bm \Delta^{z=j}_1} \, ,
\end{align}
\end{widetext}
where, as noted, the cell dependence in $\{\xi\}$ and $\{s\}$ is suppressed. 
The parameters are set to realize the time evolution, {\it e.g.} $\xi = \pm \delta t /g$ in the first step. 
The unitaries in the parenthesis above realizes the one step in the time evolution up to the outcome-dependent operators, which are additional byproduct operators. 
They are combined with $\mathcal{O}^{(j)}_{\rm bp}$ to form a new byproduct operator $\mathcal{O}^{(j+1)}_{\rm bp}$ in $|\phi\rangle_{\bm\Delta_0^{z=j+1}\cup \bm\Delta_1^{z=j+1}}=\mathcal{O}^{(j+1)}_{\rm bp} |\phi(t+\delta t)\rangle$. 
As usual in MBQC, one can adaptively adjust the signs in parameters $\{\xi\}$ according to the former measurement outcomes so that the desired time evolution can be obtained deterministically.
Detail of such a technique in related models can be found in~\cite{2023ScPP...14..129S}, where a protocol was presented for $\mathbb{Z}_N$ lattice higher-form gauge theories in general dimensions.
After performing the MBQS protocol over the entire bulk, we have a state $|\phi\rangle_{\bm\Delta^{z=L_z}_0 \cup \bm\Delta^{z=L_z}_1  }= \mathcal{O}^{(L_z)}_{\rm bp} |\phi(T)\rangle$ ($T = L_z \delta t$) at the boundary $z=L_z$, and the byproduct operator $\mathcal{O}^{(L_z)}_{\rm bp}$ can be removed by an appropriate processing.

\subsection{Protection of the gauge invariance}

Physical states in Hamiltonian lattice gauge theory are required to obey the
gauge invariance condition. 
In experiments on noisy quantum devices, it is expected to be important to minimize the effects of errors that violate gauge invariance~\cite{2020Natur.587..392Y}.
In this subsection, we present an error correction procedure in the MBQS of the Fradkin-Shenker model against noises that violate the Gauss law constraint, generalizing the scheme in Ref.~\cite{2023ScPP...14..129S}. 
See also Refs.~\cite{2020PhRvL.125c0503H,2021PRXQ....2d0311H,
2022PhRvR...4c3120H, 2022arXiv220413709H} for studies on protection of the Gauss law constraint in quantum simulation of lattice gauge theories.

In MBQS of the (2+1)d pure lattice gauge theory without a matter field~\cite{2023ScPP...14..129S}, it was proposed to use the dual 1-form symmetry of gCS$_{(3,2)}$ generated by $X(\bm \partial^* \bm \sigma_0)$ 
to diagnose the location of errors relevant to violation of the Gauss law constraint. 
Here, we propose to use a `symmetry' of the resource state for the Fradkin-Shenker model. 
For any vertex $\bm \sigma_0 \in \bm \Delta_0$, we have the following eigenvalue equation derived from the stabilizer conditions:
\begin{align}
&E(\bm \sigma_0)|\text{CS}_\text{FS} \rangle 
=
|\text{CS}_\text{FS} \rangle \, , \\
&E(\bm \sigma_0)
:= X(\bm \sigma_0) \widetilde{X}(\bm \partial^* \bm \sigma_0) \, .
\end{align}
We call this an $E$-symmetry of the noiseless cluster state.
Note that this is different from the Gauss law operator of the Fradkin-Shenker model, $X(\sigma_0)\widetilde{X}(\partial^* \sigma_0)$.

We assume the following noise model.
Let $\{ \bm e^X_0, \bm e^Z_0, \bm e^X_1, \bm e^Z_1,  \tilde{\bm e}^X_1, \tilde{\bm e}^Z_1, \tilde{\bm e}^X_2, \tilde{\bm e}^Z_2 \} $ be chains that represent the location of errors, and let $|\text{CS}^{\text{error}}_\text{FS}\rangle $ be a resource state with errors:
\begin{align}
&|\text{CS}^{\text{error}}_\text{FS}\rangle = \mathbf{E} 
|\text{CS}_\text{FS}\rangle \, , \\
&\mathbf{E}  
= X(\bm e^X_0) Z(\bm e^Z_0) X(\bm e^X_1) Z(\bm e^Z_1)\nonumber \\ &\qquad \times \widetilde{X}(\tilde{\bm e}^X_1) \widetilde{Z}(\tilde{\bm e}^Z_1) \widetilde{X}(\tilde{\bm e}^X_2) \widetilde{Z}(\tilde{\bm e}^Z_2) \, .
\end{align}
We assume that measurements are noiseless. 
We claim that {\it the operator $Z(\bm e^Z_0) \widetilde{Z}(\tilde{\bm e}^Z_1)$ contributes as an error that leads to violation of the Gauss law constraint. }
Other errors lead to a faulty time evolution of the Fradkin-Shenker model, but they do not contribute to violation of the Gauss law constraint. 
Furthermore, we show that {\it we can use the measurement outcomes in MBQS to infer errors through the $E$-symmetry of the cluster state, the result of which can be used to construct counter operators that enforce the Gauss law contraint.}
This is a type of measurement-based feedforward operation.

\subsubsection{Analysis of errors}

First, we study how the error chain affects the time evolution.
To be concrete, we consider a state in the middle of the protocol, with the simulated state at $z=j$, and trace how error operators evolve under measurements from $z=j$ to $z=j+1$. 
We write the simulated state under noises as
\begin{align} \label{eq:simulated-state-form-error-bp}
|\phi\rangle_{\bm\Delta_0^{z=j}\cup \bm\Delta_1^{z=j}}= \pm \mathcal{E}^{(j)} \cdot  \mathcal{O}^{(j)}_{\rm bp} |\phi^{\text{error}}(t)\rangle \, ,
\end{align}
(up to unimportant overall phases $\pm 1$)
where $\mathcal{E}^{(j)}$ is a product of Pauli operators that arise from errors and $|\phi^{\text{error}}(t)\rangle$ is  a state evolved with the exponential operators whose parameters are affected by noises, whose meaning
we will explain below.
We also write the reduced, noisy cluster state at the middle of the protocol as 
\begin{widetext}
\begin{align}  \label{eq:CS-error-reduced}
|{\rm CS}^{(j)\text{error}}_{\rm FS} \rangle 
=  &
\mathbf{E} \Big|_{j \leq z \leq L_z} \cdot 
\Bigg(
\prod_{{\bm\sigma_1} \in {\bm \Delta}_1} C\widetilde{Z}_{\bm\sigma_1, \bm\sigma_1} 
\times 
\prod_{\substack{ \bm\sigma_0 \in \bm \Delta_0 \\  \bm\sigma_1 \in \bm \Delta_1 }  } CZ^{a(\bm \partial \bm \sigma_1; \bm \sigma_0)}_{\bm\sigma_0, \bm\sigma_1} 
\times
\prod_{\substack{ \bm\sigma_1 \in \bm \Delta_1 \\  \bm\sigma_2 \in \bm \Delta_2 }  } \widetilde{CZ}^{a(\bm \partial \bm \sigma_2; \bm \sigma_1)}_{\bm\sigma_1, \bm\sigma_2} \Bigg) \nonumber \\
& \qquad \times 
|+\rangle^{\bm\Delta^{j < z \leq L_z}_0}
|+\rangle^{\bm\Delta^{j \leq z <L_z}_1}
|+\rangle\!\rangle^{\bm\Delta^{j < z  \leq L_z}_1}
|+\rangle\!\rangle^{\bm\Delta^{j \leq z <L_z}_2}
|\phi\rangle_{\bm\Delta^{z=j}_0 \cup \bm\Delta^{z=j}_1} \ . 
\end{align}
\end{widetext}

\begin{table*}
\centering 
\renewcommand{\arraystretch}{1.3}
\begin{tabular}{|l|| c | c | c |} 
    \hline
     & $A$-type &  $B$-type & $X$-type \\ \hline 
 measurement & \makecell{ (1) (2) \\ multi-body interaction} &
 \makecell{(3) (4) (6) (7) \\  (gate) teleportation} & \makecell{(5) \\ checking the Gauss law}  \\ \hline 
 $X$ error   & $(\xi,s+1)$ & $(-\xi,s)$  & $(\text{NA}, s)$ \\  \hline
 $Z$ error   & $(-\xi,s)$ &  $(\xi,s+1)$ & $(\text{NA}, s+1)$ \\ \hline 
\end{tabular}
\caption{Effect of errors as changes from $(\xi, s)$.}
\label{table:effect-of-errors}
\end{table*}

Let us first study how the errors affect measurements individually.
For the $A$-type measurement, an error operator $Z^p X^q$ $(p,q \in \{0,1\})$ affects a simulation step as follows:
\begin{align} \label{eq:A-error}
&\Big(\langle s | e^{i\xi X}\Big)_a \textcolor{red}{Z_a^p X_a^q}
\prod_{b \in Q(a)} CZ_{a,b} |+\rangle_a |\psi\rangle_{\text{others}}\nonumber \\
&= \frac{\textcolor{red}{(-1)^{sp}}}{\sqrt{2}} \Big( \prod_{b \in Q(a)} Z_b \Big)^{s\textcolor{red}{+q}} 
e^{i \textcolor{red}{(-1)^p}\xi \prod_{b \in Q(a)} Z_b } | \psi\rangle_{\text{others}} \, .
\end{align}
For the $B$-type measurement, on the other hand, we find
\begin{align} \label{eq:B-error}
&\Big( \langle s| H e^{i \xi Z}\Big)_a
\textcolor{red}{Z_a^p X_a^q}
CZ_{a,b}
\sum_{z_a=0,1} \psi(z_{a}) |z_a\rangle_a \otimes |+\rangle_b \nonumber \\
&= \Big(\frac{\textcolor{red}{(-1)^{(s+p)q}}}{\sqrt{2}} H e^{i \textcolor{red}{(-1)^q} \xi Z } Z^{s \textcolor{red}{+p} } \Big)_b \sum_{z_a = 0,1} \psi(z_a) |z_a\rangle_b  \, .
\end{align}
Finally, for the $X$-type measurement, we find
\begin{align} \label{eq:X-meas-error}
&\Big(\langle s | H \Big)_a \textcolor{red}{Z_a^p X_a^q} 
\prod_{b \in Q(a)} CZ_{a,b} |+\rangle_a |\psi\rangle_{\text{others}} \nonumber \\
&= \frac{ \textcolor{red}{(-1)^{(s+p)q}} }{2} 
\Big(1 +  (-1)^{s \textcolor{red}{+p} } \prod_{b \in Q(a)} Z_b \Big) 
| \psi\rangle_{\text{others}} \, .
\end{align}
These effects of errors are summarized in Table~\ref{table:effect-of-errors}.
Below, the overall $(-1)$ factors will be omitted.

We study in Appendix~\ref{app:time-evolution-error} 
how the simulated state~\eqref{eq:simulated-state-form-error-bp} evolves by measurements on the cluster state with errors~\eqref{eq:CS-error-reduced}. 
There, we explicitly show that measurements on qubits in $j \leq z < j+1$ evolves the simulated state~\eqref{eq:simulated-state-form-error-bp} such that the updated error operator $\mathcal{E}^{(j+1)}$ acting on it satisfies
\begin{align} \label{eq:error-evolution}
&\mathcal{E}^{(j+1)}\nonumber \\
&=\prod_{\sigma_1} \widetilde{X}(\sigma_1)^{a( \tilde{\bm e}^Z_2; \sigma_1 \times [j,j+1]) }
\nonumber \\
&\qquad \times
\prod_{\sigma_1} X(\sigma_1)^{a(\bm e^Z_1; \sigma_0 \times [j,j+1]) }
\prod_{\sigma_1} \nonumber \\
&\qquad \times
\textcolor{NavyBlue}{ 
\widetilde{Z}(\sigma_1)^{a(\tilde{ \bm e}^Z_1; \sigma_1 \times \{j\}) } }  \prod_{\sigma_0} \textcolor{NavyBlue}{  Z(\sigma_0)^{a(\bm e^Z_0; \sigma_0 \times \{j\}) } }\nonumber \\
&\qquad \times \prod_{\sigma_1} \big( \widetilde{Z}(\sigma_1) Z(\partial \sigma_1) \big)^{a(\bm e^X_1 ; \sigma_1 \times \{j\}) }\nonumber \\
&\qquad \times \prod_{\sigma_2} \widetilde{Z}(\partial \sigma_2)^{a(\tilde{\bm e}^X_2; \sigma_2 \times \{j\})}
\cdot 
\mathcal{E}^{(j)} \, . 
\end{align}
The $Z(\sigma_0)$ and $\widetilde{Z}(\sigma_1)$ operators anti-commute with the Gauss law operator, and hence these operators induced by errors may lead to violation of the Gauss law constraint.
Note that the $Z(\sigma_0)$ and $\widetilde{Z}(\sigma_1)$ operators in the byproduct operator may also anti-commute with the Gauss law operator (see Appendix~\ref{app:time-evolution-error} for the explicit from of $\mathcal{O}^{(j+1)}_\text{bp}$), but any effect caused by byproduct operators is completely negated by feedforward processings as in the usual MBQC. 
We also explicitly show in Appendix~\ref{app:time-evolution-error} that the updated simulated wave function $|\phi^{\text{error}} (t + \delta t) \rangle$ satisfies the Gauss law assuming that the state $|\phi^{\text{error}} (t) \rangle$ satisfies the Gauss law constraint. 
Thus, we arrive at the key observation that {\it only the error chain $\bm e^Z_0 \oplus \tilde{\bm e}^Z_1$ is relevant to violation of the gauge invariance}.
Here the symbol $\oplus$ is used to distinguish spaces which chains live in.

\subsubsection{Detection of errors}

The measurement pattern in our MBQS naturally has a capability to detect the violation of the gauge invariance.
Namely, the measurements in step (5) at $\bm \sigma_1 = \sigma_0 \times [j,j+1]$ and $\bm \sigma_1 = \sigma_0 \times [j-1,j]$ give outcomes that depend on the eigenvalues of the simulated state~\eqref{eq:simulated-state-form-error-bp} with respect to the Gauss law operator.\footnote{
The error represented by $\bm e^Z_0 \oplus \tilde{\bm e}^Z_1$ at $z=j$ cannot be detected directly by each measurement outcome on matter 0-cells or gauge 1-cells in step (3) and (4). It is because the measurement outcomes are not deterministic even when there is no error.
}
However, these measurement outcomes are affected not only by the Pauli operators acting on the simulated wave function~\eqref{eq:simulated-state-form-error-bp}---as we described below~\eqref{eq:error-evolution}---but also by the error operators acting on qubits at cells $\bm \sigma_1 = \sigma_0 \times [j,j+1]$ and $\bm \sigma_1 = \sigma_0 \times [j-1,j]$ themselves---errors also contained in $\bm e^Z_0 \oplus \tilde{\bm e}^Z_1$---via~\eqref{eq:X-meas-error}. 
All the above correlations among the errors and measurement outcomes are integrated into the following relation:
\begin{align} \label{eq:error-outcome-correlation-3d}
&s(\bm \sigma_0)
+ \sum_{\bm \sigma_1} \tilde{s}(\bm \sigma_1) a(\bm \partial^* \bm \sigma_0; \bm \sigma_1) \nonumber \\
&= 
a( \bm e^Z_0 ; \bm \sigma_0)
+ \#( \tilde{\bm e}^Z_1 \cap \bm \partial^* \bm \sigma_0) \qquad \forall \bm \sigma_0 \in \bm \Delta_0\, ,
\end{align}
where 
$s(\bm\sigma)=0,1$ and $\tilde{s}(\bm\sigma)=0,1$ are measurement outcomes at $\bm \sigma$ for matter and gauge qubits, respectively, and
$\bm \sigma_0$ is identified with a dual cube $\bm \sigma^*_3$ to calculate the intersection $\#( \tilde{\bm e}^Z_1 \cap \bm \partial^* \bm \sigma_0)=\#( \tilde{\bm e}^Z_1 \cap \bm \partial^* \bm \sigma^*_3)$.
This is easily derived as the eigenvalue equation (the stabilizer condition) 
\begin{align}\label{eq:E-error-stabilizer}
E(\bm \sigma_0) |\text{CS}^{\text{error}}_\text{FS}\rangle = (-1)^{a( \bm e^Z_0 ; \bm \sigma_0)
+ \#( \tilde{\bm e}^Z_1 \cap \bm \partial^* \bm \sigma_0)} |\text{CS}^{\text{error}}_\text{FS}\rangle 
\end{align}
of $E(\bm\sigma_0)= X(\bm \sigma_0) \widetilde{X}(\bm \partial^* \bm \sigma_0)$ for $|\text{CS}^{\text{error}}_\text{FS}\rangle$ such that the eigenvalue is modified due to the error $Z(\bm e^Z_0) \widetilde{Z}(\tilde{\bm e}^Z_1)$.\footnote{ 
Taking an overlap of \eqref{eq:E-error-stabilizer} with $ \langle S(\bm \sigma_0) |:= \big( \langle s(\bm \sigma_0)|_{\bm \sigma_0} H_{\bm \sigma_0} \big) \otimes \big( \bigotimes_{\bm \sigma_1 \subset \bm \partial^* \bm \sigma_0} \langle \tilde{s}(\bm \sigma_1)|_{\bm \sigma_1} H_{\bm \sigma_1} \big)$, we get $(-1)^{ s(\bm \sigma_0)
+ \sum_{\bm \sigma_1} \tilde{s}(\bm \sigma_1) a(\bm \partial^* \bm \sigma_0; \bm \sigma_1) } $ $ \langle S(\bm \sigma_0) |\text{CS}^{\text{error}}_\text{FS}\rangle $ $= (-1)^{a( \bm e^Z_0 ; \bm \sigma_0)
+ \#( \tilde{\bm e}^Z_1 \cap \bm \partial^* \bm \sigma_0) }$ $\langle S(\bm \sigma_0) |\text{CS}^{\text{error}}_\text{FS}\rangle  $. 
}
We also provide an alternative derivation using an explicit form of the simulated state in Appendix~\ref{app:outcome-error-correlation}.
Hence, we arrive at the following key observation:
{\it We can infer the $\mathbb{Z}_2$ parity of the chain $\bm e^Z_0 \oplus \tilde{\bm e}^Z_1$ around $\bm\sigma_0$ through the relation~\eqref{eq:error-outcome-correlation-3d} by examining measurement outcomes in our MBQS. }

\subsubsection{Correction of errors}

Having understood the usefulness of the $E$-symmetry to detect errors, we present our feedforward correction procedure to enforce the Gauss law constraint.\footnote{
Our proposal derives from the following intuition. When the $\mathbb{Z}_2$ parity of the chain $\bm e^Z_0 \oplus \tilde{\bm e}^Z_1$ around $\bm\sigma_0$ is odd, this can be understood as having a `gauge-variant operator' around it, if we interpret $E(\bm\sigma_0)$ as a Gauss law operator in three dimensions. 
For example, if $\bm e^Z_0 \oplus \tilde{\bm e}^Z_1 = \bm \sigma_0 \oplus \bm 0$, then the operator $Z(\bm \sigma_0)$ can be seen as an isolated charge without any gauge field attached. 
For another example, if $\bm e^Z_0 \oplus \tilde{\bm e}^Z_1 = \bm 0 \oplus \bm \sigma_1$, then the chain $\widetilde{Z}(\bm \sigma_1)$ can be seen as an isolated gauge field without a matter field attached at its endpoints. A simple prescription to remedy such gauge non-invariance would be to place the matter field $Z(\bm \sigma_0)$ whenever we observe $E(\bm\sigma_0) = -1$. 
This enforces $E(\bm\sigma_0) = +1$, which is an eigenvalue equation for the cluster state {\it without $\bm e^Z_0 \oplus \tilde{\bm e}^Z_1$}.
In term of the MBQS protocol, this implies a prescription applying additional $Z(\sigma_0)$ operator to the simulated state. }
Recall that the eigenvalue of the operator~$E( \bm \sigma_0)$  reflects presence of the noises through the relation~(\ref{eq:E-error-stabilizer}).
On the other hand, the eigenvalue of the operator~$E( \sigma_0  \times \{j-1\})$ can be inferred from measurement outcomes in the LHS of~\eqref{eq:error-outcome-correlation-3d}. 
We construct a chain $r^{(j)}_0 \in C_0$ from the measurement outcomes such that
\begin{align} \label{eq:counter-op-determination}
E( \sigma_0  \times \{j-1\}) = (-1)^{a(r^{(j)}_0;\sigma_0)} \, .
\end{align}
Let us introduce a counter operator 
\begin{align}
\mathcal{C}^{(j)}= \textcolor{OliveGreen}{Z(r^{(j)}_0)}
\end{align}
with a 0-chain $r^{(j)}_0$, which will be applied to the 2d state at $z=j$.
Then, we correct the simulated state as
\begin{align}\label{eq:simulated-state-form-error-bp-new}
 &|\phi\rangle_{\bm\Delta_0^{z=j}\cup \bm\Delta_1^{z=j}}\nonumber \\ &\longrightarrow 
|\phi'\rangle_{\bm\Delta_0^{z=j}\cup \bm\Delta_1^{z=j}}= \pm
\mathcal{C}^{(j)} \cdot 
\mathcal{E}^{(j)} \cdot  \mathcal{O}^{(j)}_{\rm bp} |\phi^{\text{error}}(t)\rangle \, ,
\end{align}
which can be implemented by directly applying 
\begin{align}
\mathbf{C}^{(j)}:=Z(r^{(j)}_0 \times \{j\})
\end{align}
to the cluster state as it commutes through $CZ$ operators in~(\ref{eq:CS-error-reduced}).
If we wish, as an equivalent way, we can also implement this correction indirectly by pretending that $\mathcal{C}^{(j)}$ is a part of the byproduct operator. As such, the counter operator $\mathcal{C}^{(j)}$ affects the time evolution in the following time steps just as the byproduct operator $\mathcal{O}^{(j)}_\text{bp}$ does.

\begin{figure*}
    \includegraphics[width=0.7\linewidth]{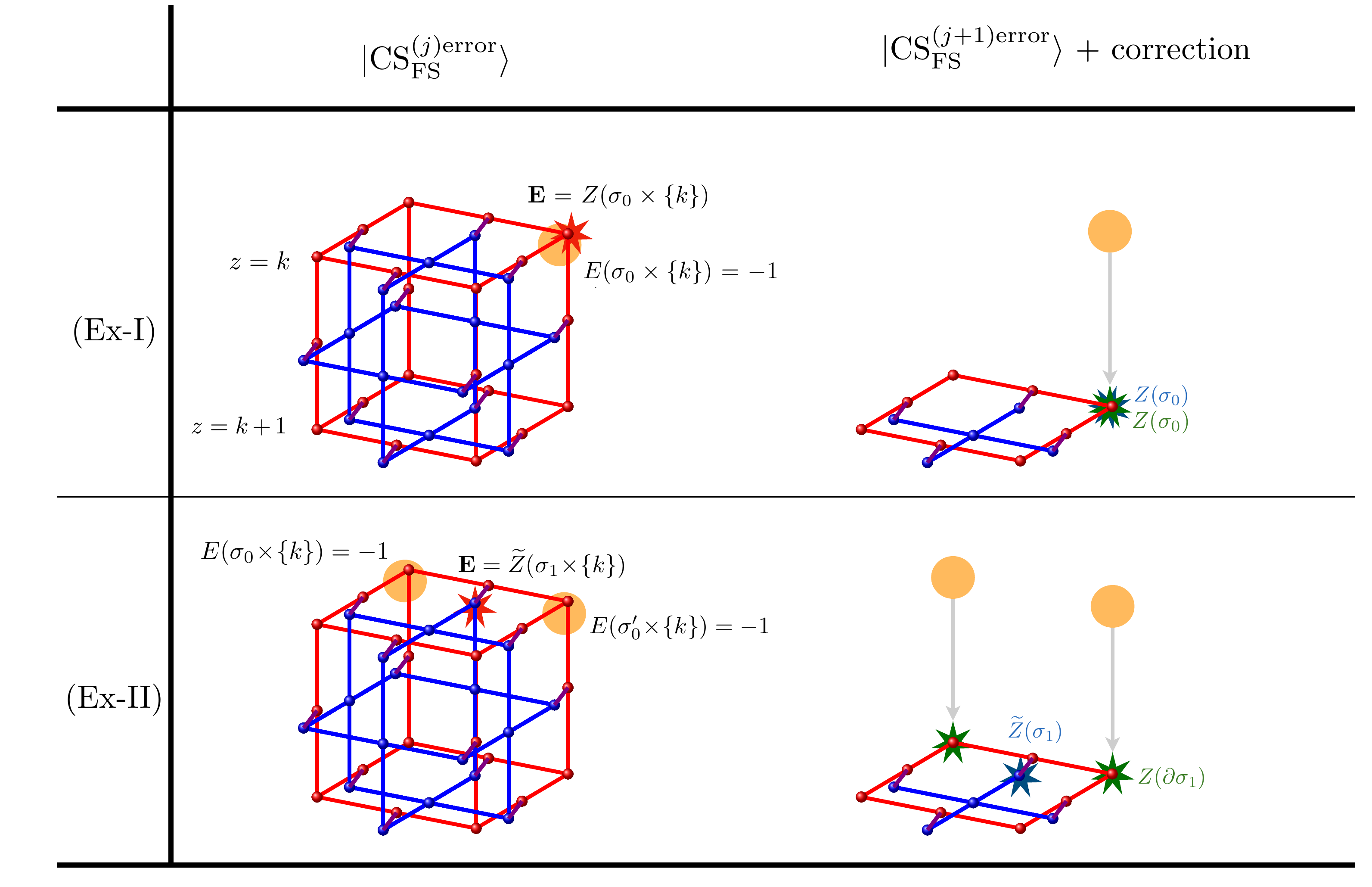}
    \caption{Illustration of the error correction procedure for the examples (Ex-I) and (Ex-II). Errors (red stars) are detected through the operator $E(\bm \sigma_0)$ ($\bm \sigma_0$ such that $E(\bm \sigma_0)=-1$ is represented by an orange circle). The error operators (blue stars) propagated to the layer $z=j+1$ are either canceled or made gauge invariant by counter operators (green stars). }
    \label{fig:error-correction}
\end{figure*}

Let us study a few examples with simple error chains to show how our feedforward correction works.
(This also demonstrates that our correction works for general error chains by linearity.)
\begin{itemize}
\item[(Ex-I)] $\mathbf{E} = Z( \sigma_0 \times \{k\})$:  After measuring qubits up to $z<k+1$, one finds that $E( \sigma_0 \times \{k\} ) =-1$. 
The simulated state induced at $z=k+1$ takes the form 
$|\phi\rangle_{\bm\Delta_0^{z=k+1}\cup \bm\Delta_1^{z=k+1}}= \pm \textcolor{NavyBlue}{Z(\sigma_0)} \cdot  \mathcal{O}^{(k+1)}_{\rm bp} |\phi^{\text{error}}(t+\delta t)\rangle$.
We are instructed to apply $\mathbf{C}^{(k+1)} = Z(\sigma_0 \times \{k+1\})$, which cancels the error operator precisely:
\begin{align}
&|\phi\rangle_{\bm\Delta_0^{z=k+1}\cup \bm\Delta_1^{z=k+1}} \nonumber \\ 
&\longrightarrow \pm 
\textcolor{OliveGreen}{Z(\sigma_0)}\textcolor{NavyBlue}{Z(\sigma_0)} \cdot  \mathcal{O}^{(k+1)}_{\rm bp} |\phi^{\text{error}}(t+\delta t)\rangle \nonumber \\
&\quad = \pm \mathcal{O}^{(k+1)}_{\rm bp} |\phi^{\text{error}}(t+\delta t)\rangle \, . 
\end{align}
The state is gauge invariant up to the byproduct operators. See Figure~\ref{fig:error-correction} for illustration.
\item[(Ex-II)] $\mathbf{E} = \widetilde{Z}( \sigma_1 \times \{k\})$:  After measuring qubits up to $z<k+1$, one finds that $E( \sigma_0 \times \{k\} ) =-1$ and $E( \sigma'_0 \times \{k\} ) =-1$, where $\partial \sigma_1 = \sigma_0 + \sigma'_0$. 
The simulated state induced at $z=k+1$ takes the form 
$|\phi\rangle_{\bm\Delta_0^{z=k+1}\cup \bm\Delta_1^{z=k+1}}= \pm \textcolor{NavyBlue}{\widetilde{Z}(\sigma_1) } \cdot  \mathcal{O}^{(k+1)}_{\rm bp} |\phi^{\text{error}}(t+\delta t)\rangle$.
We are instructed to apply $\mathbf{C}^{(k+1)} = Z(\sigma_0 \times \{k+1\})Z(\sigma'_0 \times \{k+1\})$, which leaves us a state with an insertion of an open Wilson line operator, which is gauge invariant:
\begin{align}|&\phi\rangle_{\bm\Delta_0^{z=k+1}\cup \bm\Delta_1^{z=k+1}} \nonumber \\
&\longrightarrow
\pm \textcolor{OliveGreen}{ Z(\partial \sigma_1) } \textcolor{NavyBlue}{\widetilde{Z}(\sigma_1)} \cdot  \mathcal{O}^{(k+1)}_{\rm bp} |\phi^{\text{error}}(t+\delta t)\rangle \,. 
\end{align}
The state is gauge invariant up to the byproduct operator.
(The additional open Wilson line operator remains as a physical error, however.) See Figure~\ref{fig:error-correction} for illustration.
\item[(Ex-III)] $\mathbf{E} = \widetilde{Z}( \sigma_0 \times [k,k+1])$:
After measuring qubits up to $z<k+1$, one finds that $E( \sigma_0 \times \{k\} ) =-1$. 
The simulated state induced at $z=k+1$ takes the form 
$|\phi\rangle_{\bm\Delta_0^{z=k+1}\cup \bm\Delta_1^{z=k+1}}= \pm \mathcal{O}^{(k+1)}_{\rm bp} |\phi^{\text{error}}(t+\delta t)\rangle$.
In other words, the syndrome measurement misinforms us of an error although there is none acting on the simulated state.
We are instructed to apply $\mathbf{C}^{(k+1)} = Z(\sigma_0 \times \{k+1\})$. 
Then, after further measuring qubits up to $z<k+2$, one finds that $E( \sigma_0 \times \{k+1\} ) =-1$.
We are instructed to apply $\mathbf{C}^{(k+2)} = Z(\sigma_0 \times \{k+2\})$.
The two counter operators cancel each other, leaving some sign flips in the exponents in the unitary time evolution during $k+1<z<k+2$.
Such a side effect is implicitly included in $|\phi^{\text{error}}(t+2\delta t)\rangle$. 
The simulated state after corrections, $\pm \textcolor{OliveGreen}{\mathcal{C}^{(k+2)} \mathcal{C}^{(k+1)} } \mathcal{O}_\text{bp}^{(k+2)}|\phi^{\text{error}}(t+2\delta t)\rangle = \pm  \mathcal{O}_\text{bp}^{(k+2)}|\phi^{\text{error}}(t+2\delta t)\rangle$, is gauge invariant up to the byproduct operator.
\end{itemize}

The method presented in this section is a generalization of the one for the (2+1)d $\mathbb{Z}_2$ pure lattice gauge theory given in Ref~\cite{2023ScPP...14..129S}, which builds upon the idea used in topological quantum memories~\cite{2002JMP....43.4452D, 2006AnPhy.321.2242R,2007NJPh....9..199R, 2007PhRvL..98s0504R}.
The correction procedure to protect the gauge invariance in the pure lattice gauge theory involved a task of finding optimal paths connecting points in spacetime to make open line operators made of errors closed, where a classical side-processing with the so-called Minimum-Weight Perfect Matching algorithm~\cite{Edmonds_1965} was suggested to use.
In the method proposed in this section, on the other hand, the counter operator is determined uniquely by~\eqref{eq:counter-op-determination}, and it does not necessitate use of non-trivial classical side-processings.

\section{Dualities and phases in the Fradkin-Shenker model} \label{sec:duality-phases-FS}

\subsection{Self-duality in classical partition function}

Let us present a Van den Nest-D\"{u}r-Briegel-type formula~\cite{PhysRevLett.98.117207}  for the Fradkin-Shenker model.
We impose the periodic boundary conditions in all directions.
Consider a product (bra) state
\begin{align}
\begin{split}
&\langle \Omega(J,K) | \\
&:= \bigotimes_{{\bm \sigma_0 \in \bm \Delta_0 } } \langle + | 
 \bigotimes_{{\bm \sigma_1 \in \bm \Delta_1 } } \langle 0 | e^{JX} 
 \bigotimes_{{\bm \sigma_1 \in \bm \Delta_1 } } \langle\!\langle + | 
 \bigotimes_{{\bm \sigma_2 \in \bm \Delta_2 } } \langle\!\langle 0 | e^{KX} \ . 
 \end{split}
\end{align}
Then, we find that the classical partition function~(\ref{eq:FS-partition-function}) can be written as
\begin{align}
Z_{\rm FS}(J,K) = \mathcal{N} \times \langle \Omega(J,K) | \rm{CS}_{\rm FS} \rangle \, 
\end{align}
with a normalization factor $\mathcal{N}$.
We consider the state $|\text{CS}_\text{FS}\rangle$ on a periodic lattice.

\begin{figure*}
\includegraphics[scale=0.9]{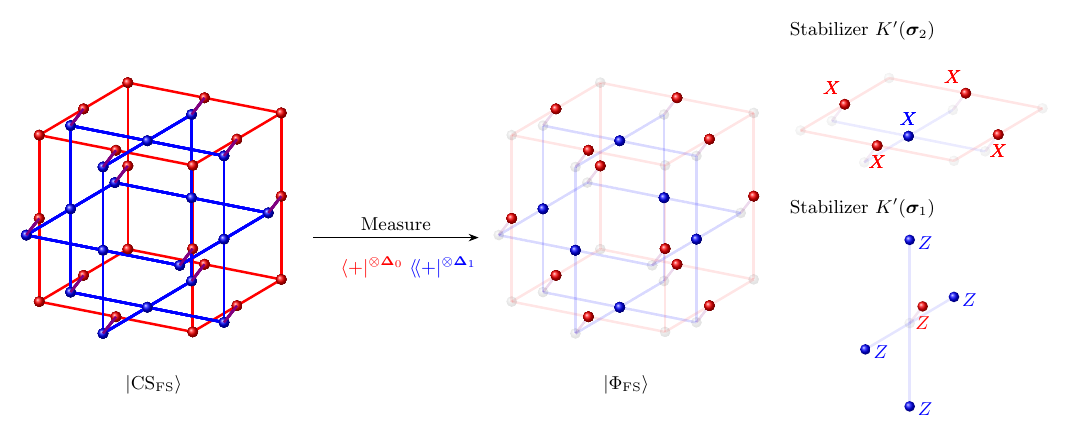}
    \caption{Partially measuring the state $|\text{CS}_\text{FS}\rangle$, we obtain the state $|\Phi_\text{FS}\rangle$ in \eqref{eq:partially-measuring-FS}. The state is stabilized by $K'(\bm \sigma_1)$ and $K'(\bm \sigma_2)$, which are the stabilizers for the Raussendorf-Bravyi-Harrington state up to Hadamard transformations.}
    \label{fig:FS-partial-meas}
\end{figure*}

We now demonstrate the self-duality of the Euclidean path integral using the Van den Nest-D\"{u}r-Briegel-type formula. 
We first note that, by partially taking the overlap with $\langle +|$ states, the cluster state reduces to
another cluster state
\begin{equation} \label{eq:partially-measuring-FS}
|\Phi_{\rm FS}\rangle :=
\langle +|^{\otimes \bm \Delta_0}
\langle\!\langle +|^{\otimes\bm\Delta_1}
| \rm{CS}_{\rm FS} \rangle
\end{equation}
supported on 2- and 1-cells 
characterized by the stabilizers
\begin{align}
K'({\bm \sigma_1}) = Z({\bm \sigma_1}) \widetilde{Z}({\bm \partial^* \bm \sigma_1}) \, , \quad 
K'({\bm \sigma_2}) = \widetilde{X}({\bm \sigma_2}) X({\bm \partial \bm \sigma_2}) \,   , 
\end{align}
which determines a unique eigenstate as the number of $K'$ operators matches the number of unmeasured qubits. See Figure~\ref{fig:FS-partial-meas} for illustration.
Indeed, $|\Phi_{\rm FS}\rangle$ is the cluster state gCS$_{(3,2)}$ up to Hadamard transformations on 1-cells.\footnote{\label{footnote:FS-unitary-gauge}
It is a manifestation of the fact that the Fradkin-Shenker model can be rewritten in terms of link variables $z(\bm\sigma_1)=U(\bm\sigma_1)s(\bm\partial\bm\sigma_1)$ by taking the unitary gauge~\cite{PhysRevD.19.3682,fradkin_2013}. 
Namely, we obtain the partition function with the action consisting of a plaquette term $z(\bm\partial\bm\sigma_2)$ and an additional single-link term $z(\bm\sigma_1)$ if we evaluate the overlap between the remaining rotated bases and gCS$_{(3,2)}$.}
Next, we note that applying the Hadamard 
transformations on the qubits on 1- and 2-cells, $\mathsf{H} = \prod_{\bm \sigma \in \bm \Delta_1 \cup \bm \Delta_2}H_{\bm \sigma}$, gives us a stabilizer state with
\begin{equation}
\begin{split}
&\mathsf{H}  K'({\bm \sigma_1}) \mathsf{H}  = X({\bm \sigma_1}) \widetilde{X}({\bm \partial^* \bm \sigma_1}) \, , \quad  \\
&\mathsf{H}  K'({\bm \sigma_2}) \mathsf{H}  = \widetilde{Z}({\bm \sigma_2}) Z({\bm \partial \bm \sigma_2}) \,   .
\end{split}
\end{equation}
This state can also be obtained by measuring 3- and 2-cell on a ``dual resource state" 
\begin{align}
|\rm{CS}_\text{FS,dual} \rangle = \mathsf{T}_{\frac{1}{2}} |\rm{CS}_{\rm FS} \rangle \, ,
\end{align}
which is related to the state $|\rm{CS}_{\rm FS} \rangle$ by $\mathsf{T}_{\frac{1}{2}}$, the shift of the lattice by $(\frac{1}{2},\frac{1}{2},\frac{1}{2})$ ({\it i.e.,} dualizing). 
Let us define an operator $\mathsf{S}$ that exchanges between the gauge and matter qubits: 
\begin{align} \label{eq:S-def}
& \mathsf{S} 
\bigotimes_{\bm \sigma_1} |s_{\bm \sigma_1} \rangle\!\rangle_{\bm \sigma_1} 
\bigotimes_{\bm \sigma_2} |t_{\bm \sigma_2} \rangle_{\bm \sigma_2}
=
\bigotimes_{\bm \sigma_1} |s_{\bm \sigma_1} \rangle_{\bm \sigma_1} 
\bigotimes_{\bm \sigma_2} |t_{\bm \sigma_2} \rangle\!\rangle_{\bm \sigma_2} 
\, .
\end{align}
Then, our claim is: 
\begin{align} 
\begin{split}\label{eq:state_duality_FS}
&\mathsf{H} \cdot \langle +|^{\otimes \bm \Delta_0} \langle\!\langle +|^{\otimes \bm \Delta_1} 
 |\text{CS}_\text{FS}\rangle \\
& = 
 \mathsf{S} \cdot  
 \langle +|^{\otimes \bm \Delta_3} \langle\!\langle +|^{\otimes \bm \Delta_2} 
 |\text{CS}_\text{FS,dual}\rangle  \, .
 \end{split}
\end{align}
Unlike the corresponding relations~(\ref{eq:H-Phi-Phi-star}) and (\ref{eq:Phi-star-H-Phi}) in the case of the Wegner model, there is no summation over defects.
Taking the inner product for both sides in~\eqref{eq:state_duality_FS} with the bra vector $\bigotimes_{\bm \sigma_2 \in \bm \Delta_2 } \langle\!\langle 0 | e^{KX}
\bigotimes_{\bm \sigma_1 \in \bm \Delta_1 } \langle 0 | e^{JX} = 
\bigotimes_{\bm \sigma_2 \in \bm \Delta_2 }\langle 0 | e^{KX}
\bigotimes_{\bm \sigma_1 \in \bm \Delta_1 }  \langle\!\langle 0 | e^{JX} \cdot \mathsf{S}^{\dagger}$, we arrive at the equality,
\begin{widetext}
\begin{align}
&\Big( \bigotimes_{\bm \sigma_2 \in \bm \Delta_2 } \langle\!\langle 0 | e^{KX}
\bigotimes_{\bm \sigma_1 \in \bm \Delta_1 } \langle 0 | e^{JX} 
\Big)
\cdot 
\mathsf{H}
\cdot 
\Big(
\bigotimes_{\bm \sigma_1 \in \bm \Delta_1 }  \langle\!\langle  +| 
\bigotimes_{\bm \sigma_0 \in \bm \Delta_0 } \langle + | 
\Big)   |{\rm CS}_{\rm FS} \rangle \nonumber \\ 
&=
\Big( \bigotimes_{\bm \sigma_2 \in \bm \Delta_2 } \langle\!\langle 0 | e^{KX}
\bigotimes_{\bm \sigma_1 \in \bm \Delta_1 } \langle 0 | e^{JX} 
\Big)
 \cdot \mathsf{S} \cdot 
\Big(
\bigotimes_{\bm \sigma_3 \in \bm \Delta_3 }  \langle  +| 
\bigotimes_{\bm \sigma_2 \in \bm \Delta_2 } \langle\!\langle + | 
\Big) 
|{\rm CS}_\text{FS,dual} \rangle \, \nonumber \\  
&=
\Big( \bigotimes_{\bm \sigma_2 \in \bm \Delta_2 } \langle 0 | e^{KX}
\bigotimes_{\bm \sigma_1 \in \bm \Delta_1 } \langle\!\langle 0 | e^{JX} 
\Big)
\cdot \mathsf{S}^{\dagger} \mathsf{S} \cdot 
\Big(
\bigotimes_{\bm \sigma_3 \in \bm \Delta_3 }  \langle  +| 
\bigotimes_{\bm \sigma_2 \in \bm \Delta_2 } \langle\!\langle + | 
\Big) 
|{\rm CS}_\text{FS,dual} \rangle \, \nonumber \\
&=
\Big( \bigotimes_{\bm \sigma_2 \in \bm \Delta_2 } \langle 0 | e^{KX}
\bigotimes_{\bm \sigma_1 \in \bm \Delta_1 } \langle\!\langle 0 | e^{JX} 
\Big)
\Big(
\bigotimes_{\bm \sigma_3 \in \bm \Delta_3 }  \langle  +| 
\bigotimes_{\bm \sigma_2 \in \bm \Delta_2 } \langle\!\langle + | 
\Big) 
|{\rm CS}_\text{FS,dual} \rangle \,
, \label{eq:FS-SD-intermed}
\end{align}
where we used $\mathsf{S}^\dagger \mathsf{S}={\rm id}$, which can be derived from the definition~\eqref{eq:S-def}.
\end{widetext}

We further notice a set of relations
\begin{align}
&\langle\!\langle + | e^{KZ} = 
(\sinh 2K)^{1/2}
\langle\!\langle 0 | e^{K^* X} \nonumber\\
&\qquad {\rm when~} K^* = - \frac{1}{2} \ln \tanh (K) \, , \\ 
&\langle + | e^{JZ} = 
(\sinh 2J)^{1/2}
\langle 0 | e^{J^* X} \nonumber\\
&\qquad {\rm when~} J^* = - \frac{1}{2} \ln \tanh (J) \, .
\end{align}
Therefore, the left-hand side of \eqref{eq:FS-SD-intermed} becomes $Z_{\rm FS}(J^*,K^*)$ up to normalization, and the right-hand side becomes $Z_{\rm FS}(K,J)$. 
This is the renowned self-duality of the model under the relations between couplings, $(J^*,K^*)=(K,J)$.

\subsection{Self-duality in  quantum states}

The $(2+1)$d Fradkin-Shenker model is self-dual as we saw above.
Here we will demonstrate that such a self-dual transformation can be physically realized by $CZ$ entanglers and local measurements along the line of~\cite{https://doi.org/10.48550/arxiv.2112.01519}.

We recall from the beginning of this section that the degrees of freedom of the model reside in 0- and 1-cells of the square lattice.
Let us introduce additional qubits on 1- and 2-cells and call them {\it dualized qubits};
qubits that describe the original degrees of freedom shall be called {\it undualized qubits}.
We note that gauge and matter degrees of freedom live on 1- and 2-cells on dualized qubits, respectively. 
We use the prime ($'$) for cells and/or the bold font for Pauli operators to distinguish between them.
Consider the following map:
\begin{align}
\mathsf{D} 
= \langle +|^{\Delta_1} \langle +|^{\Delta_0} 
\mathcal{U}_{{\rm SD}} |+\rangle^{\Delta'_1} |+\rangle^{\Delta'_2}  
\end{align}
with the entangler 
\begin{align}
\begin{split}
\mathcal{U}_{\rm SD} 
= & \Big( \prod_{\substack{\sigma_0 \in \Delta_0 \\ \sigma_1 \in \Delta_1} }  C{\bm Z}^{a(\partial^* \sigma_0; \sigma_1)}_{\sigma_0, \sigma'_1} \Big) \\
&\times 
\Big(
\prod_{\substack{\sigma_1 \in \Delta_1 \\ \sigma_2 \in \Delta_2} }  C{\bm Z}^{a(\partial^* \sigma_1; \sigma_2)}_{\sigma_1, \sigma'_2} C{\bm Z}_{\sigma_1, \sigma'_1} \Big) \, . 
\end{split}
\end{align}
One can also express this as 
\begin{align}
\mathsf{D} 
= \langle +|^{\Delta_0} 
\mathcal{U}^{(2) }_{{\rm SD}} |+\rangle^{\Delta'_2}  
\end{align}
with the entangler 
\begin{align}
\mathcal{U}^{(2) }_{\rm SD} 
= & \Big( \prod_{\substack{\sigma_0 \in \Delta_0 \\ \sigma_1 \in \Delta_1} }  C{\bm Z}^{a(\partial^* \sigma_0; \sigma_1)}_{\sigma_0, \sigma'_1} \Big) \nonumber \\
&\times  \Big(
\prod_{\sigma_1 \in \Delta_1} \frac{1}{\sqrt 2} H_{\sigma_1} \Big)
\times
\Big(
\prod_{\substack{\sigma_1 \in \Delta_1 \\ \sigma_2 \in \Delta_2} }  C{\bm Z}^{a(\partial^* \sigma_1; \sigma_2)}_{\sigma_1, \sigma'_2} \Big) 
\end{align}
upon identifying the undualized and dualized qubits on $\sigma_1$ and $\sigma'_1$.
We note that the map $\mathcal{U}^{(2) }_{{\rm SD}} |+\rangle^{\Delta'_2}$ brings a state subject to the Gauss law constraint~(\ref{eq:FS-Gauss}) to a state with the condition $X(\sigma_0)=+1$. 
Therefore, even without an explicit projection by $\langle +|^{\Delta_0}$, the vertex degrees of freedom are disentangled, so its role 
is simply to remove the qubits from the output.
Each term in the Hamiltonian~ (\ref{eq:FS-Hamiltonian}) gets transformed as follows:
\begin{align}
&\mathsf{D} \cdot X(\sigma_0) 
=  \widetilde{ {\bm Z} }(\partial^* \sigma^*_2) \cdot \mathsf{D}  \qquad (\sigma_0 \simeq \sigma^*_2) \, ,\\ 
&\mathsf{D} \cdot \widetilde{X}(\sigma_1) =  \widetilde{{\bm Z}}(\sigma^*_1) {\bm Z}(\partial^* \sigma^*_1) \cdot \mathsf{D} \qquad (\sigma_1 \simeq \sigma^*_1) \, , \\ 
&\mathsf{D} \cdot \widetilde{Z}(\partial \sigma_2) =  {\bm X}(\sigma^*_0) \cdot \mathsf{D} \qquad (\sigma_2 \simeq \sigma^*_0) \, , \\ 
&\mathsf{D} \cdot \widetilde{Z}(\sigma_1) Z(\partial \sigma_1) =  \widetilde{{\bm X}}(\sigma^*_1) \cdot \mathsf{D} \qquad (\sigma_1 \simeq \sigma^*_1) \, , \label{eq:FS-duality-4}
\end{align}
and the generator of the gauge transformation gets trivialized, 
\begin{align}
\mathsf{D} \cdot X(\sigma_0) \widetilde{X}(\partial^* \sigma_0) = {\bm I} \cdot \mathsf{D} \ . 
\end{align}
In the image of the map $\mathsf{D}$, the gauge symmetry is emergent as we also have $\mathsf{D}  \cdot {I}  = {\bm X}(\sigma^*_0) \widetilde{{\bm X}}(\partial \sigma^*_0) \cdot  \mathsf{D}$.

We note that the duality operator $\mathsf{D}$ and its reverse 
\begin{align}
\mathsf{D}^* 
= \langle +|^{\Delta'_1} \langle +|^{\Delta'_2} 
\mathcal{U}_{{\rm SD}} |+\rangle^{\Delta_1} |+\rangle^{\Delta_0} 
\end{align}
obeys a non-invertible fusion rule. 
We find by a calculation similar to the one given for Wegner's model that 
\begin{align}
&\langle t_{\sigma_1} , t_{\sigma_0} |  \mathsf{D}^*  \circ \mathsf{D} | s_{\sigma_1} , s_{\sigma_0} \rangle \nonumber \\
& = \frac{1}{2^{|\Delta_0|+|\Delta_1|}}
 \prod_{\sigma_1 \in \Delta_1}  \delta^{\rm mod~2}_{s_{\sigma_1}  + t_{\sigma_1}  + \sum_{ \sigma_0 \subset \sigma_1 } (s_{\sigma_0} + t_{\sigma_0})  , 0 } \nonumber \\
 &\qquad \times \prod_{\sigma_2 \in \Delta_2}  \delta^{\rm mod~2}_{\sum_{ \sigma_1 \subset \sigma_2 } (s_{\sigma_1} + t_{\sigma_1})  , 0 } 
 \end{align}
 which implies
 \begin{align}
 \mathsf{D}^*  \circ \mathsf{D}  & = \frac{1}{2^{|\Delta_0|+|\Delta_1|}}\prod_{\sigma_0 \in \Delta_0} (1 + X(\sigma_0) \widetilde{X}(\partial^* \sigma_0)) \, .
\end{align}

\subsection{Phases of Fradkin-Shenker model}

The resource state $|{\rm CS}_{\rm FS}\rangle$ does not belong to the SPT order of either gCS${}_{(3,2)}$ or gCS${}_{(3,1)}$, as the minimal coupling entangler $\mathcal{U}_{\rm m.c.}$ explicitly breaks some of the symmetries that protect the SPT order. 
We argue that, if we introduce boundaries to the resource state, taking an overlap between the resource state and a product state with parameters defined on the bulk, the remaining boundary state is projected to a state which may possess a topological order, a nontrivial SPT order or a trivial SPT order, depending on the parameters. 
The aim of this subsection is to present an interplay of symmetries and (symmetry-protected) topological orders between the bulk resource state and the boundary induced state related through measurements.
As a remark, although the product state $\langle \Omega (J,K)|$ that we consider is not a measurement basis that we can practically use in MBQS,\footnote{An imaginary-time quantum simulation for the Fradkin-Shenker model using the resource state $|\text{CS}_\text{FS}\rangle$ and the measurement basis that contains $|\Omega(J,K)\rangle$, {\it e.g.}
$\{ e^{KX}|0\rangle , \, e^{-KX}|1\rangle\}$, is not practical. This is because the measurement with such orthogonal bases gives a time evolution in the opposite time direction, thus we may need post-selection for each measurement with high probability. In our previous work \cite{2023ScPP...14..129S}, we presented how to modify cluster states and measurement bases to implement an imaginary-time evolution more practically.} taking an overlap may be seen as an analogue of the imaginary-time evolution implemented by measurement and post-selection, where the boundary state belongs to different phases depending on coupling constants.\footnote{The basis $\langle 0|e^{KX}$ in $\langle \Omega(J,K)|$ can be seen as the imaginary-time version of the $A$-type measurement basis $\langle 0|e^{i\xi X}$ as it can be obtained by taking $K = i \xi$. It can be also seen as the imaginary-time version of the $B$-type measurement basis $\langle +|e^{i\xi Z}$, as $\langle +|e^{kZ} = \sqrt{\sinh(k)} \langle 0| e^{k^*X}$ with $k^* =-\frac12 \log \tanh k$. The same applies to the part with $J$.}

Let us first look at symmetries of the state $|{\rm CS}_{\rm FS}\rangle$ on a periodic lattice.
The state $|{\rm CS}_{\rm FS}\rangle$ obeys the following symmetries that transcend from gCS${}_{(3,2)}$ or gCS${}_{(3,1)}$:
\begin{align}
P := \prod_{ {\bm \sigma_0} \in \bm \Delta_0 } X({\bm \sigma_0}) = 1 \,  , \\ 
M({\bm z_2}) := \widetilde{X}({\bm z_2}) = 1 \, ,
\end{align}
where $\bm z_2$ is a 2-cycle.
The former is the 0-form symmetry that carries over from gCS${}_{(3,1)}$, and the latter is the 1-form symmetry that comes from gCS${}_{(3,2)}$. 
In addition, the state is characterized by stabilizers labeled by arbitrary 0- and 2-chains:
\begin{align}
\Gamma({\bm c_2}) := X({\bm \partial \bm c_2}) \widetilde{X}({\bm c_2}) =1   \, , \\ 
E({\bm c_0}) := \widetilde{X}({\bm \partial^* \bm c_0}) X({\bm c_0})= 1   \, \ . 
\end{align}
Below, we consider introducing boundaries and how the symmetries terminate on them.

\begin{figure*}
\includegraphics[width=0.9\linewidth]{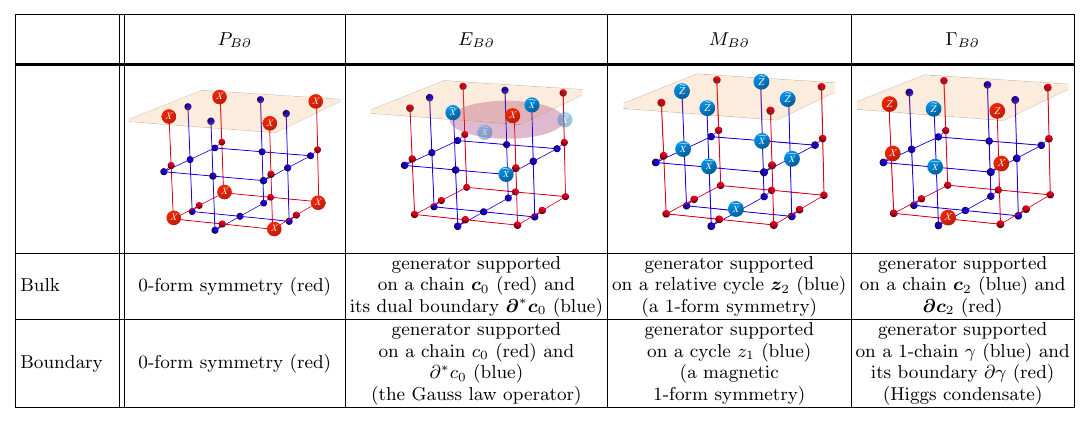}
\caption{Bulk and boundary symmetries in the resource state for the MBQS of the Fradkin-Shenker model.}
\label{tab:FS-model-sym}
\end{figure*}

For concreteness, we take the open boundary condition for the $z$ direction, and the periodic boundary condition for both $x$ and $y$ directions. 
The lattice has two boundaries at $z=0$ and $z=L_z$.
On the boundary at $z=L_z$, qubits are placed on 0-cells $\Delta_0$ and 1-cells $\Delta_1$, as depicted in Figure~\ref{tab:FS-model-sym}.
On the other boundary at $z=0$ and in the bulk $0<z<L_z$, qubits are placed on 0-cells $\bf \Delta_0$ and 1-cells $\bf\Delta_1$ as in gCS${}_{(3,1)}$, as well as on 1-cells $\bf\Delta_1$ and 2-cells $\bf\Delta_2$ as in gCS${}_{(3,2)}$.
We write the resource state as
\begin{widetext}
\begin{align}
|{\rm CS}_{\rm FS} \rangle_{B \partial}
=  &
\Bigg(
\prod_{{\bm\sigma_1} \in {\bm \Delta}_1} C\widetilde{Z}_{\bm\sigma_1, \bm\sigma_1} 
\times 
\prod_{\substack{ \bm\sigma_0 \in \bm \Delta_0 \\  \bm\sigma_1 \in \bm \Delta_1 }  } CZ^{a(\bm \partial \bm \sigma_1; \bm \sigma_0)}_{\bm\sigma_0, \bm\sigma_1} 
\times
\prod_{\substack{ \bm\sigma_1 \in \bm \Delta_1 \\  \bm\sigma_2 \in \bm \Delta_2 }  } \widetilde{CZ}^{a(\bm \partial \bm \sigma_2; \bm \sigma_1)}_{\bm\sigma_1, \bm\sigma_2} \Bigg) \nonumber \\
& \qquad \times 
|+\rangle^{\bm\Delta^{0 \leq z \leq L_z}_0}
|+\rangle^{\bm\Delta^{0 \leq z <L_z}_1}
|+\rangle\!\rangle^{\bm\Delta^{0 \leq z  \leq L_z}_1}
|+\rangle\!\rangle^{\bm\Delta^{0 \leq z <L_z}_2}  \ . 
\end{align}
It is a special state $|{\rm CS}^{(0)}_{\rm FS}\rangle$ in eq.~\eqref{eq:FS-MBQS-resource} with $|\phi\rangle_{\Delta^{z=0}_0 \cup \Delta^{z=0}_1} = |+\rangle^{\Delta^{z=0}_0 \cup \Delta^{z=0}_1}$.

We denote bulk-boundary symmetry generators by subscript $B\partial$. 
For symmetry generators $\{M,\Gamma,E \}$, it suffices for our discussion to focus on those in the vicinity of the future boundary $z=L_z$.
We find the following is the symmetry of the state $|{\rm CS}_{\rm FS} \rangle_{B \partial}$:
\begin{align}
P_{B\partial} &:= \underbrace{\prod_{ {\bm \sigma_0} \in \bm \Delta_0 } X({\bm \sigma_0})\times  \prod_{ {\sigma_0} \in  \Delta^{z=0}_0 } X({ \sigma_0})   }_{=: P_B} \times \underbrace{\prod_{ {\sigma_0} \in  \Delta^{z=L_z}_0 } X({ \sigma_0}) }_{=: P_{\partial}} \,  , \\ 
M_{B\partial} ({\bm z_2}) &:= 
\underbrace{\widetilde{X}({\bm z_2}) }_{=: M_{B} ({\bm z_2})} \times  \underbrace{\widetilde{Z}(z_1)}_{=: M_{\partial} ({z_1})}  \quad {\rm such~that~}  {\bm \partial \bm z_2} =  z_1  \, , \\
\Gamma_{B\partial}({\bm c_2}) &:= 
\underbrace{ X({\bm \partial \bm c_2}) \widetilde{X}({\bm c_2})}_{=:\Gamma_{B}({\bm c_2}) }  \times
\underbrace{ \widetilde{Z}(\gamma) Z(\partial \gamma) }_{=:\Gamma_{\partial}({ \gamma}) } \quad {\rm such~that~}    {\bm \partial \bm c_2}\big|_{\partial} = \gamma  \, , \\ 
E_{B\partial}({\bm c_0}) &:= 
\underbrace{
\widetilde{X}( {\bm \partial^* \bm c_0} ) X({\bm c_0}) }_{ =: E_B({\bm c_0})}
\times 
\underbrace{\widetilde{X}(\partial^* c_0) X(c_0) }_{=: E_\partial(c_0)}  \quad {\rm such~that~} {\bm c_0}\big|_{\partial} =c_0 \, \ ,
\end{align}
and they are depicted in Figure~\ref{tab:FS-model-sym}.
\end{widetext}


\begin{figure*}
\includegraphics[width=0.6\linewidth]{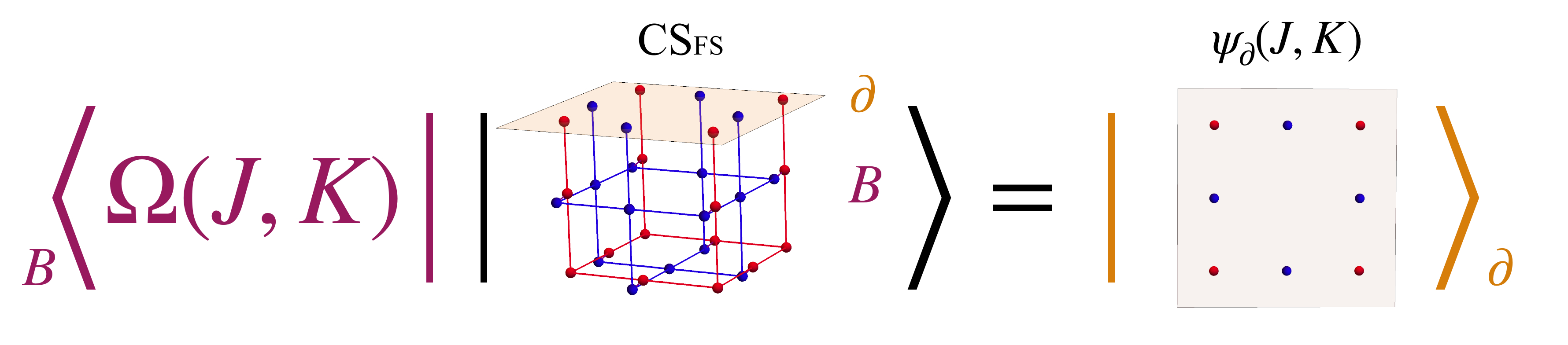} \\
\begin{tabular}{|c||c | c | c | c ||c|}
\hline
\makecell{bulk overlap vector \\ $\langle \Omega(J,K)|_B$}& $P_{B\partial}$ &  $E_{B\partial}$ &  $M_{B\partial}$ &   $\Gamma_{B\partial}$ & \makecell{induced boundary state \\ $|\psi_\partial(J,K)\rangle$}  \\
\Xhline{4\arrayrulewidth}
$\Omega(\infty,\infty)$ & \checkmark & \checkmark & \checkmark &  \checkmark & \makecell{the cluster state $\text{gCS}_{(2,1)}$ \\ (SPT, Higgs)}\\
\hline
$\Omega(0,\infty)$ & \checkmark & \checkmark  & \checkmark  & broken & \makecell{the toric code state \\ (topologically ordered, deconfining)} \\
\hline
$\Omega(0,0)$ & \checkmark  & \checkmark  & broken & broken & \makecell{the product state \\ (trivially ordered, confining)}  \\
\hline
\end{tabular}
\caption{When bulk qubits of the resource state are projected by $\langle \Omega(J,M)|$, the state at the boundary after such partial measurement has correlations according to the bulk-boundary symmetries of the pre-measurement resource state.
In the middle columns, we indicate whether the overlap state is an eigenstate of each bulk-boundary symmetry generator (`\checkmark') or not (`broken'). The boundary state induced by the overlap has corresponding boundary symmetries listed in Figure~\ref{tab:FS-model-sym}. }
\label{fig:FS-model-VDB}
\end{figure*}

We take an overlap with $\langle \Omega (J,K)|_B$ in the bulk $0<z<L_z$ as well as the past boundary $z=0$, as depicted in Figure~\ref{fig:FS-model-VDB}. 
Below, we wish to consider a wave function 
\begin{align}
|\psi_{\partial }(J,K)\rangle := \langle \Omega(J,K) |_B |{\rm CS}_{\rm FS}\rangle_{B\partial}  \ , 
\end{align}
and investigate the state $|\psi_{\partial }(J,K)\rangle$ induced at $z=L_z$.

Note that
\begin{align} \label{eq:prod_state_expansion-J}
\langle 0 | e^{J(X-I)} = e^{-2J} \langle 0 | + (1- e^{-2J}) \frac{\langle + |}{ \sqrt{2}} \ ,  \\ 
\langle 0 | e^{K(X-I)} = e^{-2K} \langle 0 | + (1- e^{-2K}) \frac{\langle + |}{ \sqrt{2}} 
\label{eq:prod_state_expansion-K} \ . 
\end{align}
Thus, in interesting limits, the overlap bra state $\langle \Omega(J,K)|$ becomes
\begin{widetext}
\begin{align} \label{eq:prod_state_limits}
\langle \Omega(J,K) | \nonumber 
\propto 
\begin{cases}
 \bigotimes_{{\bm \sigma_0 \in \bm \Delta_0 } } \langle + | 
 \bigotimes_{{\bm \sigma_1 \in \bm \Delta_1 } } \langle 0 |
 \bigotimes_{{\bm \sigma_1 \in \bm \Delta_1 } } \langle\!\langle + | 
 \bigotimes_{{\bm \sigma_2 \in \bm \Delta_2 } } \langle\!\langle 0 |   & (J,K) \rightarrow (0,0) \, , \\
 \bigotimes_{{\bm \sigma_0 \in \bm \Delta_0 } } \langle + | 
 \bigotimes_{{\bm \sigma_1 \in \bm \Delta_1 } } \langle + |
 \bigotimes_{{\bm \sigma_1 \in \bm \Delta_1 } } \langle\!\langle + | 
 \bigotimes_{{\bm \sigma_2 \in \bm \Delta_2 } } \langle\!\langle + |   & (J,K) \rightarrow (\infty,\infty) \, ,\\
 \bigotimes_{{\bm \sigma_0 \in \bm \Delta_0 } } \langle + | 
 \bigotimes_{{\bm \sigma_1 \in \bm \Delta_1 } } \langle 0 |
 \bigotimes_{{\bm \sigma_1 \in \bm \Delta_1 } } \langle\!\langle + | 
 \bigotimes_{{\bm \sigma_2 \in \bm \Delta_2 } } \langle\!\langle + |   & (J,K) \rightarrow (0,\infty) \, . 
\end{cases}
\end{align}
\end{widetext}
We recall that bra states are defined on cells in $0\leq z <L_z$. 
We use the following type of relations to extract symmetries of the boundary state at $z=L_z$.
For example, we have
\begin{align}
\langle \Omega (J, K)| {\rm CS}_{\rm FS}\rangle_{B\partial} 
&= \Big( \langle \Omega (J, K)| P_B \Big) \Big( P_{B\partial}  |{\rm CS}_{\rm FS}\rangle_{B\partial} \Big) \nonumber \\
&= P_{\partial} \langle \Omega (J, K)| {\rm CS}_{\rm FS}\rangle_{B\partial} \ ,
\end{align}
and this implies
\begin{align}
 P_{\partial} |\psi_{\partial }(J,K)\rangle = |\psi_{\partial }(J,K)\rangle \ . 
\end{align}
With a similar analysis, we also find that
\begin{align}
E_{\partial}(c_0)  |\psi_{\partial }(J,K)\rangle = |\psi_{\partial }(J,K)\rangle \qquad \forall c_0 \in C_0\ . 
\end{align}
Namely, the boundary state is gauge symmetric.
Below, we extend this analysis to $M$ operators and $\Gamma$ operators to determine symmetries of the boundary state.\footnote{In the other limit $(J,K) \rightarrow (\infty,0)$, we don't find more symmetry generators for the boundary state as the regime corresponds to a transition from the confinement and Higgs phases in $\mathbb{Z}_2$ Fradkin-Shenker model, which is smooth when the spatial manifold does not have a boundary.}

\subsubsection{$(J,K)\rightarrow (\infty,\infty)$: boundary SPT phase}
In the limit $(J,K)\rightarrow (\infty,\infty)$, the bulk resource state is projected by $\langle \Omega(\infty,\infty)|$, and it satisfies $\langle \Omega(\infty,\infty)| \Gamma_{B}(\bm c_2) = \langle \Omega(\infty,\infty)|$ and $\langle \Omega(\infty,\infty)| M_{B}(\bm z_2) = \langle \Omega(\infty,\infty)|$.
Therefore, on the boundary we have the following symmetries:
\begin{align}
M_{\partial}(z_1) |\psi_{\partial }(\infty,\infty)\rangle &= |\psi_{\partial }(\infty,\infty)\rangle  \quad (z_1 \in Z_1)\, , \\
\Gamma_{\partial}(\gamma) |\psi_{\partial }(\infty,\infty)\rangle &= |\psi_{\partial }(\infty,\infty)\rangle \quad (\gamma \in C_1)\, .  
\end{align}
The latter implies the former. 
We also have $P_{\partial}$ and $E_{\partial}(c_0)$ symmetries. 
In particular, $\Gamma_\partial(\gamma)=1$ and $E_\partial(c_0)=1$ contain the stabilizer conditions for the cluster state, so the boundary induced state is uniquely determined.
This is the two-dimensional cluster state, 
\begin{align} \label{eq:boundary-inf-inf}
|\psi_{\partial} (\infty, \infty) \rangle \sim  |{\rm gCS}_{(2,1)}\rangle \, ,
\end{align}
up to Hadamard transforms.
This state possesses an SPT order protected by the $\mathbb{Z}^{[0]}_2 \times \mathbb{Z}^{[1]}_2$ symmetry, as explained in Section~\ref{sec:SPT} (see also Refs.~\cite{Yoshida:2015cia,2023ScPP...14..129S}).
This state is also the ground state of the Hamiltonian~\eqref{eq:FS-Hamiltonian} in the limit $(\lambda,g) \longrightarrow (\infty, 0)$ (or $(\infty, \text{finite})$), which corresponds to the Higgs phase~\cite{PhysRevD.19.3682, fradkin_2013,verresen2022higgs}.

\subsubsection{$(J,K)\rightarrow (0,\infty)$: boundary deconfining phase and effective bulk SPT}

In the limit $(J,K)\rightarrow (0,\infty)$, stabilizer conditions in the vicinity of the boundary determines the boundary induced state as
\begin{align}
|\psi_{\partial} (0, \infty) \rangle \sim  |+\rangle^{\Delta_0} \otimes |TC_{3,2}\rangle \, .
\end{align}
This can be shown by first noting that the projector $\langle 0|$ on the bulk matter 1-cells forces the boundary 0-cells to be the product state. 
Then, we notice that the $M_\partial$ and $E_\partial$ symmetries fix the boundary induced state to
one of the toric code ground states.
The boundary state
is one of the ground states of the Hamiltonian~\eqref{eq:FS-Hamiltonian} in the limit $(\lambda,g) \longrightarrow (0,0)$, which is in the deconfining phase (the topologically ordered phase) of the Fradkin-Shenker model.

The appearance of an anomalous state at the boundary indicates existence of an SPT state in the bulk.
This is precisely the case in $(J,K)\rightarrow (0,\infty)$. 
In graph states, a $Z$ measurement removes the qubits from the graph, see Ref.~\cite{2004PhRvA..69f2311H} for example. 
As  $\langle \Omega(0,\infty)|$ contains $Z$ basis projection on bulk (matter) 1-cells, {\it i.e.,} $Z_{\bm \sigma_1}=1$, the bulk resource state after a partial projection becomes a state
\begin{align}
\Big(\bigotimes_{\bm \sigma_1 \in \bm \Delta_1} \langle 0| \Big) |{\rm CS}_{\rm FS} \rangle_{B\partial} 
\sim   |+\rangle^{\bm \Delta_0} \otimes | {\rm gCS}_{(3,2)}\rangle\!\rangle \, .
\end{align}
The state $ | {\rm gCS}_{(3,2)}\rangle\!\rangle$ possesses the $\mathbb{Z}^{[1]}_2 \times \mathbb{Z}^{[1]}_2$ SPT order, and a projection of the bulk wave function with the $X$ basis produces the toric code at the boundary~\cite{RBH}.

\subsubsection{$(J,K)\rightarrow (0,0)$: boundary confining phase}
Finally, as  $\langle \Omega(0,0)|$ contains $\widetilde{Z}$ projections on bulk (gauge) 2-cells and $Z$ projections on bulk (matter) 1-cells, {\it i.e.,} $\widetilde{Z}_{\bm \sigma_2}=1$ and ${Z}_{\bm \sigma_1}=1$, all the qubits in the graph state are disconnected. 
In the limit, the boundary state is thus the product state
\begin{align}
|\psi_{\partial} (0, 0) \rangle \sim  |+\rangle^{\Delta_0} \otimes |+\rangle^{\Delta_1}  \, .
\end{align}
Thus the boundary state belongs to the confining phase (the trivially ordered phase). 
As we bring the parameters finite, we still have the $P_{\partial}$ and $E_{\partial}$ symmetries.

\subsubsection{Measurement-induced phases}

Here, we interpret the projection by $\langle\Omega (J,K)|$ as a measurement. We argue that appearance of different boundary induced states by projections in the bulk can be seen as a type of measurement-induced phases, where different phases from the entanglement perspective appears depending on different measurement patterns in the bulk.

First, the limit $(J,K)\rightarrow (0,\infty)$ motivates us to consider the following procedure involving measurements in the bulk. We first prepare the state $|\text{CS}_\text{FS}\rangle_{B\partial}$. Then, according to~\eqref{eq:prod_state_limits}, we measure the matter 0-cells and the gauge 1- and 2-cells in the $X$ basis, while we measure the matter 1-cells in the $Z$ basis. 
We obtain a state at the boundary, which is the state $|\psi_\partial (0,\infty)\rangle$ up to by-product operators.  
The output state is a stabilizer state, differences in bulk measurement outcomes can be accounted for by modifying $\pm 1$ phase factors in stabilizers,
and the entanglement entropy in a stabilizer state does not depend on the phase factor of each stabilizer.
Hence, the topological entanglement entropy~\cite{2006PhRvL..96k0404K, 2006PhRvL..96k0405L} for the boundary induced states with any measurement outcome will exhibit the topological constant term. 
One could instead clean up the by-product operators by a post processing. 

The limit $(J,K)\rightarrow (\infty, \infty)$ motivates us to measure all the bulk qubits in the $X$ basis. 
The boundary state induced by measurements then is the 2d SPT state up to by-product operators. 
One can clean up the by-product operators and compute, for example, the SPT string order parameter. 
The other limit $(J,K)\rightarrow (0, 0)$ is related to measuring the matter 1-cells and the gauge 2-cells in the $Z$ basis instead, and the boundary measurement-induced state will have no entanglement.

\section{Summary and discussion}
\label{sec:discussion}

In this paper, we demonstrated an anomaly inflow mechanism between the deconfining phase (generalized toric code) of the lattice gauge theory~$M_{d,n}$ (Wegner model) and a symmetry-protected topological (SPT) state in one higher dimension.
The SPT state is a cluster state~$\mathrm{gCS}_{d,n}$ that arises as a resource state for the measurement-based quantum simulation (MBQS) scheme introduced in~\cite{2023ScPP...14..129S}.
 See Table~\ref{tab:models} for a summary of the models.
We formulated anomaly inflow directly on the lattice rather than in the continuum limit, and in terms of defects, whose world-volumes consist of excitations (violations of stabilizer conditions) and operators that generate symmetries.
Such a formulation is useful for bridging between the study of anomaly inflow in continuum field theories (much of high energy theory) to that in lattice models (lattice gauge theory and condensed matter physics.)
Our demonstration of anomaly inflow involved relating the bulk and boundary statement by bulk measurement.\footnote{
A field-theoretic description of how anomalous states are obtained from measuring a part (subgroup) of SPT states was given in Ref.~\cite{2023arXiv231017740L}.
}
 In addition, we used the cluster state $\mathrm{gCS}_{d,n}$ and its defining entangler as tools to study the Kramers-Wannier dualities in Wegner models in classical and quantum formulations.

We also presented an MBQS scheme for a gauge theory with matter (Fradkin-Shenker model), which extends the simulation protocols for pure gauge theories and a free fermion model given in~\cite{2023ScPP...14..129S}.
 We also proposed a scheme to protect the gauge invariance using measurement outcomes in our MBQS.
Remarkably, our method does not necessitate non-trivial classical side processings to determine correction operators, which would be an advantage in an experiment where the run time is a critical metric that determines the fidelity of the simulation under decoherence.
Formulation of MBQS schemes for more general lattice models and their implementation on experimental systems are important future problems.

The Wegner models $M_{d,n}$ and the Fradkin-Shenker model exhibit non-trivial phase structures in the space of coupling constants~\cite{Wegner,PhysRevD.19.3682, 2010PhRvB..82h5114T,fradkin_2013,verresen2022higgs}.
It would be interesting to study the phases in more detail using the strange correlator representations involving cluster states, which may be constructed experimentally.

{\green Furthermore,}
we analyzed the overlap formula for the Fradkin-Shenker model and interpreted it as a type of measurement-indeced phases. 
{\green Let us offer}
a potential measurement-induced phase transition problem.
Motivated by our overlap formula, one can study a problem where 
(i) we prepare the MBQS resource state \eqref{eq:FS-resource-state},
(ii) we measure the bulk matter 0- and gauge 1-cells in the $X$ basis with the unit probability, (iii) motivated by \eqref{eq:prod_state_expansion-J}, we measure bulk matter 1-cells in the $Z$ basis with probability $p_J$ and in the $X$ basis with probability $1-p_J$,
(iv) motivated by \eqref{eq:prod_state_expansion-K}, we measure bulk gauge 2-cells in the $Z$ basis with probability $p_K$ and in the $X$ basis with probability $1-p_K$, (v) we compute measures to detect SPT phases (such as the SPT string order parameter) or a topological ordered state (such as the topological entanglement entropy~\cite{2006PhRvL..96k0404K, 2006PhRvL..96k0405L}), and (vi) we repeat (i)-(v) to sample the values of entanglement measures. Finally, we average them over samples.
We remark that the series (i)-(v) is a constant-depth procedure and the entanglement entropy
does not depend on measurement outcomes.\footnote{
Each output state is a stabilizer state with outcome-dependent phase factors, and the entanglement entropy in a stabilizer state does not depend on the phase factor of each stabilizer.
}
We speculate that the parameter space with $(p_J,p_K)$ has an interesting phase diagram involving the trivial, topological, and SPT ordered phases, possibly with phase boundaries.\footnote{
In \cite{2022PhRvB.106n4311L}, authors considered measuring a cluster state (which is different from ours) with measurement bases randomly chosen from $X$ or $Z$ in the bulk. 
The boundary state was shown to experience an entanglement transition between a volume-law phase and a area-law phase as the probability $p_x$ (the probability $p_z=1-p_x$) of the $X$ basis (the $Z$ basis) being selected is varied. See also Refs.~\cite{2023PhRvR...5d3069G, 2024PhRvB.109l5148N,2023arXiv231116651K}. } 
The protocol we proposed can be efficiently simulated on classical computers as well as on digital quantum computers. It would be interesting to implement numerical simulations and compare the results against the phase diagram of the Fradkin-Shenker model~\cite{2010PhRvB..82h5114T, verresen2022higgs}.

{\green Finally, the construction of the MBQS resource state in~\cite{2023ScPP...14..129S} and this paper can be generalized~\cite{our-fracton-paper} to arbitrary lattice models based on Calderbank-Shor-Steane codes~\cite{calderbank1996good,1996RSPSA.452.2551S} using the procedure called foliation~\cite{2016PhRvL.117g0501B}.}

\section*{Acknowledgements}

The research of T.\,O. was supported in part by Grant-in-Aid for Transformative Research Areas (A) ``Extreme Universe'' No.\,21H05190 and by JST PRESTO Grant Number JPMJPR23F3.
APM and HS were partially supported by the Materials Science and Engineering Divisions, Office of Basic Energy Sciences of the U.S. Department of Energy under Contract No. DESC0012704.
We are grateful to the long-term workshop YITP-T-23-01 held at YITP, Kyoto University, where a part of this work was done.

\appendix

\section{Anomaly inflow in the continuum description}
\label{sec:inflow-continuum}

In this appendix, we study the $\mathbb{Z}_N$ generalization of the Wegner model~$M_{d,n}$. 
In the appendix of~\cite{Burnell:2021reh}, the continuum description of the anomaly inflow for the Wegner model~$M_{3,2}^{(N)}$ was given.
Here we review and extend it to $M_{d,n}$ with general $d$ and $n$.
The continuum description is compared with the corresponding lattice models in Section~\ref{sec:SPT}.
Our discussion is schematic because the continuum is not the main focus of this work.
We refer the reader to~\cite{Burnell:2021reh} for more details.

It is expected that the low-energy physics of the deconfining phase of the $\mathbb{Z}_N$ Wegner model~$M_{d,n}^{(N)}$ is described by the BF theory whose action is
\begin{equation}\label{eq:BF-action}
S_{\rm BF} = \frac{iN}{2\pi}\int A_{n-1} \wedge dA_{d-n} \,,
\end{equation}
where $A_{n-1}$ and $A_{d-n}$ are (distinct even when $n-1=d-n$) gauge fields given locally by $(n-1)$- and $(d-n)$-forms.
The theory has two types of defects corresponding to the symmetry generators in (\ref{eq:sym-generators-Wegner}):
\begin{equation} \label{eq:BF-defects}
\exp  i    \oint_{C_{n-1}} A_{n-1} \,,
\quad
\exp  i    \oint_{C_{d-n}} A_{d-n} \,.
\end{equation}
The equations of motion show that one type of defect induces a non-trivial phase for the other when the cycles $C_{n-1}$ and $C_{d-n}$ are linked, implying a mixed 't Hooft anomaly between the symmetries $\mathbb{Z}_N^{[n-1]}$ and $\mathbb{Z}_N^{[d-n]}$ that the defects generate.

We characterized the 't Hooft anomaly in terms of defects above.
Another common characterization is in terms of background gauge transformations.
We recall that the two characterizations are equivalent.
To see this, let us consider the functional
\begin{equation}
\begin{split}
\label{eq:Z-B-B-boundary}
&Z_{\rm bdry}[B_{n},B_{d-n+1}] \\
&\sim \int \mathcal{D}A_{n-1} \mathcal{D}A_{d-n} e^{-S_{\rm BF}}  \\
&\qquad \times \exp \frac{iN}{2\pi} \int (A_{n-1}\wedge B_{d-n+1} + A_{d-n}\wedge B_{n}) \,,
\end{split}
\end{equation}
where $B_{n}$ and $B_{d-n+1}$ are the background gauge fields for $\mathbb{Z}_N^{(n-1)}$ and $\mathbb{Z}_N^{(d-n)}$, respectively.
The mixed anomaly manifests itself in the non-invariance of partition function under simultaneous background gauge transformations.
The two characterizations are equivalent because the background gauge fields are Poincar\'e dual to the world-volumes of the defects: 
\begin{equation}
    B_n  = \frac{2\pi}{N}\delta(C_{d-n}) 
\quad\text{and}\quad
B_{d-n+1}  = \frac{2\pi}{N}\delta(C_{n-1}) \,,
\end{equation}
where $\delta(C_{d-n})$ and $\delta(C_{n-1})$ are the differential forms that have delta function supports along the cycles $C_{d-n}$ and $C_{n-1}$.
The background transformations then deform the world-volumes and cause the defects to pass each other, producing a phase.

In the continuum description, the mixed anomaly can be canceled via anomaly inflow~\cite{Callan:1984sa} by coupling the BF theory to an SPT phase in one-higher dimension.
The bulk SPT phase that cancels the anomaly of the BF theory via anomaly inflow is roughly given by
\begin{equation}\label{eq:Z-SPT-Wegner-partition-function}
Z_{\rm bulk}[ \bm B_{n}, \bm B_{d-n+1}] \sim \exp \frac{i N}{2\pi} \int   \bm B_{n}\wedge \bm B_{d-n+1}  \,,
\end{equation}
where  $\bm B_{n}$ and $\bm B_{d-n+1}$ are the bulk extensions of the corresponding fields on the boundary denoted by the same symbols.
In the absence of a boundary, the bulk partition function~(\ref{eq:Z-SPT-Wegner-partition-function}) is invariant under background gauge transformations.
When there is a boundary and the bulk theory is coupled to the boundary theory, the bulk partition varies by a boundary term, which cancels the variation of the boundary partition function.

In the bulk, the background fields are Poincar\'e dual to the topological defects that generate the generalized symmetry~$\mathbb{Z}_N^{[n-1]}\times \mathbb{Z}_N^{[d-n]}$ protecting the bulk system: 
\begin{equation}\label{eq:B-C-bulk}
 \bm B_n =\frac{2\pi}{N}\delta(\bm C_{d-n+1})
\quad\text{and}\quad
\bm B_{d-n+1} = \frac{2\pi}{N}\delta(\bm C_n)\,, 
\end{equation}
where  $\bm C_{d-n+1}$ and $\bm C_n$ are the world-volumes of the defects in the bulk.
Then, the partition function~(\ref{eq:Z-SPT-Wegner-partition-function}) reduces to a phase given by the intersection number between the world-volumes of the two types of defects.

Coming back to the boundary theory, it is easy to see that the partition function is given by the linking number~\cite{Putrov:2016qdo} , as in the lattice model, when the defect world-volumes are homologically trivial.
Indeed, when $C_{d-n}=\partial C_{d-n+1}$ and $C_{n-1} = \partial C_n$, we have $\delta(C_{d-n}) = d \delta(C_{d-n+1})$ and $\delta(C_{n-1}) = d \delta(C_{n})$ as can be proved by considering local descriptions using coordinates.
Field re-definition $A_{n-1} \rightarrow A_{n-1}+ \frac{2\pi}{N} \delta(C_{d-n+1})$ and $A_{d-n} \rightarrow A_{d-n} + \frac{2\pi}{N} \delta(C_n)$ shows that
\begin{equation}
\begin{split}
&\frac{Z_\text{bdry}[B_n,B_{d-n+1}]}{Z_\text{bdry}[0,0]} \\
&= \exp \frac{2\pi i}{N} \int \delta(C_{d-n+1})\wedge d \delta(C_n) \\
&= \exp \frac{2\pi i}{N} \#( C_{d-n+1}\cap \partial C_n) \,.
\end{split}
\end{equation}

\input{error-correction-appendix}

\bibliography{refs}

\end{document}

%% file: error-correction-appendix.tex
\section{Technical details in protecting the gauge invariance of the Fradkin-Shenker model}
\label{sec:technical-FS-gauge-protection}

In the presence of error, the simulated state at the time slice $z=j$ evolves by measurements up to $z=j+1$ as 
\begin{widetext}
\begin{align} \label{eq:state-evolution-error-explicit}
|\phi \rangle_{\bm \Delta^{z=j+1}_0 \cup \bm \Delta^{z=j+1}_1}
=
& (\pm 1) \times 
\Bigg( 
\prod_{\sigma_1 \in \Delta_1 }
\widetilde{X}(\sigma_1 )^{\tilde{s}(\sigma_1 \times [j,j+1]) + a( \tilde{\bm e}^Z_2; \sigma_1 \times [j,j+1]) } 
\exp\big[i (-1)^{ a( \tilde{\bm e}^X_2; \sigma_1 \times [j,j+1]) } \xi \widetilde{X}(\sigma_1 ) \big] \nonumber \\
&\times \prod_{\sigma_0 \in \Delta_0} X(\sigma_0 )^{s(\sigma_0 \times [j,j+1]) + a( \bm e^Z_1; \sigma_0 \times [j,j+1]) } \exp\big[i (-1)^{ a( \bm e^X_1; \sigma_0 \times [j,j+1]) }\xi  X(\sigma_0 ) \big]   \, 
\nonumber \\
& 
\times \prod_{\sigma_0 \in \Delta_0} \frac{1+ (-1)^{\tilde{s}(\sigma_0 \times [j,j+1]) + a( \tilde{\bm e}^Z_1; \sigma_0 \times [j,j+1]) } 
X(\sigma_0) \widetilde{X}(\partial^*\sigma_0)}{2}
\, 
\nonumber \\ 
& \times 
\prod_{\sigma_0 \in \Delta_0} 
Z(\sigma_0)^{s(\sigma_0 \times \{j\}) + a(\bm e^Z_0;\sigma_0 \times \{j\})  }
\prod_{\sigma_1 \in \Delta_1} 
\widetilde{Z}(\sigma_1)^{\tilde{s}(\sigma_1 \times \{j\}) + a( \tilde{\bm e}^Z_1;\sigma_1 \times \{j\})  } \, 
\nonumber \\
& \times \prod_{\sigma_2 \in \Delta_2 } 
\widetilde{Z}(\partial \sigma_2 )^{\tilde{s}(\sigma_2 \times \{j\}) +a(\tilde{\bm e}^X_2; \sigma_2 \times \{j\}) }
\exp \big[ i (-1)^{ a(\tilde{\bm e}^Z_2; \sigma_2 \times \{j\})} \xi  \widetilde{Z}(\partial \sigma_2 ) \big]
\, 
\nonumber \\
& \times 
\prod_{\sigma_1 \in \Delta_1 }
\big( \widetilde{Z}(\sigma_1)Z(\partial \sigma_1 ) \big)^{s(\sigma_1 \times \{j\}) +a({\bm e}^X_1; \sigma_1 \times \{j\}) } 
\exp \big[i (-1)^{ a({\bm e}^Z_1; \sigma_1 \times \{j\})} \xi\widetilde{Z}(\sigma_1)Z(\partial \sigma_1 ) \big]
\Bigg) \, 
\nonumber \\
&  \times |\phi \rangle_{\bm \Delta^{z=j}_0 \cup \bm \Delta^{z=j}_1} \, ,
\end{align}
where the cell dependence of $\{\xi\}$ is omitted as in the main text, but that of $\{s\}$ is explicitly written. 
Note that the first two lines involving $X$ operators come from the measurement steps (6) and (7), where the errors lead to extra $Z$ operators due to \eqref{eq:B-error}, and the conjugation by the Hadamard transform.
\end{widetext}

\subsection{Details of the time evolution}
\label{app:time-evolution-error}

\begin{widetext}
Equation~\eqref{eq:state-evolution-error-explicit} teaches us in detail how each piece in $|\phi\rangle_{\bm \Delta^{z=j}_{0} \cup \bm \Delta^{z=j}_{1}} = \pm \mathcal{E}^{(j)} \mathcal{O}^{(j)}_\text{bp} | \phi^{\text{error}} (t)\rangle$ evolves.
The right hand side can be arranged as 
\begin{align} 
&  
\Bigg( 
\prod_{\sigma_1 \in \Delta_1 }
\widetilde{X}(\sigma_1 )^{\tilde{s}(\sigma_1 \times [j,j+1]) + 
a( \tilde{\bm e}^Z_2; \sigma_1 \times [j,j+1]) } 
\exp\big[i (-1)^{ a( \tilde{\bm e}^X_2; \sigma_1 \times [j,j+1]) } \xi \widetilde{X}(\sigma_1 ) \big] \nonumber \\
&\times \prod_{\sigma_0 \in \Delta_0} X(\sigma_0 )^{s(\sigma_0 \times [j,j+1]) + a( \bm e^Z_1; \sigma_0 \times [j,j+1]) } \exp\big[i (-1)^{ a( \bm e^X_1; \sigma_0 \times [j,j+1]) }\xi  X(\sigma_0 ) \big]   \, 
\nonumber \\
& 
\times \prod_{\sigma_0 \in \Delta_0} \frac{1+ (-1)^{\tilde{s}(\sigma_0 \times [j,j+1]) + a( \tilde{\bm e}^Z_1; \sigma_0 \times [j,j+1]) } 
X(\sigma_0) \widetilde{X}(\partial^*\sigma_0)}{2}
\, 
\nonumber \\ 
& \times 
\prod_{\sigma_0 \in \Delta_0} 
Z(\sigma_0)^{s(\sigma_0 \times \{j\}) + a(\bm e^Z_0;\sigma_0 \times \{j\})  }
\prod_{\sigma_1 \in \Delta_1} 
\widetilde{Z}(\sigma_1)^{\tilde{s}(\sigma_1 \times \{j\}) + a( \tilde{\bm e}^Z_1;\sigma_1 \times \{j\})  } \, 
\nonumber \\
& \times \prod_{\sigma_2 \in \Delta_2 } 
\widetilde{Z}(\partial \sigma_2 )^{\tilde{s}(\sigma_2 \times \{j\}) +a(\tilde{\bm e}^X_2; \sigma_2 \times \{j\}) }
\exp \big[ i (-1)^{ a(\tilde{\bm e}^Z_2; \sigma_2 \times \{j\})} \xi  \widetilde{Z}(\partial \sigma_2 ) \big]
\, 
\nonumber \\
& \times 
\prod_{\sigma_1 \in \Delta_1 }
\big( \widetilde{Z}(\sigma_1)Z(\partial \sigma_1 ) \big)^{s(\sigma_1 \times \{j\}) +a({\bm e}^X_1; \sigma_1 \times \{j\}) } 
\exp \big[i (-1)^{ a({\bm e}^Z_1; \sigma_1 \times \{j\})} \xi\widetilde{Z}(\sigma_1)Z(\partial \sigma_1 ) \big]
\Bigg) \times \mathcal{E}^{(j)} \mathcal{O}^{(j)}_\text{bp}|\phi^\text{error}(t) \rangle \nonumber 
\end{align}
\begin{align}
&= 
\pm 1 \times 
\Bigg( 
\prod_{\sigma_1 \in \Delta_1 }
\widetilde{X}(\sigma_1 )^{ a( \tilde{\bm e}^Z_2; \sigma_1 \times [j,j+1]) } 
\prod_{\sigma_0 \in \Delta_0} X(\sigma_0 )^{a( \bm e^Z_1; \sigma_0 \times [j,j+1]) }  \prod_{\sigma_0 \in \Delta_0} 
Z(\sigma_0)^{a(\bm e^Z_0;\sigma_0 \times \{j\})  }
\prod_{\sigma_1 \in \Delta_1} 
\widetilde{Z}(\sigma_1)^{a( \tilde{\bm e}^Z_1;\sigma_1 \times \{j\})  } \, 
\nonumber \\
& \times \prod_{\sigma_2 \in \Delta_2 } 
\widetilde{Z}(\partial \sigma_2 )^{a(\tilde{\bm e}^X_2; \sigma_2 \times \{j\}) }
\prod_{\sigma_1 \in \Delta_1 }
\big( \widetilde{Z}(\sigma_1)Z(\partial \sigma_1 ) \big)^{a({\bm e}^X_1; \sigma_1 \times \{j\}) } \mathcal{E}^{(j)}
\Bigg) \nonumber \\
&\times 
\Bigg( 
\prod_{\sigma_1 \in \Delta_1 }
\widetilde{X}(\sigma_1 )^{\tilde{s}(\sigma_1 \times [j,j+1])  } 
\prod_{\sigma_0 \in \Delta_0} X(\sigma_0 )^{s(\sigma_0 \times [j,j+1])  }\prod_{\sigma_0 \in \Delta_0} 
Z(\sigma_0)^{s(\sigma_0 \times \{j\})   }
\prod_{\sigma_1 \in \Delta_1} 
\widetilde{Z}(\sigma_1)^{\tilde{s}(\sigma_1 \times \{j\})   } \, 
\nonumber \\
& \times \prod_{\sigma_2 \in \Delta_2 } 
\widetilde{Z}(\partial \sigma_2 )^{\tilde{s}(\sigma_2 \times \{j\})  }
\prod_{\sigma_1 \in \Delta_1 }
\big( \widetilde{Z}(\sigma_1)Z(\partial \sigma_1 ) \big)^{s(\sigma_1 \times \{j\})  } \mathcal{O}^{(j)}_\text{bp}
\Bigg) \nonumber \\ 
& 
\times \Bigg( \prod_{\sigma_1 \in \Delta_1} \exp\big[i (\pm 1) \xi \widetilde{X}(\sigma_1 ) \big] 
\prod_{\sigma_0 \in \Delta_0}  \exp\big[i (\pm 1)\xi  X(\sigma_0 ) \big]  \nonumber \\
&
\times \prod_{\sigma_2 \in \Delta_2 } 
\exp \big[ i (\pm 1)   \xi \widetilde{Z}(\partial \sigma_2 ) \big] 
\prod_{\sigma_1 \in \Delta_1 }
\exp \big[i (\pm 1) \xi \widetilde{Z}(\sigma_1)Z(\partial \sigma_1 ) \big]
\Bigg)
|\phi^\text{error}(t) \rangle \nonumber \\
& = \pm 1 \times  \mathcal{E}^{(j+1)} \cdot \mathcal{O}^{(j+1)}_\text{bp} |\phi^\text{error}(t+\delta t) \rangle \, ,
\end{align}
\end{widetext}
where we moved the error-dependent Pauli operators to the first parenthesis and combined them with the former error $\mathcal{E}^{(j)}$, the $s$-dependent Pauli operators to the second parenthesis and combined them with the former byproduct operator $\mathcal{O}^{(j)}_\text{bp}$.
We also dropped the projectors in the first equality as in any non-vanishing wave function the projectors have to be one.
(We don't observe measurement outcomes that give rise to a vanishing wave function.)
In the second expression, the first parentheses are the new error operator $\mathcal{E}^{(j+1)}$ 
and the second parenthesis is the new byproduct operator $\mathcal{O}^{(j+1)}_\text{bp}$.
The last parenthesis in the second expression consists of exponential operators and they are combined with $|\phi^\text{error}(t) \rangle$ to give $|\phi^\text{error}(t+\delta t) \rangle$.
The signs $(\pm 1)$ in the exponents are due to 
direct changes by the errors (namely, the error-induced factors $(-1)^{ a( \tilde{\bm e}^X_2; \sigma_1 \times [j,j+1])}$,  $(-1)^{ a( \bm e^X_1; \sigma_0 \times [j,j+1]) }$, $(-1)^{ a(\tilde{\bm e}^Z_2; \sigma_2 \times \{j\})}$, and $(-1)^{ a({\bm e}^Z_1; \sigma_1 \times \{j\})}$ in the first expression)
as well as the signs that arise from passing both the error and byproduct Pauli operators to the front.
In the noiseless case, these $(\pm 1)$ signs are completely harmless, as one can absorb them into the adaptive choice of $\{\xi\}$ in measurements according to former measurement outcomes --- a standard technique in MBQC (see {\it e.g.} Section~2.1.2 of \cite{2018AdPhX...361026W} or Section~3.1.1 of \cite{2023ScPP...14..129S}).
Regardless of these signs, it is straightforward to see that $|\phi^\text{error}(t+\delta t) \rangle$ is gauge invariant assuming that $|\phi^\text{error}(t) \rangle$ is also gauge invariant.

\subsection{The outcome-error correlation from the perspective of the simulated state} \label{app:outcome-error-correlation}

Here, we re-derive the relation~\eqref{eq:error-outcome-correlation-3d} using the above expression of the simulated state.
Consider the evolution from $z=j-1$ to $z=j+1$.
Using \eqref{eq:state-evolution-error-explicit} twice, we note that the operator acting on $|\phi \rangle_{\bm \Delta^{z=j-1}_0 \cup \bm \Delta^{z=j-1}_1}$ contains a portion,
\begin{widetext}
{\allowdisplaybreaks
\begin{align}
\cdots &\times\prod_{\sigma_0 \in \Delta_0} \frac{1+ (-1)^{\tilde{s}(\sigma_0 \times [j,j+1]) + a( \tilde{\bm e}^Z_1; \sigma_0 \times [j,j+1]) } 
X(\sigma_0) \widetilde{X}(\partial^*\sigma_0)}{2}
\, 
\nonumber \\ 
& \times 
\prod_{\sigma_0 \in \Delta_0} 
Z(\sigma_0)^{s(\sigma_0 \times \{j\}) + a(\bm e^Z_0;\sigma_0 \times \{j\})  }
\prod_{\sigma_1 \in \Delta_1} 
\widetilde{Z}(\sigma_1)^{\tilde{s}(\sigma_1 \times \{j\}) + a( \tilde{\bm e}^Z_1;\sigma_1 \times \{j\})  } \, 
\nonumber \\
& \times \prod_{\sigma_2 \in \Delta_2 } 
\widetilde{Z}(\partial \sigma_2 )^{\tilde{s}(\sigma_2 \times \{j\}) +a(\tilde{\bm e}^X_2; \sigma_2 \times \{j\}) }
\exp \big[ -i (-1)^{ a(\tilde{\bm e}^Z_2; \sigma_2 \times \{j\})} \xi  \widetilde{Z}(\partial \sigma_2 ) \big]
\, 
\nonumber \\
& \times 
\prod_{\sigma_1 \in \Delta_1 }
\big( \widetilde{Z}(\sigma_1)Z(\partial \sigma_1 ) \big)^{s(\sigma_1 \times \{j\}) +a({\bm e}^X_1; \sigma_1 \times \{j\}) } 
\exp \big[-i (-1)^{ a({\bm e}^Z_1; \sigma_1 \times \{j\})} \xi\widetilde{Z}(\sigma_1)Z(\partial \sigma_1 ) \big] \nonumber \\
&\times \prod_{\sigma_1 \in \Delta_1 }
\widetilde{X}(\sigma_1 )^{\tilde{s}(\sigma_1 \times [j-1,j]) + a( \tilde{\bm e}^Z_2; \sigma_1 \times [j-1,j]) } 
\exp\big[-i (-1)^{ a( \tilde{\bm e}^X_2; \sigma_1 \times [j-1,j]) } \xi \widetilde{X}(\sigma_1 ) \big] \nonumber \\
&\times \prod_{\sigma_0 \in \Delta_0} X(\sigma_0 )^{s(\sigma_0 \times [j-1,j]) + a( \bm e^Z_1; \sigma_0 \times [j-1,j]) } \exp\big[-i (-1)^{ a( \bm e^X_1; \sigma_0 \times [j-1,j]) }\xi  X(\sigma_0 ) \big]   \, 
\nonumber \\
& 
\times \prod_{\sigma_0 \in \Delta_0} \frac{1+ (-1)^{\tilde{s}(\sigma_0 \times [j-1,j]) + a( \tilde{\bm e}^Z_1; \sigma_0 \times [j-1,j]) } 
X(\sigma_0) \widetilde{X}(\partial^*\sigma_0)}{2} \times \cdots \, .
\end{align}}
The wave function vanishes unless the two projectors upon commuting through the operators in the second line are identical.
To understand a constraint on the measurement outcomes and error chains, we move the projectors and fuse them together. 
As we move one of them, the operator $X(\sigma_0) \widetilde{X}(\partial^*\sigma_0)$ picks up a phase from the commutation with the operators $Z(\sigma_0)$ and $\widetilde{Z}(\sigma_1)$.
It follows that 
\begin{align}
&\tilde{s}(\sigma_0 \times [j,j+1]) 
+ a( \tilde{\bm e}^Z_1; \sigma_0 \times [j,j+1]) 
+ \tilde{s}(\sigma_0 \times [j-1,j]) + a( \tilde{\bm e}^Z_1; \sigma_0 \times [j-1,j]) \nonumber \\ 
&= 
+ s(\sigma_0 \times \{j\}) + a(\bm e^Z_0;\sigma_0 \times \{j\})  
+ \sum_{ \sigma_1 \in \Delta_1} 
\tilde{s}(\sigma_1 \times \{j\}) a (\partial^*\sigma_0; \sigma_1)
+ \sum_{ \sigma_1 \in \Delta_1}  a( \tilde{\bm e}^Z_1;\sigma_1 \times \{j\})  
a (\partial^*\sigma_0; \sigma_1) \, 
\end{align}
for all $\sigma_0 \in \Delta_0$.
This coincides with the relation~\eqref{eq:error-outcome-correlation-3d}.
\end{widetext}